\documentclass[aip,reprint]{revtex4-1}

\usepackage[utf8]{inputenc}
\usepackage[T1]{fontenc}
\usepackage{mathptmx}
\usepackage{etoolbox}

\usepackage{graphicx}% Include figure files
\graphicspath{ {./Images/} }
\usepackage{caption}
\usepackage{subcaption}
\usepackage{bm}
\usepackage{amsmath}
\usepackage[version=4]{mhchem}
\usepackage{tabularx,ragged2e}
\usepackage{booktabs}
\usepackage{rotating}
\usepackage{adjustbox,dcolumn}
\usepackage{threeparttable}
\usepackage{hyperref}

\draft % marks overfull lines with a black rule on the right

\begin{document}

%\preprint{AIP/123-QED}

%\preprint{}

\title{A novel numerical tool to study electron energy distribution functions of spatially-anisotropic and non-homogeneous ECR plasmas} %Title of paper



\author{B. Mishra}
\email[Corresponding address:]{mishra@lns.infn.it}
\affiliation{INFN - LNS, via S. Sofia 62, 95123 Catania, Italy}
\affiliation{Dipartimento di Fisica e Astronomia "Ettore Majorana", Universit\`{a} degli Studi di Catania, 95123 Catania, Italy}

\author{A. Pidatella}
%\thanks{}
\affiliation{INFN - LNS, via S. Sofia 62, 95123 Catania, Italy}

\author{S. Biri}
\affiliation{Institute for Nuclear Research (ATOMKI), Bem t\'{e}r 18/C, H-4026 Debrecen, Hungary}

\author{A. Galat\`{a}}
\affiliation{INFN - LNL, Viale dell'Universit\`{a}, 2, 35020 Legnaro, Italy}

\author{E. Naselli}
\affiliation{INFN - LNS, via S. Sofia 62, 95123 Catania, Italy}

\author{R. R\'{a}cz}
\affiliation{Institute for Nuclear Research (ATOMKI), Bem t\'{e}r 18/C, H-4026 Debrecen, Hungary}

\author{G. Torrisi}
\author{D. Mascali}
\affiliation{INFN - LNS, via S. Sofia 62, 95123 Catania, Italy}


\date{\today}

\begin{abstract}
A numerical tool for analysing spatially anisotropic electron populations in electron cyclotron resonance (ECR) plasmas has been developed, using a trial-and-error electron energy distribution function (EEDF) fitting method. The method has been tested on space-resolved warm electrons in the energy range $2-20\,\mathrm{keV}$, obtained from self-consistent simulations modelling only electron dynamics in ECR devices, but lacked real-world validation. For experimentally benchmarking the method, we attempted to numerically reproduce the experimental X-ray emission spectrum measured from an argon plasma. Results of this analysis have provided crucial information about density and temperature of warm electrons, and competing distributions of warm and hot electron components. This information can be fed back to simulation models to generate more realistic data. Subsequent application of the numerical tool as described to the improved simulation data can result in continuous EEDFs that reflect the nature of charge distributions in anisotropic ECR plasmas. These functions can be also applied to electron dependent reactions, in order to reproduce experimental results, like those concerning space-dependent K$\alpha$ emissions.     
\end{abstract}


\maketitle %\maketitle must follow title, authors, abstract and \pacs

%\section{}
%\label{}
%\subsection{}
%\subsubsection{}
\section{Introduction}
\label{intro}
\begin{figure}
\resizebox{1.00\columnwidth}{!}{\includegraphics{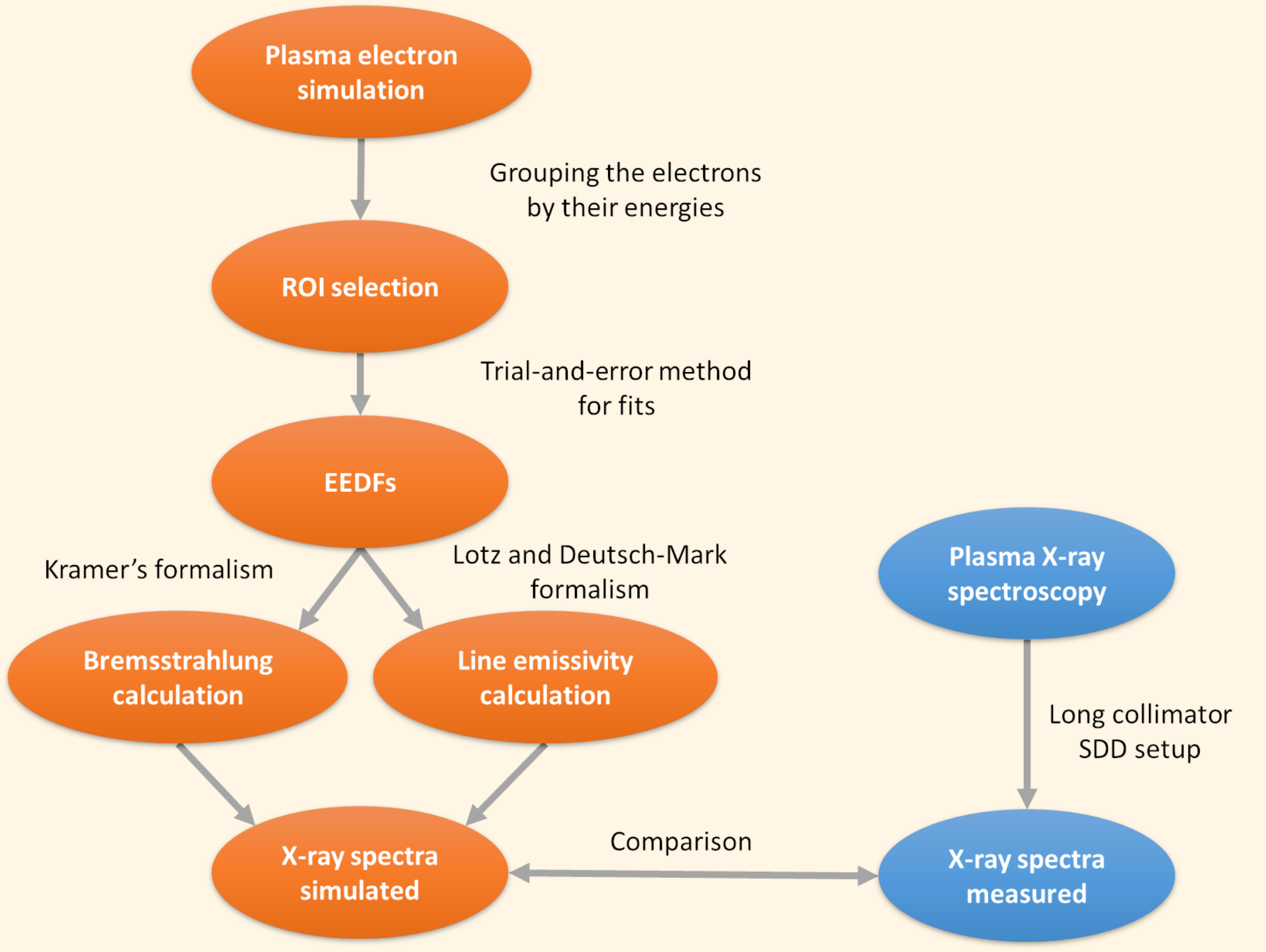}}
\caption{Flowchart outlining the methodology adopted.}
\label{fig:FlowChart}
\end{figure}

Electron Cyclotron Resonance Ion Sources (ECRIS) are some of the most versatile devices used to generate and supply highly-charged ion beams of variable intensity to high-energy accelerators \cite{Ref1,Ref2}. They are based on the ECR heating of plasma electrons via interactions with a microwave radiation field, as a consequence of which plasmas might be highly inhomogeneous and have strongly anisotropic energy distribution. These plasmas are characterized by an EEDF usually consisting of two to three components: cold (average energy 10$\div$100 eV), warm (1$\div$10 keV), and hot electrons (10 keV$\div$1 MeV). Since the EEDF and the electron density directly control reaction rates, understanding their link with electron dynamics is of primary importance to improve the performance of these devices. In this context, several \emph{direct} and \emph{indirect} diagnostic techniques have been developed over the years like microwave interferometry \cite{Ref3}, X-ray spectroscopy \cite{Ref4,Ref5}, visible light observations \cite{Ref6}, and small-size electrostatic probes \cite{Ref7,Ref8}.
While direct diagnostics, for instance Langmuir probes, are a convenient means to measure the EEDF and plasma parameters, they always run the risk of perturbing the plasma, producing uncertainties in measurements. Indirect diagnostics, like those built around deduction of plasma parameters through analysis of emitted radiation, are obviously free of plasma perturbations, but are capable of looking at only a selected energy range of the system at a time. X-ray bremsstrahlung spectroscopy and electron cyclotron emission are established indirect diagnostic approaches to probe the hot electron population ($T_{e}>50\,\mathrm{keV}$) \cite{Ref9,Ref10,Ref11}, while information about the cold electrons ($T_{e}\lesssim 10\,\mathrm{eV}$) has been deduced through optical emission spectroscopy \cite{Ref12,Ref13}. However, data on the warm component is harder to come across, and inconclusive statements on the shape of the EEDF of strongly non-Maxwellian plasmas are offered by using complicated spectra deconvolution. This is quite unfortunate since these electrons are primarily responsible for the sequential ionization process that leads to high charge states in an ECRIS.
\\\\
Recently, efforts have been put into analysis of energetic electrons spanning the boundary between warm and hot, using high-resolution and spatially-resolved X-ray spectroscopy involving quasi-optical methods like pinhole cameras. The utility of these techniques was first demonstrated in a series of pioneering experiments at ATOMKI, Debrecen (Hungary) in 2002-03 \cite{Ref5,Ref15}, and later employed in joint measurements by the ATOMKI and INFN-LNS groups \cite{Ref16,Ref17,Ref18} to better understand the structural evolution of intermediate energy electrons as a function of wave-to-plasma coupling and resonance actions in the plasma chamber. If the contribution of warm electrons to the measured spectra can be ascertained, the study can be used as a powerful tool to verify density, temperature and EEDF of said electrons as a function of their position in the plasma.\\\\
% Data analysis of the K$\alpha$ X-ray fluorescence emission map, acquired from the aforementioned X-ray imaging experiments \cite{Ref17}, lacked a comprehensive theoretical analysis, capable of modelling the plasma EEDF to correlate the emission map with the plasma electron spatial structure and properties. 
The present paper deals with a new method for step-by-step determination of possible continuous EEDFs that can effectively describe the space-dependent properties of a minimum-B ECRIS plasma electron population. The use of the method is demonstrated through EEDF estimation of warm electrons ($T_{e}\sim1-10\,\mathrm{keV}$) from self-consistent numerical simulations performed with a code jointly developed by the ion source groups of INFN-LNS and LNL \cite{Ref23}. Additionally, analysis of line and bremsstrahlung emissivity density measured from an argon ECR plasma during the aforementioned experimental campaign \cite{Ref17} has been also made, to establish whether the contribution these warm electrons is sufficiently large. As it turns out, the \emph{volumetric} density and temperature as estimated from this secondary analysis lies outside the energy range of the simulated electrons, implying a need to update the latter to compatible energy intervals. This will help deduce space-resolved analytical EEDFs representing the emitting plasma, which can then be used for calculating all reaction rates involving electrons and eventually reproduce observed K$\alpha$ emission maps. A brief outline of the entire procedure is shown in the flowchart of Fig. \ref{fig:FlowChart}.
\\\\
The general structure of the paper is as follows: in Sec. \ref{sec:2} we briefly recall the self-consistent numerical methods to model the plasma studied, and to obtain the related three-dimensional (3D) electron energy and density data. In Sec. \ref{sec:3} we discuss the necessity to slice the plasma into regions-of-interest (ROIs) to make progress on the EEDF analysis, and on the ROI selection scheme. The method to derive the most suitable EEDF for each of the plasma ROI is presented in Sec. \ref{sec:4}. In Sec. \ref{sec:5} we present and discuss results from theoretical and numerical analysis of X-ray spectrum of the argon plasma, reported in \cite{Ref17}. We conclude with Sec. \ref{sec:6}, discussing the results obtained with possible near-future outlook.

\section{Self-Consistent Numerical Modelling of ECRIS Plasmas}
\label{sec:2}
\begin{figure*}
\centering
{\includegraphics[width=1\textwidth]{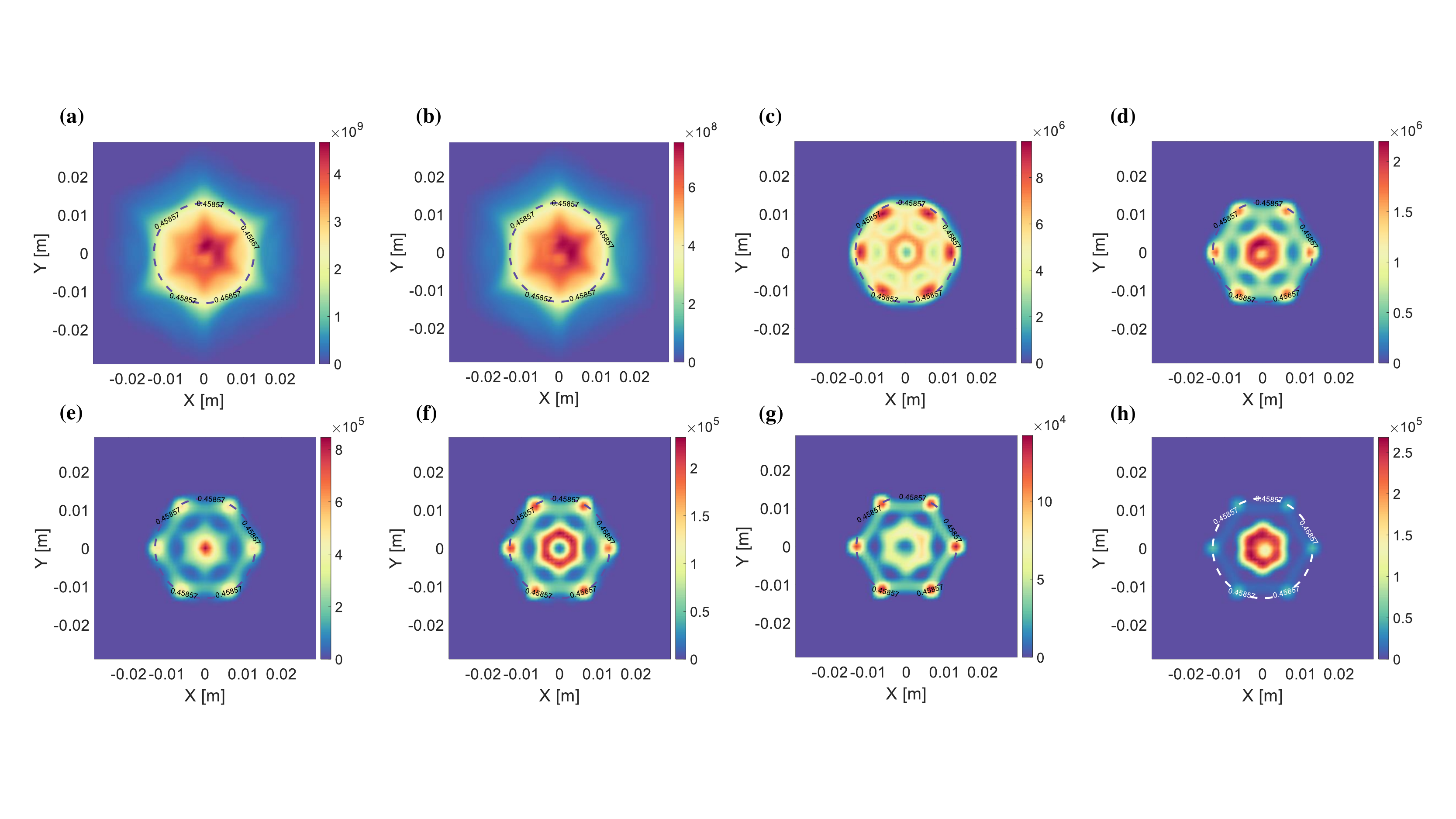}}
\caption{XY-plane projections of the occupation maps $\bm{\rho}_i$ for electrons \textbf{(a)} summed over all the seven energy intervals, and in each of the intervals: \textbf{(b)} $0-2~\mathrm{keV}$, \textbf{(c)} $2-4~\mathrm{keV}$, \textbf{(d)} $4-6~\mathrm{keV}$, \textbf{(e)} $6-8~\mathrm{keV}$, \textbf{(f)} $8-10~\mathrm{keV}$, \textbf{(g)} $10-12~\mathrm{keV}$, and \textbf{(h)} $12-\infty~\mathrm{keV}$. The contour of the points corresponds to the ECR surface (dashed line) with a magnetic field $B=0.4587~\mathrm{T}$.}
\label{fig:dens}
\end{figure*}

\begin{figure*}
\centering
{\includegraphics[width=1\textwidth]{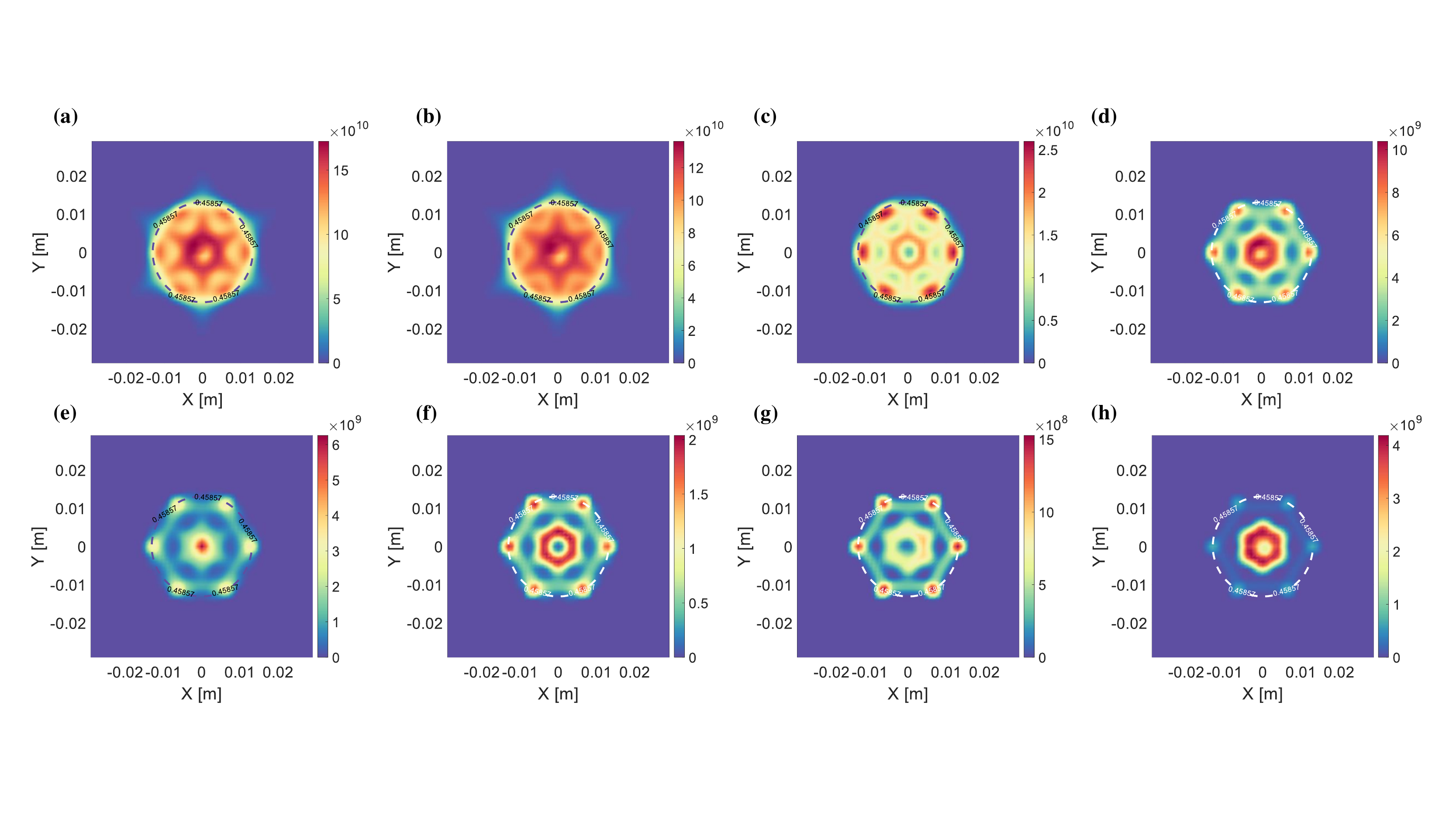}}
\caption{XY-plane projections of the total energy maps $\mathbf{E_{\emph{tot}}}$ for electrons \textbf{(a)} summed over all the seven energy intervals, and in each of the intervals: \textbf{(b)} $0-2~\mathrm{keV}$, \textbf{(c)} $2-4~\mathrm{keV}$, \textbf{(d)} $4-6~\mathrm{keV}$, \textbf{(e)} $6-8~\mathrm{keV}$, \textbf{(f)} $8-10~\mathrm{keV}$, \textbf{(g)} $10-12~\mathrm{keV}$, and \textbf{(h)} $12-\infty~\mathrm{keV}$. The contour of the points corresponds to the ECR surface (dashed line) with a magnetic field $B=0.4587~\mathrm{T}$.}
\label{fig:endens}
\end{figure*}

Numerical simulations can be a predictive tool to determine the spatial and energy density distributions of both electrons and ions. In the case of magnetized plasmas, as for ECRIS plasmas, a self-consistent (SC) approach is necessary to solve the \emph{collisional} Vlasov-Boltzman equation, as described in  \cite{Ref19}.  The two INFN Laboratories, LNL and LNS, developed an iterative procedure to obtain an SC description of ECR \emph{stationary} plasmas, by joining the FEM electromagnetic solver of COMSOL Multiphysics\textsuperscript{\textcopyright}, and a kinetic code written in MATLAB\textsuperscript{\textcopyright} for solving particles' equation of motion. An SC method is required because propagating electromagnetic (EM) field affects electrons' motion and energy through resonant interactions. Beyond this, the plasma - being an anisotropic and dispersive medium - presents a 3D dielectric tensor that needs to be in turn included in the calculation of the EM field \cite{Ref25}. The developed code has already proven its validity describing the so-called frequency tuning effect \cite{Ref23}, and reproducing experimental results for light and heavy ion dynamics in ECR-based charge breeding devices \cite{Ref26,Ref27}. When applied to electrons, the kinetic code follows the evolution of $N$ \emph{particles} for a given simulation time T$_{span}$, with an integration step T$_{step}$. 
The numerical routine describes the stationary structure of the plasma in the phase space: local charge densities are computed through a density accumulation in a 3D simulation's domain. Particles' paths evolve simultaneously, with the local density accumulation arising as single particles move inside the single cells in the simulation's volume of the plasma chamber. The SC loop is run until the achievement of convergence among the $(k-1)$-th and $k$-th step, checking both for 3D density and EM field maps.
Further details about the method can be found in Ref. \cite{Ref24}.\\ Therefore, \emph{occupation} and energy maps are generated. Dividing the latter by the former, the distribution of the energy density can be obtained. The occupation map can then be scaled to a \emph{density} map by assuming an equivalent total number of particles as those provided by a 3D plasmoid-halo density scheme \cite{Ref24}. It should be noted, however, that correct density scaling requires corroboration with experiments - the absolute particle density can only be extracted from some plasma measurement. This further extols the need for experimental benchmarking as attempted in Sec. \ref{sec:5}.  \\\\
The same numerical code was used to investigate the distribution of the warm electron component, by storing seven pairs of occupation and energy maps, associated to electrons in seven energy ranges. In particular, we chose the intervals $[0,2]$, $[2,4]$, $[4,6]$, $[6,8]$, $[8,10]$,$[10,12]$ and $[12,\infty]\,\mathrm{keV}$. The division of the range $[2,12]\,\mathrm{keV}$ in five sub-intervals enhances the energy resolution for a more accurate estimation of the EEDF. We simulated the evolution of $N=40000$ electrons for a total simulation time of T$_{span}=40~\mu$s and with an integration step T$_{step}=1~\rm{ps}$. A Maxwell-Boltzmann (MB) distribution with a temperature $k_{B}T_e =5~ \mathrm{eV}$ was imposed as initial electrons energy distribution. Microwave field at 12.84 GHz and power of 30 W, sustaining the plasma, was considered in the simulations. An example of the results obtained is shown in Figures \ref{fig:dens} and \ref{fig:endens}, where the XY-plane projection of the occupation and energy maps, respectively, are displayed for all the energy ranges considered. It is worth noting how electrons of different energies locate at different spatial domains, which foreshadows the necessity to subdivide the plasma into finer regions for in-depth analysis, as is discussed in the next section. 
%\subsection{Subsection title}

\section{ROI Selection}
\label{sec:3}
\begin{figure}
\resizebox{1.00\columnwidth}{!}{\includegraphics{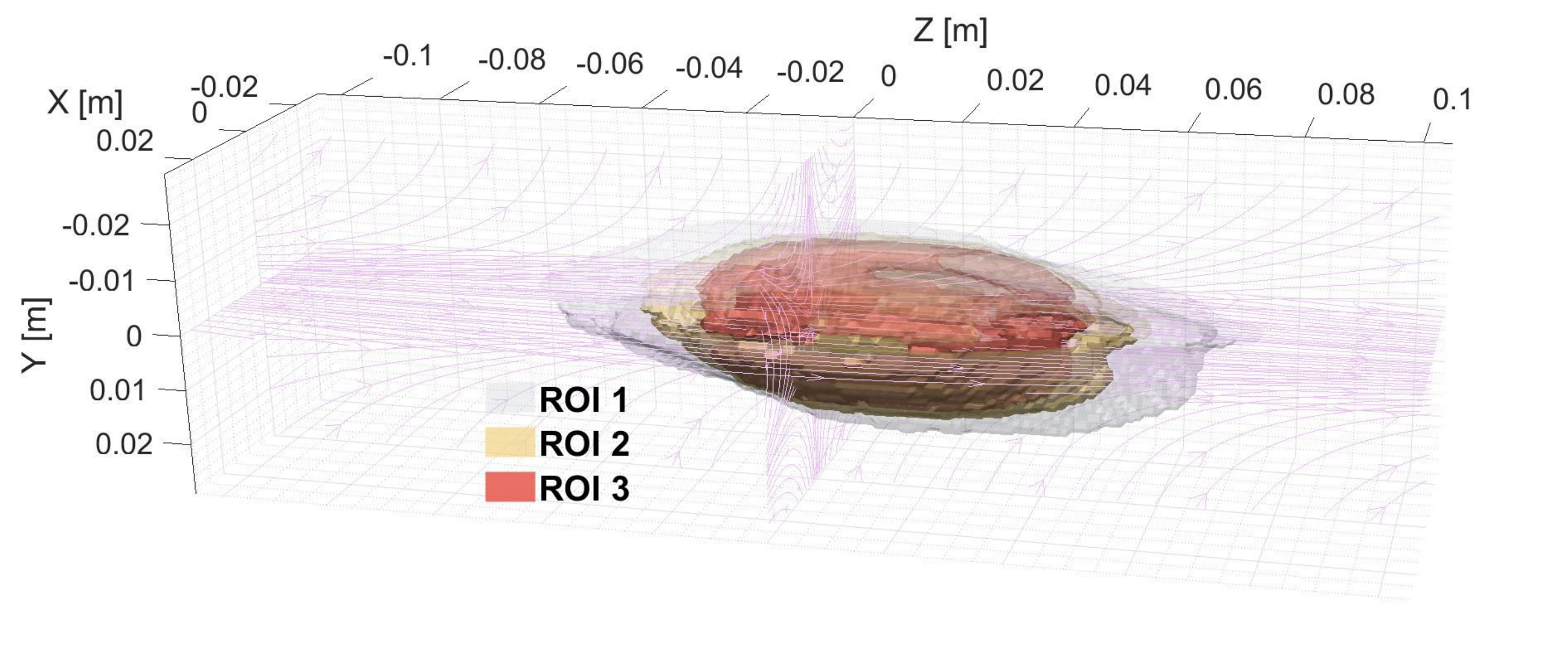}}
\caption{Spatial structure and displacement of AVE-based ROIs 1,2 and 3 according to Eq. (\ref{eq1}). Electrons at higher $\langle \mathbf{E} \rangle$ are found deep inside the plasmoid egg and at the ECR zone. Magnetic field $B(0,0,0)$ lines are also shown (violet solid lines).}
\label{fig:1}
\end{figure}

\begin{figure}
  \begin{subfigure}{0.49\columnwidth}
  \includegraphics[width=\textwidth]{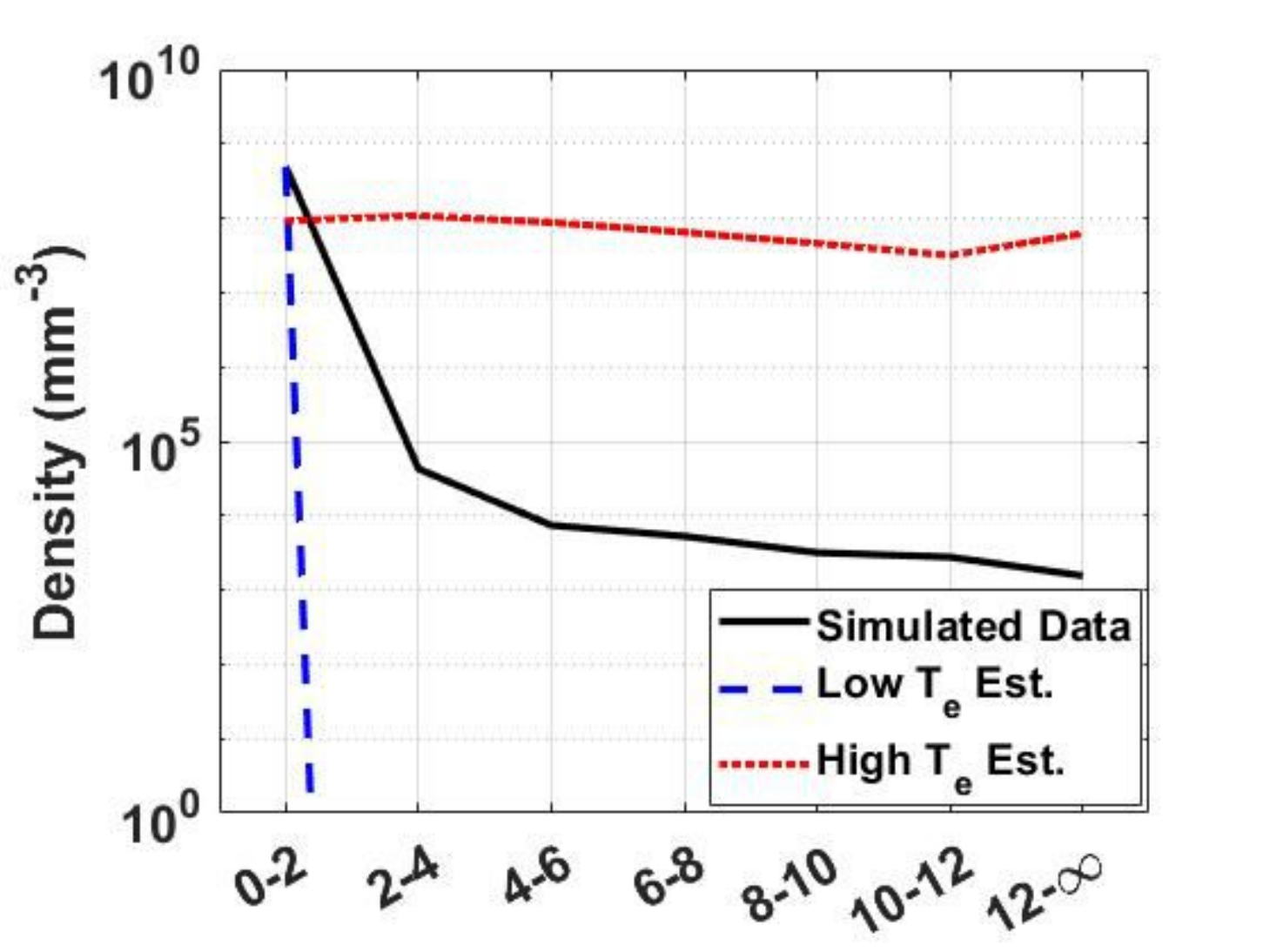}
  \end{subfigure}
  \hfill
  \begin{subfigure}{0.49\columnwidth}
  \includegraphics[width=\textwidth]{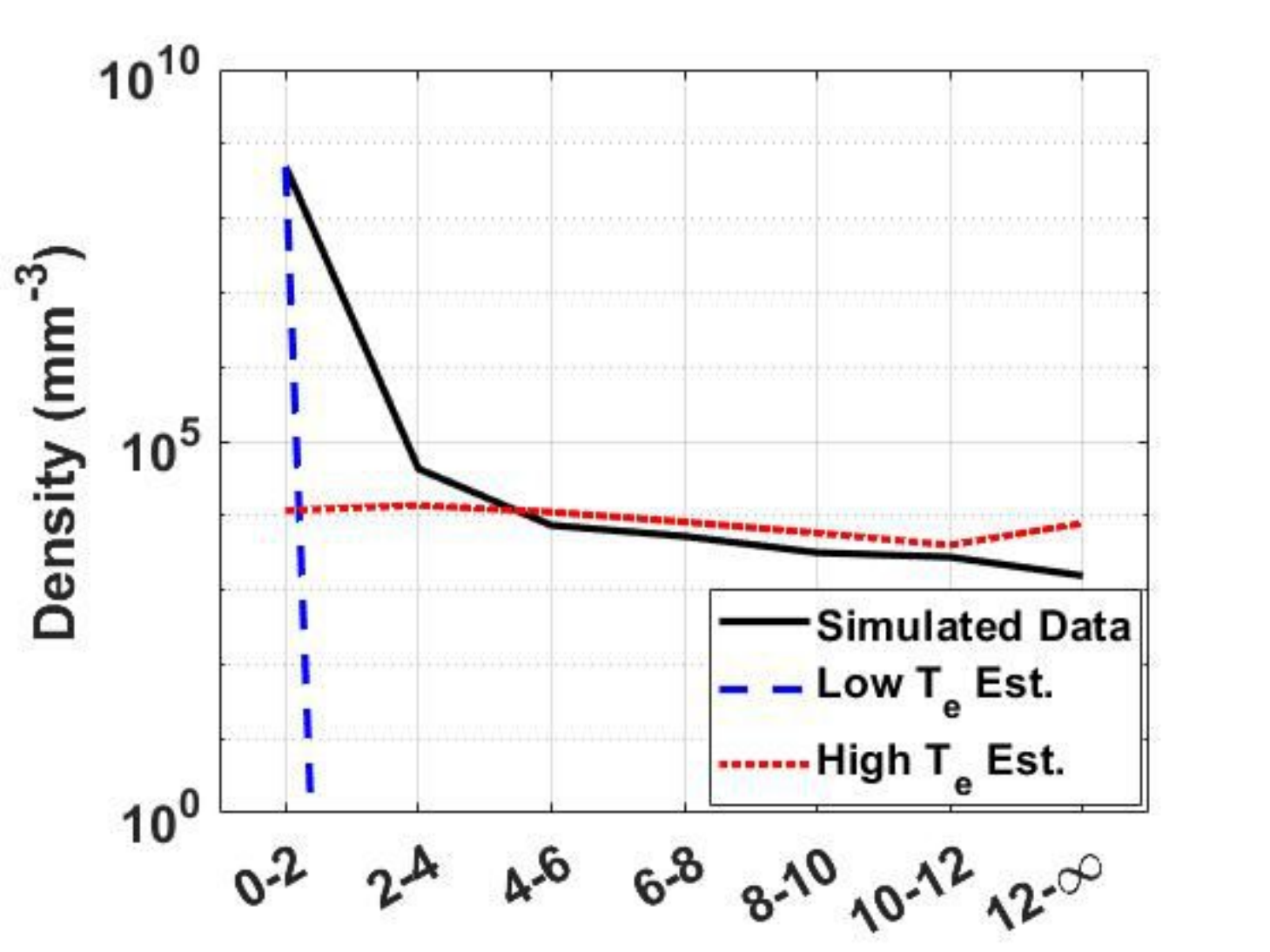}
  \end{subfigure} 
  \begin{subfigure}{0.49\columnwidth} 
  \includegraphics[width=\textwidth]{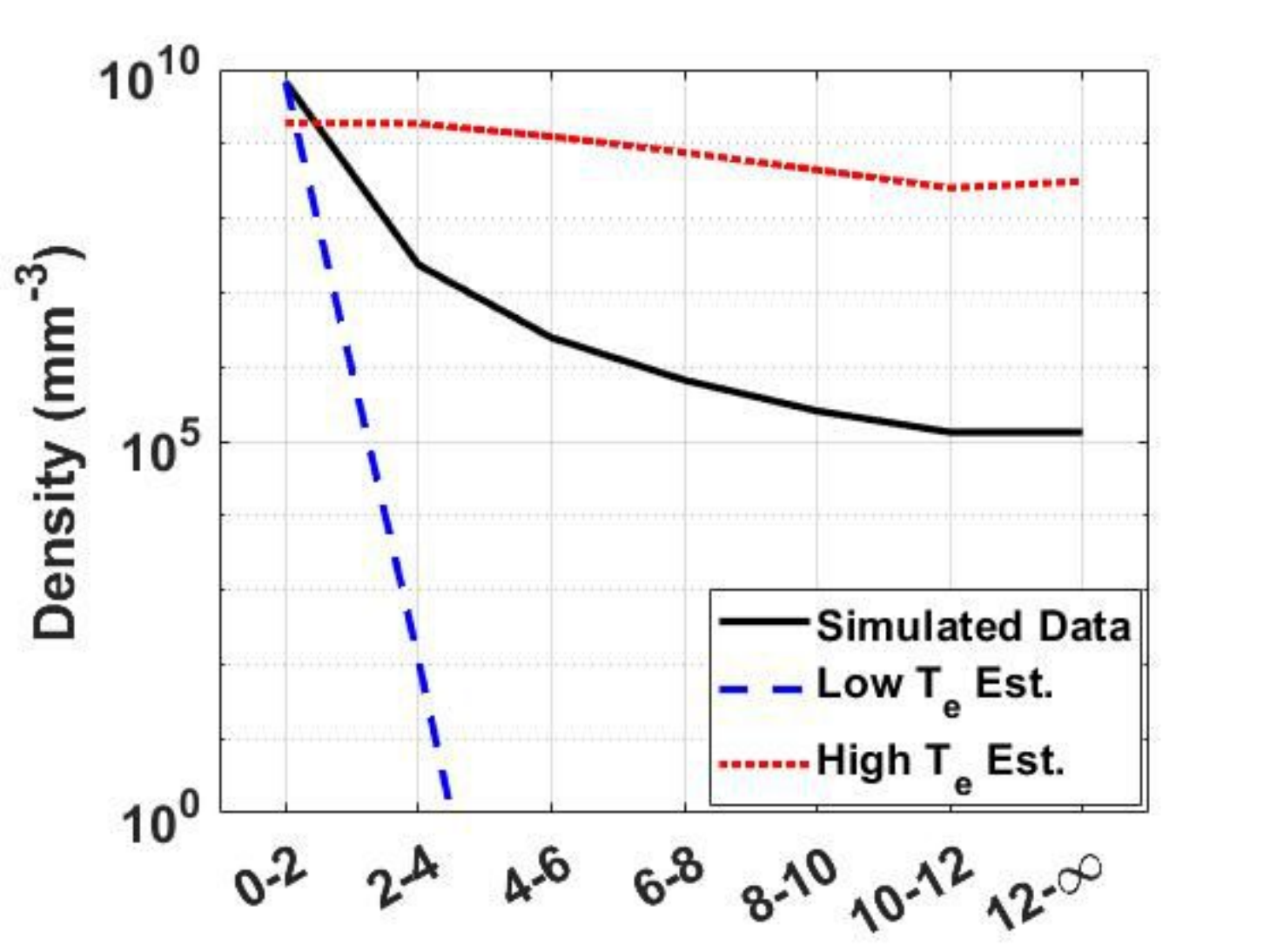} 
  \end{subfigure}  
  \hfill 
  \begin{subfigure}{0.49\columnwidth} 
  \includegraphics[width=\textwidth]{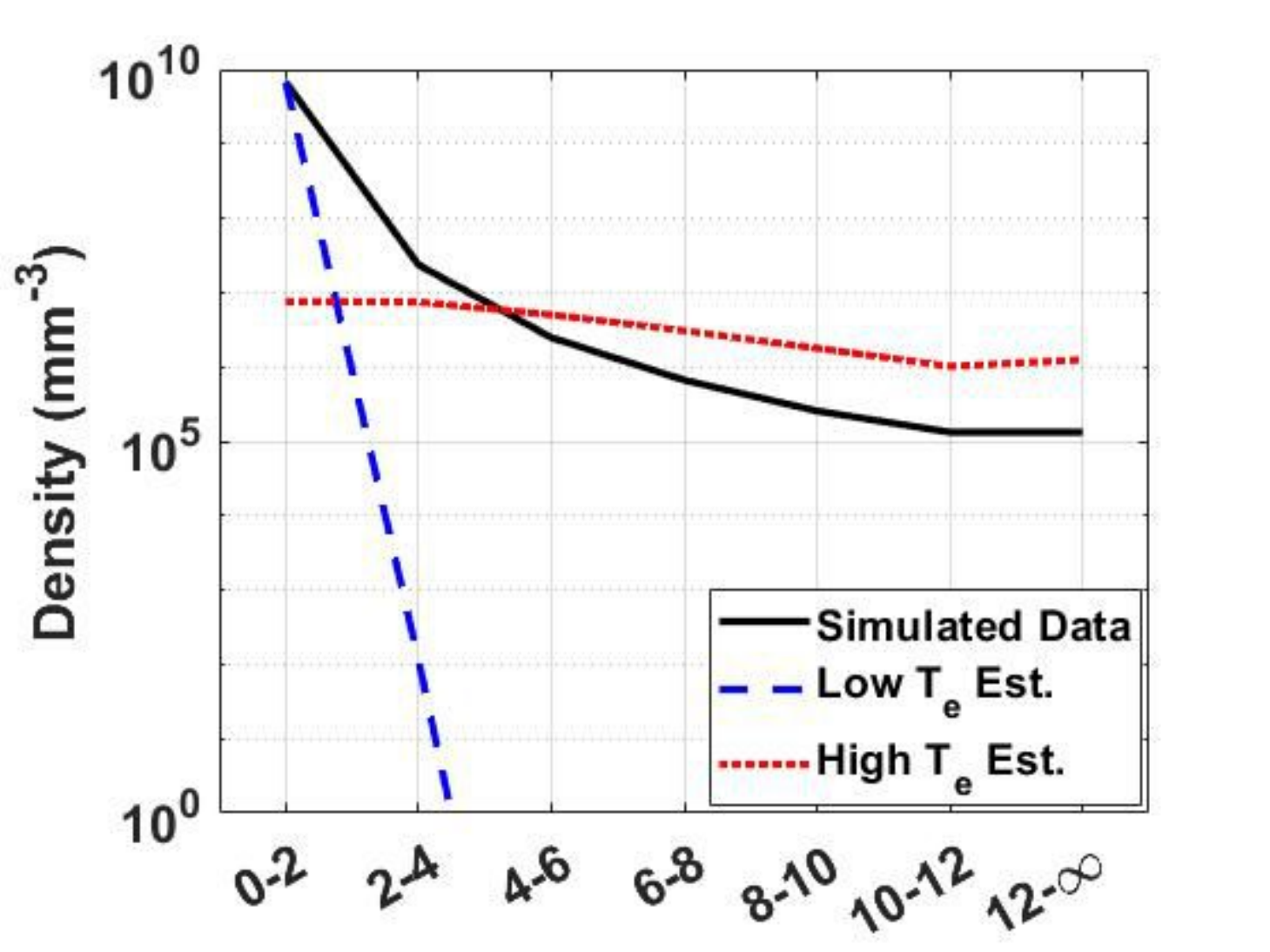}  
  \end{subfigure}
  \begin{subfigure}{0.49\columnwidth} 
  \includegraphics[width=\textwidth]{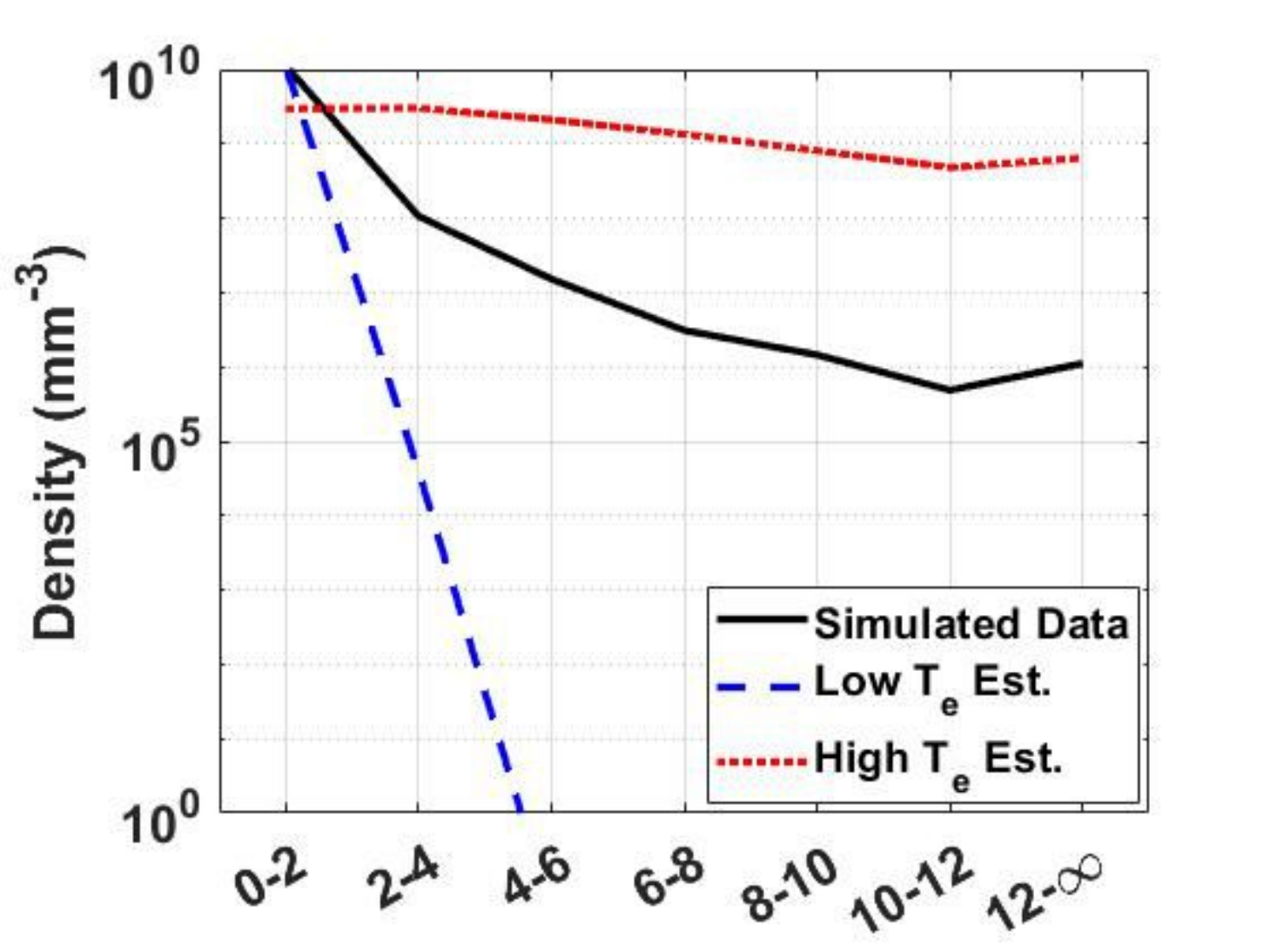} 
  \end{subfigure}  
  \hfill 
  \begin{subfigure}{0.49\columnwidth} 
  \includegraphics[width=\textwidth]{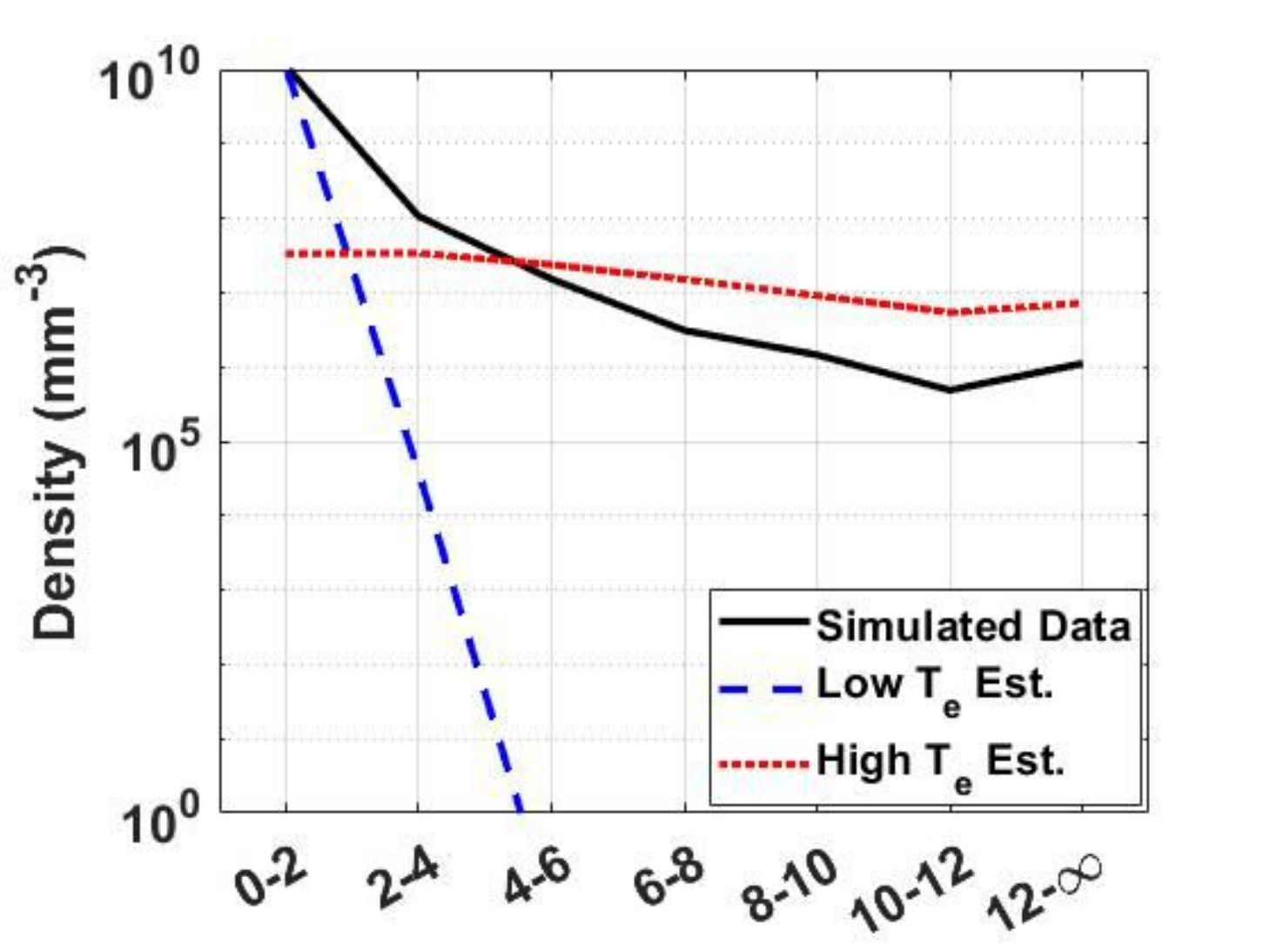}  
  \end{subfigure}
  \caption{ROI-averaged simulated electron density data vs. estimated results numerically calculated using low- and high-$T_e$ single-MB distributions (left), and the same corrected for total electron density (right), in the seven energy intervals for each of the ROIs displayed in Fig. \ref{fig:1}.}
\label{fig:2}
  \end{figure}

\begin{figure}
\centering
  \begin{subfigure}{0.65\columnwidth}
  \includegraphics[width=\textwidth]{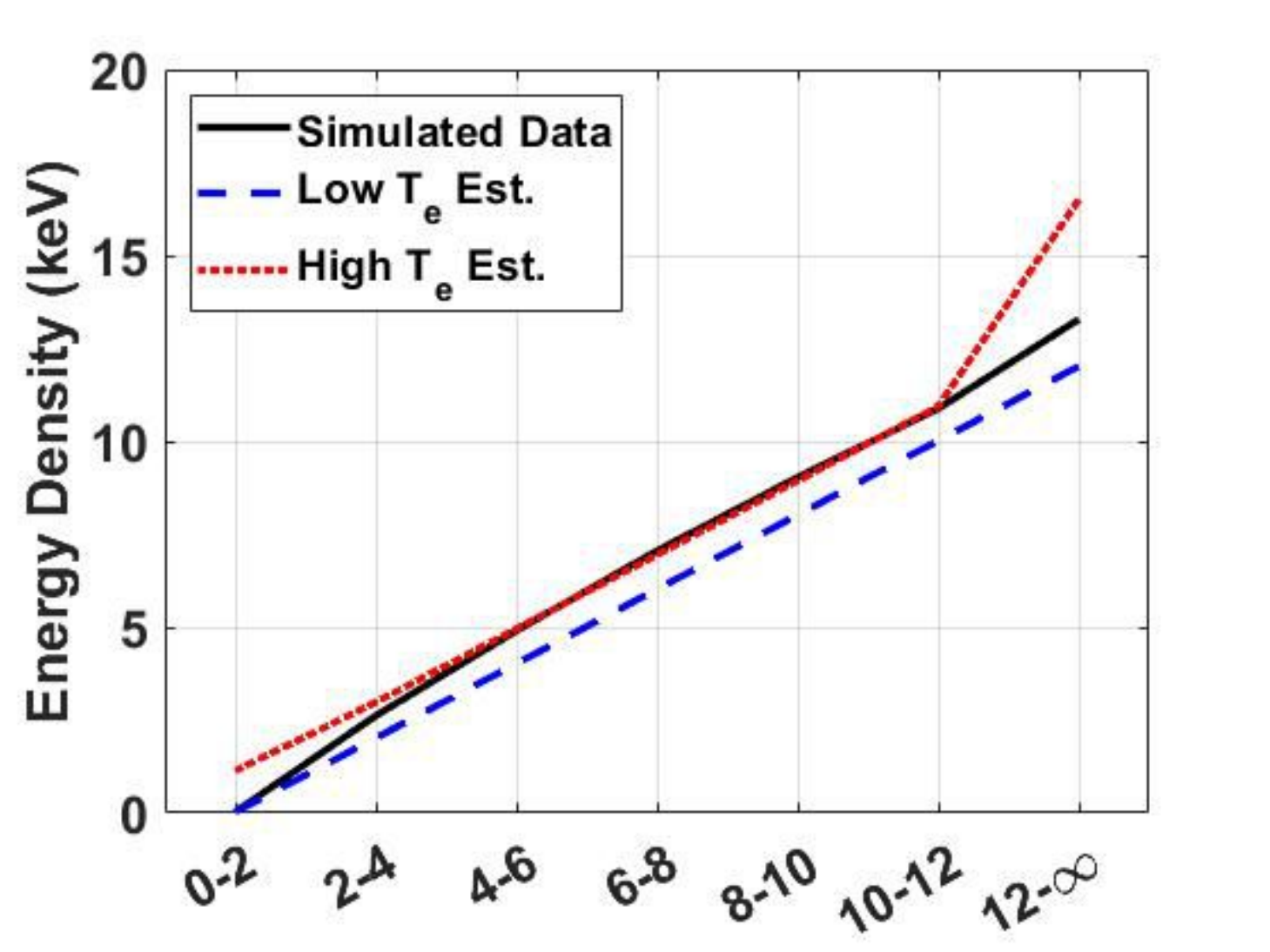}
  \end{subfigure}
  \begin{subfigure}{.65\columnwidth}
  \includegraphics[width=\textwidth]{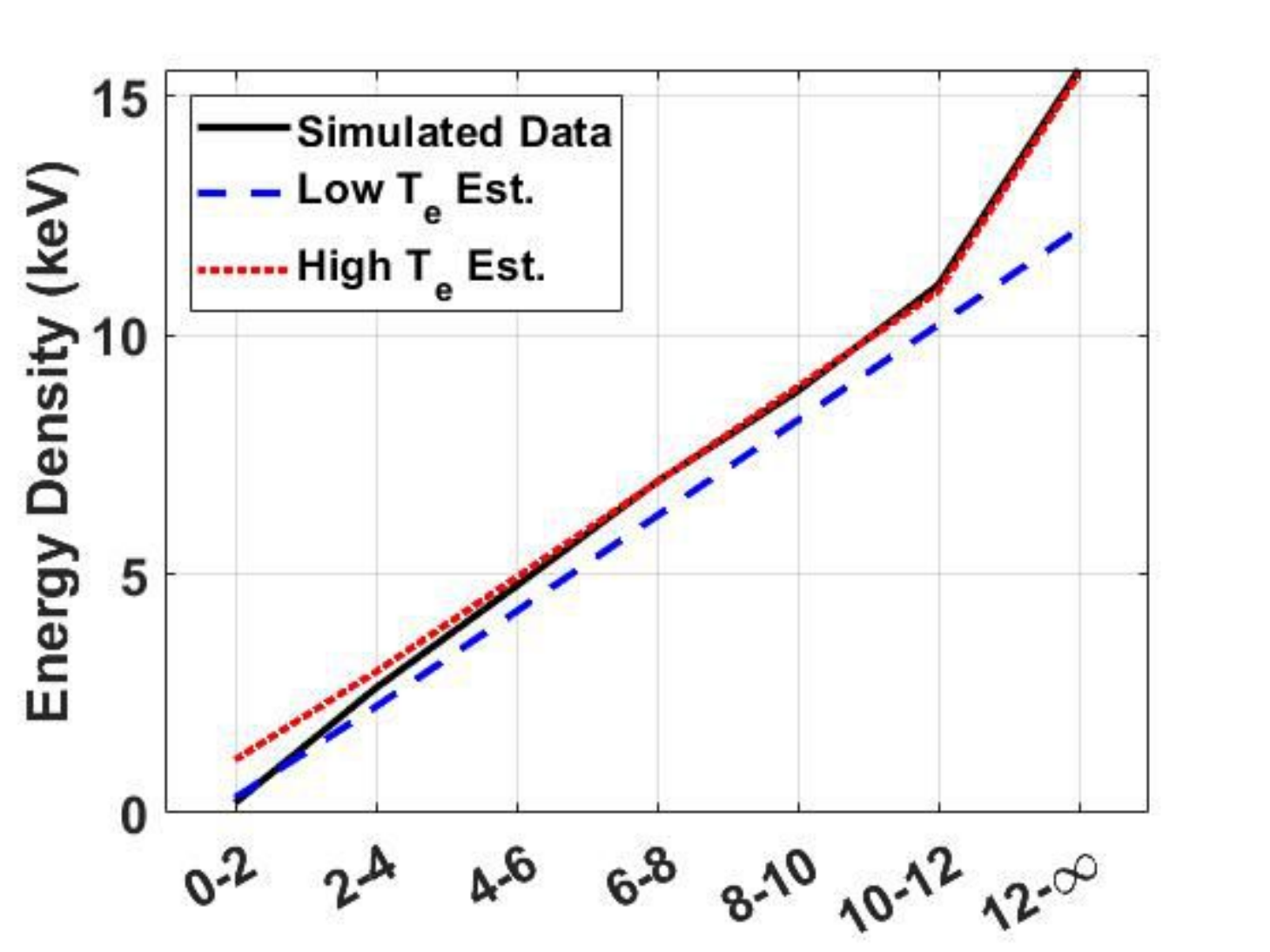}
  \end{subfigure} 
  \begin{subfigure}{.65\columnwidth} 
  \includegraphics[width=\textwidth]{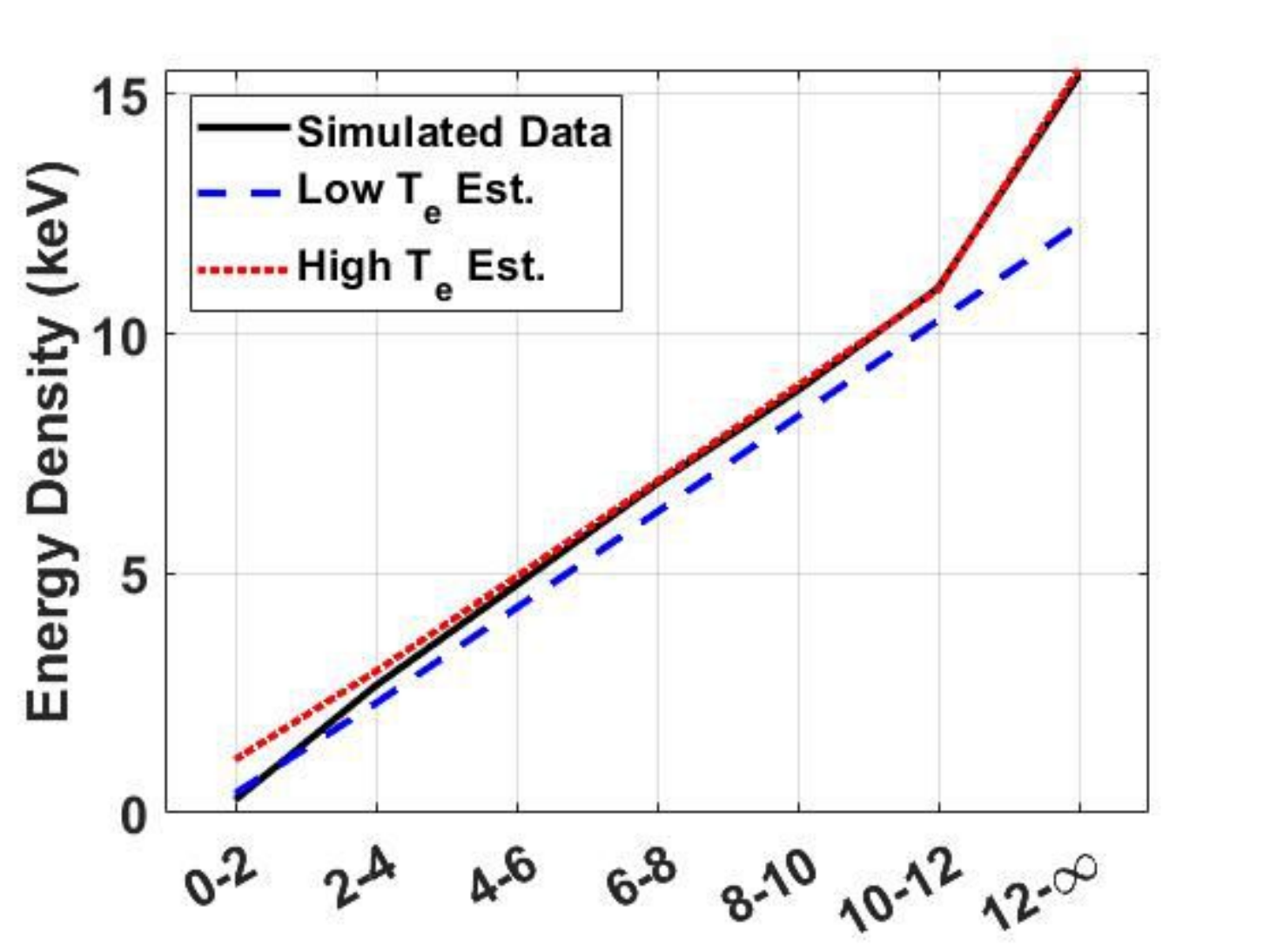} 
  \end{subfigure}  
  \caption{ROI-averaged simulated energy density data vs. estimated results numerically calculated using low- and high-$T_e$ single-MB distributions in the seven energy intervals for each of the ROIs displayed in Fig. \ref{fig:1}}
\label{fig3}
  \end{figure}

The data on occupation and energy density maps presented in Sec. \ref{sec:2}, are structured as pairs of 3D matrices, one pair for each one of the seven energy intervals chosen.  Each 3D matrix is made of 59 x 59 x 211 cells, corresponding to a cuboidal plasma chamber of length, width and depth 59 \emph{mm}, 59 \emph{mm} and 211 \emph{mm} respectively. 
Henceforth, for the sake of consistency and clarity, we represent the electron (number) density as $\bm{\rho}_{i}$ and energy density as $\mathbf{E_{\emph{i}}}$, when referring to the complete 3D matrices or group of cells, and as $\rho_{i}$ and $E_{i}$ when referring to individual cells or density and energy numbers, respectively. The index $i$ will always label the energy intervals. 
Since the basic operation of a minimum-B ECRIS results in anisotropic transfer of energy between the EM field and the plasma electrons, we first separate the plasma domain into different electron energy-based ROIs, assuming these ROIs as containing independent electron populations. Then, the nature of a local EEDF in each of those ROIs was studied. The ROIs are selected according to the \emph{average electron energy} (AVE) in the plasma. Thus, we calculate two quantities that we named \emph{total electron energy} and \emph{total electron density}, given as 
\begin{equation}
\label{eq1}
\mathbf{E_{\emph{tot}}}=\sum_{i=1}^{7}\bm{\rho}_{i}\mathbf{E_{\emph{i}}}\,,\,\,\,\,\,\bm{\rho}_{tot}=\sum_{i=1}^{7}\bm{\rho}_{i}.
\end{equation}
They actually correspond to the total energy and density of electrons at each discrete point in the plasma marked by a plasma cell. It should be noted that these quantities are obtained through \emph{element-wise multiplication} and not a matrix multiplication. The AVE is obtained from Eq. (\ref{eq1}) as
\begin{equation}
\label{eq2}
\langle \mathbf{E} \rangle=\frac{\mathbf{E_{\emph{tot}}}}{\bm{\rho}_{tot}}.
\end{equation}
By grouping together cells whose $\langle \mathbf{E}\rangle$ content lies in some defined range, we construct the different plasma ROIs. Figure \ref{fig:1} shows the isosurfaces of a few $\langle \mathbf{E} \rangle$-based ROIs contained in the plasma chamber, while Table \ref{ROI} lists the characteristics of all the ROIs considered, including the index which will be henceforth used to refer to them. 
\begin{table}
\centering
\caption{\footnotesize{$\langle \mathbf{E}\rangle$-based plasma ROIs}}
\label{ROI}
\begin{tabular*}{\columnwidth}{@{\extracolsep{\fill}}ll@{}ll@{}}
\hline
\textbf{ROI} & $\langle \mathbf{E} \rangle$ & \textbf{Index}\\
\hline
$\mathrm{ROI1}$           & $0.0-0.1\,\mathrm{keV}$		& $j=1$   \\      
$\mathrm{ROI2}$           & $0.1-0.2\,\mathrm{keV}$		& $j=2$   \\
$\mathrm{ROI3}$           & $0.2-0.3\,\mathrm{keV}$		& $j=3$   \\
$\mathrm{ROI4}$           & $0.3-0.4\,\mathrm{keV}$		& $j=4$   \\
$\mathrm{ROI5}$           & $0.4-0.5\,\mathrm{keV}$		& $j=5$   \\
$\mathrm{ROI6}$           & $0.5-0.6\,\mathrm{keV}$		& $j=6$   \\
$\mathrm{ROI7}$           & $>0.6\,\mathrm{keV}$		& $j=7$   \\
\hline
\end{tabular*}
\end{table}
To get some \emph{qualitative} information on the spatial properties of the electron distribution, and the kind of continuous EEDF we should be looking at, we first evaluate \emph{ROI-averaged} electron density and energy density matrices (Eq. (\ref{eq3})) in each of the seven aforementioned energy intervals. The idea is to generate \emph{collective} data for each ROI by adding together the density and energy contributions from the cells belonging to that ROI
\begin{equation}
\label{eq3}
\rho_{ij}=\sum_{k(j)=1}^{N(j)}\rho_{ik(j)}\,,\,\,\,\,\,E_{ij}=\frac{\sum_{k(j)=1}^{N(j)}\rho_{ik(j)}E_{ik(j)}}{\rho_{ij}}.
\end{equation}
Here $N(j)$ represents the total number of cells and $k(j)$ is the index of an individual cell, of any particular ROI $j$. Just as in Eq. (\ref{eq1}), the \emph{total electron density} in each ROI can be calculated by summing up the electron densities in each interval, i.e. $(\rho_{tot})_{j}=\sum_{i=1}^{7}\rho_{ij}$. Next, we assume an EEDF as a \emph{single-component} MB distribution of the form
\begin{equation}
\label{eq4}
f_{M}(E;k_{B}T_e) = \frac{2}{\sqrt{\pi}}\frac{\sqrt{E}}{(\sqrt{k_{B}T_{e}})^{3}}e^{-E/k_{B}T_{e}},
\end{equation}
where $T_{e}$ is the electron temperature, and $k_{B}$ is the Boltzmann constant. Then, we numerically estimate the same collective data as generated in Eq. (\ref{eq3}) using the equations 
\begin{align}
\label{eq5}
(\rho_{ij})_{est}=(\rho_{tot})_{j}\int^{b}_{a}f(E;k_{B}T_{e})\mathrm{d}E\,\\\nonumber 
(E_{ij})_{est}=\frac{\int^{b}_{a}f(E;k_{B}T_{e})E\mathrm{d}E}{\int^{b}_{a}f(E;k_{B}T_{e})\mathrm{d}E}.
\end{align}
Here there is an equivalence between the energy intervals marked by $i$ and $[a,b]$. The $i=1$ interval implies $[a,b]=[0,2]$ keV, $i=2$ implies $[a,b]=[2,4]$ keV, and so on.
%\begin{equation}
%\label{eq6}
%\big[E_{i}\big]_{Est}=\frac{\int^{b}_{a}f(E)E\mathrm{d}E}{\int^{b}_{a}f(E)\mathrm{d}E},
%\end{equation}
Then, we plot the simulated data for electron density from Eq. (\ref{eq3}) against its numerically approximated counterpart from Eqs. (\ref{eq4}-\ref{eq5}), for two different temperatures $T_{e}$ - a low one in the $\mathrm{eV}$ range and a higher one in the $\mathrm{keV}$ range. The numerical integration in Eqs. (\ref{eq5}) is performed using the Adaptive Quadrature Method. Results are presented in Fig. \ref{fig:2}. 
%
%
%
%
%\begin{figure*}
%\resizebox*{0.50\columnwidth}{!}{\includegraphics{Fig2ROI2DensityComp}}
%\resizebox*{.5\columnwidth}{!}{\includegraphics{Fig2ROI2EnergyComp}}
%\resizebox*{.5\columnwidth}{!}{\includegraphics{Fig2ROI4DensityComp}}
%\resizebox*{.5\columnwidth}{!}{\includegraphics{Fig2ROI4EnergyComp}}
%\resizebox*{.5\columnwidth}{!}{\includegraphics{Fig2ROI5DensityComp}}
%\resizebox*{.5\columnwidth}{!}{\includegraphics{Fig2ROI5EnergyComp}}
%\caption{ROI-averaged electron density (left) and energy density (right) in the seven energy intervals for each of the ROIs displayed in Fig. \ref{fig:1}, against approximated results numerically calculated using low- and high-$T_e$ single Maxwellian distributions}
%\label{fig:2}
%\end{figure*}
%
The first set of plots in Fig. \ref{fig:2} (left) allow us to draw some general statements on the electron properties. It can be evinced that a single-component distribution function does not fit the data well. The high-$T_{e}$ function underestimates the density in the $[0,2]\,\mathrm{keV}$ interval, while simultaneously overestimating it in the $[2,\infty]\,\mathrm{keV}$ one. The low-$T_{e}$ function does the exact opposite. This might arise from using an incorrect evaluation of $T_{e}$, density $\rho_{tot}$, or both, as the density estimation in Eq. (\ref{eq5}) involves only these parameters. To make progress, we posit the existence of two distinct populations - a "cold" population localized only at the $[0,2]\,\mathrm{keV}$ interval, and a "warm" population spread across the remaining intervals. The aim is to check the applicability of the low-$T_{e}$ and high-$T_{e}$ functions to the cold and warm populations, respectively, by replacing  $(\rho_{tot})_{j}$ in Eq. (\ref{eq5}) with $\rho_{1j}$ when using the low-$T_{e}$ function, and by $(\rho_{tot})_{j}-\rho_{1j}$ for the high-$T_{e}$ function. The results of this correction are shown in the second set of plots in Fig. \ref{fig:2} (right). Here the overestimation by the high- and the low-$T_{e}$ functions in the $[2,\infty]$ and $[0,2]\,\mathrm{keV}$, respectively, are drastically reduced, though the latter is not noticeable because of $(\rho_{tot})_{j}$ and $\rho_{1j}$ having the same order of magnitude. This serves to prove our assumption of separate populations. The remaining problem of underestimating electron densities in the different intervals could be related with incorrect $T_{e}$ assumption, although it seems more likely to be an unavoidable fallacy on part of the usage of a single-component EEDF to describe a plasma proved to contain two populations. Thus, a single-component EEDF underperforms for warm electrons, independent of the temperature chosen. This can be seen even more clearly in Fig. \ref{fig3}, where we plot the simulated data for electron energy density from Eq. (\ref{eq3}) against the numerically estimates from Eq. (\ref{eq5}).
Since the calculation of the energy density is independent of $(\rho_{tot})_{j}$, the results are purely based on the chosen EEDF, and it can be appreciated that a single-component function can describe well either the cold electrons or the warm electrons at a time, but not both. This further seems to justify the need for a \emph{multi-component} EEDF, not uncommon in plasmas. Inductively Coupled Plasmas (ICPs) \cite{Ref20} and RF-discharge \cite{Ref21} plasmas were among the first to be probed for the multi-component EEDFs. The authors found the presence of two or even three separate MB-like distributions in the cold plasma (energies in $\sim\, \mathrm{eV}$). The possibility of a non-Maxwellian distribution was detected in the Minimafios and Quadramafios ECRIS experiments  \cite{Ref11,Ref22} that focused on the high-energy Bremsstrahlung X-ray radiation as well. Although plasma electrons under study in this work are of intermediate energy, it seems that the aforementioned aspects may be true of them as well, and analytical EEDFs proposed should then be multi-component and capable of describing the spatial-dependence of the electron population.

\section{EEDF Analysis}
\label{sec:4}
Various approaches can be used to guess the EEDF, based on the physics of electron thermalisation in an ECRIS chamber. Plasma electrons in local thermodynamic equilibrium (LTE) can be thought of as a classical gas, whose thermal interactions with each other are governed by the kinetic theory of gases. The distribution of speeds (and thus energies) for such an electron gas at a certain temperature is captured by the known MB distribution of Eq. (\ref{eq4}). Deviations from such a distribution arise in non-LTE cases, which are typical conditions for an ECRIS plasma. In such cases, a single MB function often does not completely describe the electron properties (as seen in Fig. \ref{fig:2}) and to proceed, not only additional components, but also other distribution functions may need to be employed. With this in mind, we present a detailed study of different multi-component EEDFs to model the plasma structure. Since the plasma was sliced into different ROIs, every EEDF is applied to each of the ROIs, and then the best outcome bases on the robustness of the least-square regression fit function used. This is primarily an exercise in statistics because each ROI consists of many cells, and since it is quite impossible to check the goodness-of-fit of each continuous function in \emph{every} cell of the ROI, the EEDF quality is based on ROI-aggregated Mean Squared Error (MSE) estimate and the correlation coefficient to the least-square regression fit function, namely the $r^{2}$ coefficient.
In our analysis, each part of our multi-component test EEDF is either a MB, $f_M$, or a \emph{Druyvesteyn} (DR) distribution function, $f_D$, the latter given as
\begin{equation}
\label{eq6}
f_{D}(E;k_{B}T_{e})=1.04\frac{\sqrt{E}}{(\sqrt{k_{B}T_{e}})^{3}}e^{-0.55E^{2}/(k_{B}T_{e})^{2}}.
\end{equation}
The existence of DR EEDFs in plasmas has been acknowledged since a long time, but under different conditions. Studies in negative glow and positive column plasmas, using Langmuir probe techniques \cite{Ref28}, and in pulsed-RF ICP sou\-r\-ces, using a combination of direct and indirect diagnostics \cite{Ref29}, have both included the presence of one or more DR EEDFs. This function is characterized by a shorter tail as compared to a regular Maxwell distribution, which is often associated with processes resulting in a high loss-rate of energetic electrons.
When dealing with $f_{M}$, the quantity of interest is the \emph{temperature} parameter $k_{B}T_{e}$. Dealing with $f_{D}$ of Eq.~(\ref{eq6}), while in the literature the fit parameter is indicated as the average energy $W_{av}$, we choose to also treat it as a sort of "temperature" $k_{B}T_{e}$ for ease of reference.
In general, for similar values of average energy, $f_{D}$ trails off faster than $f_{M}$, making it more accurate when describing high-energy electron populations which do not completely thermalize.
As for the number of components in the EEDF, we choose to work with two- and three-component distribution functions, with the cases of study as described in Table \ref{EEDF}.
\begin{table}
\centering
\caption{\footnotesize{Test-EEDF cases of study}}
\label{EEDF}
\begin{tabular*}{\columnwidth}{@{\extracolsep{\fill}}ll@{}}
\hline
\textbf{Ref. case name} & \textbf{Type}\\
\hline
$\mathrm{EEDF1}$           & Low-$E$ $f_{M}$ + High-$E$ $f_{M}$          \\
$\mathrm{EEDF2}$           & Low-$E$ $f_{M}$ + High-$E$ $f_{D}$\\
$\mathrm{EEDF3}$           & Low-$E$ $f_{M}$ + Medium-$E$ $f_{M}$ + High-$E$ $f_{M}$\\
$\mathrm{EEDF4}$           & Low-$E$ $f_{M}$ + Medium-$E$ $f_{D}$ + High-$E$ $f_{M}$         \\
\hline
\end{tabular*}
\end{table}
A general two-component distribution function is taken to be of the form 
\begin{equation}
\label{eq7}
f(E;k_{B}T_{l},k_{B}T_{h})=A_{l}f_{l}(E;k_{B}T_{l})+
 A_{h}f_{h}({E;k_{B}T_{h}}),
\end{equation}
where $f_{l}(E;k_{B}T_{l})$ is always chosen as the normalized MB distribution function of Eq. (\ref{eq4}), to model electrons in the $[0,2]\, \mathrm{keV}$ range, while $f_{h}(E;k_{B}T_{h})$ can be both the MB or the DR function of Eq. (\ref{eq6}), to model the electrons in the remaining energy intervals, i.e. $[2,\infty]\, \mathrm{keV}$. The idea is to provide an EEDF that treats the electrons in the two intervals as separate populations. This means that the respective temperature parameters need to be calculated keeping in mind (\emph{i}) the physics behind this assumption, (\emph{ii}) minimum overlap between the functions, (\emph{iii}) optimization of the goodness-of-fit to real data, and (\emph{iv}) general restrictions on values (e.g., the cold electron energy $k_{B}T_{l}$ should always be in the $[10,100]\, \mathrm{eV}$ range, while the warm energy $k_{B}T_{h}$ should be in the $[1,10]\,\mathrm{keV}$ range). Along with temperature, the normalization coefficients $A_{l,h}$ for each component are also of great importance, because they represent the share of each function within the combined EEDF and are intrinsically linked to points (\emph{ii}) and (\emph{iii}) above mentioned. In our analysis, we address all the above concerns in a systematic manner, starting with the calculation of the cold- and warm-electron temperatures as 

\begin{equation}
\label{eq8}
\mathbf{(k_{B}T_{\emph{l}})_{\emph{j}}}=\frac{2}{3}\mathbf{C_{\emph{j}}}\mathbf{E}_{1\emph{j}},\,\,\,\,\,\,\,\mathbf{(k_{B}T_{\emph{h}})_{\emph{j}}}=\mathbf{S_{\emph{j}}}\frac{\sum_{i=2}^{7}\bm{\rho}_{ij}\mathbf{E_{\emph{ij}}}}{\sum_{i=2}^{7}\bm{\rho}_{ij}},
\end{equation}
where $\mathbf{C_{\emph{j}}}$ and $\mathbf{S_{\emph{j}}}$ are optimization factors in matrix format for all the cells of the $j$-th ROI. The evaluation of the normalization coefficients, $\mathbf{A_{\emph{lj}}}$ and $\mathbf{A_{\emph{hj}}}$, is performed according to the following constraints
\begin{align}
\label{eq9}
\frac{\bm{\rho}_{1\emph{j}}}{\sum_{i=1}^{7}\bm{\rho}_{\emph{ij}}}=\mathbf{A_{\emph{lj}}}\int_{0}^{2}f_{\emph{l}}(E;\mathbf{(k_{B}T_{\emph{l}})_{\emph{j}}})\mathrm{d}E+\\\nonumber  \mathbf{A_{\emph{hj}}}\int_{0}^{2}f_{\emph{h}}(E;\mathbf{(k_{B}T_{\emph{h}})_{\emph{j}}})\mathrm{d}E,
\end{align}
\begin{align}
\label{eq10}
\frac{\sum_{i=2}^{7}\bm{\rho}_{ij}}{\sum_{i=1}^{7}\bm{\rho}_{ij}}=\mathbf{A_{\emph{lj}}}\int_{2}^{\infty}f_{\emph{l}}(E;\mathbf{(k_{B}T_{\emph{l}})_{\emph{j}}})\mathrm{d}E+\\\nonumber  \mathbf{A_{\emph{hj}}}\int_{2}^{\infty}f_{\emph{h}}(E;\mathbf{(k_{B}T_{\emph{h}})_{\emph{j}}})\mathrm{d}E.
\end{align}
Since $\mathbf{(k_{B}T_{\emph{l}})_{\emph{j}}}$ is carefully calculated to be in the $\sim\mathrm{eV}$ range, the contribution of $f_{\emph{l}}(E;\mathbf{(k_{B}T_{\emph{l}})_{\emph{j}}})$ to the $[12,\infty]\, \mathrm{keV}$ interval can be neglected, but some \emph{overlap} may still remain due to the integral 
\begin{equation}
\label{eq11}
\mathbf{u_{\emph{j}}}=\int_{0}^{2}f_{\emph{h}}(E;\mathbf{(k_{B}T_{\emph{h}})_{\emph{j}}})\mathrm{d}E,
\end{equation}
being non-zero. Thus, the normalization coefficients can be calculated from the simulated data, correcting for the overlap and correlating with the temperature as
\begin{equation}
\label{eq12}
\mathbf{A_{\emph{lj}}}=\frac{\bm{\rho}_{1\emph{j}}}{\sum_{i=1}^{7}\bm{\rho}_{\emph{ij}}}-\frac{\mathbf{u_{\emph{j}}}\sum_{i=2}^{7}\bm{\rho}_{\emph{ij}}}{(\sum_{i=1}^{7}\bm{\rho}_{\emph{ij}})(1-\mathbf{u_{\emph{j}}})},
\end{equation}
\begin{equation}
\label{eq13}
\mathbf{A_{\emph{hj}}}=\frac{\sum_{i=2}^{7}\bm{\rho}_{\emph{ij}}}{(\sum_{i=1}^{7}\bm{\rho}_{\emph{ij}})(1-\mathbf{u_{\emph{j}}})}.
\end{equation}
The final step is the estimation of electron density and energy density in each interval using expressions similar to those used in Eq. (\ref{eq5}), as
\begin{align}
\label{eq14}
(\bm{\rho}_{ij})_{est}=(\sum_{i=1}^{7}\bm{\rho}_{\emph{ij}})\int_{a}^{b}f(E;\mathbf{(k_{B}T_{\emph{l}})_{\emph{j}}};\mathbf{(k_{B}T_{\emph{h}})_{\emph{j}}})\mathrm{d}E,
\end{align}
\begin{align}
\label{eq15}
(\mathbf{E_{\emph{ij}}})_{est}=\frac{\int_{a}^{b}f(E;\mathbf{(k_{B}T_{\emph{l}})_{\emph{j}}};\mathbf{(k_{B}T_{\emph{h}})_{\emph{j}}})E\mathrm{d}E}{\int_{a}^{b}f(E;\mathbf{(k_{B}T_{\emph{l}})_{\emph{j}}};\mathbf{(k_{B}T_{\emph{h}})_{\emph{j}}})\mathrm{d}E}.
\end{align}
The numerical integration here is performed using the Adaptive Quadrature Method. It is worth noting that element-wise estimate via Eq. (\ref{eq14},\ref{eq15}) differs from the collective  evaluation in Eq. (\ref{eq5}). 
%and the boldface estimated density and energy density values indicate an element-wise evaluation in each cell of the ROI $j$, as opposed to the collective evaluation in Eq. (\ref{eq5}). 

\begin{figure}
  \begin{subfigure}{0.49\columnwidth}
  \includegraphics[width=\textwidth]{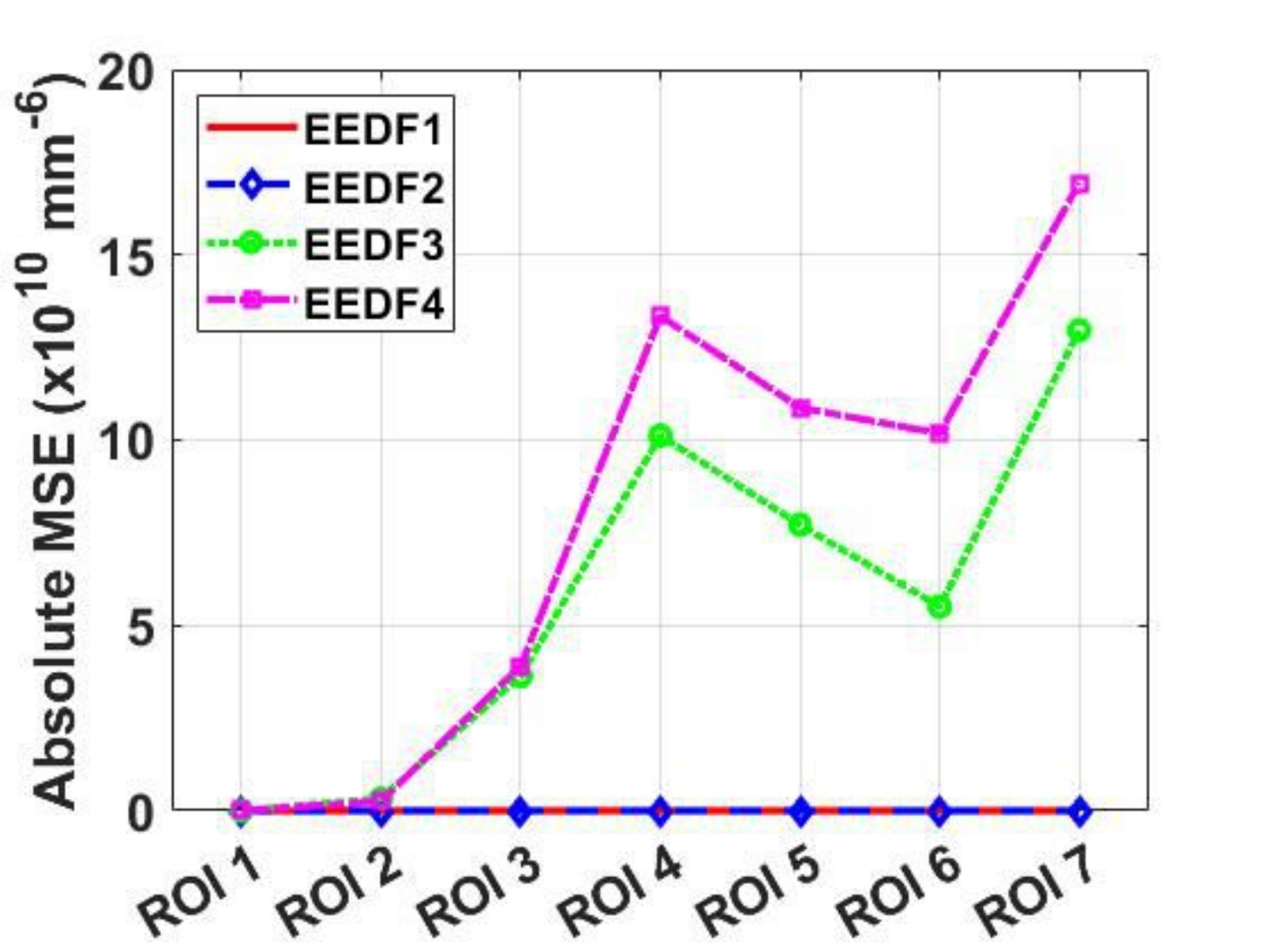}
  \end{subfigure}
  \hfill
  \begin{subfigure}{0.49\columnwidth}
  \includegraphics[width=\textwidth]{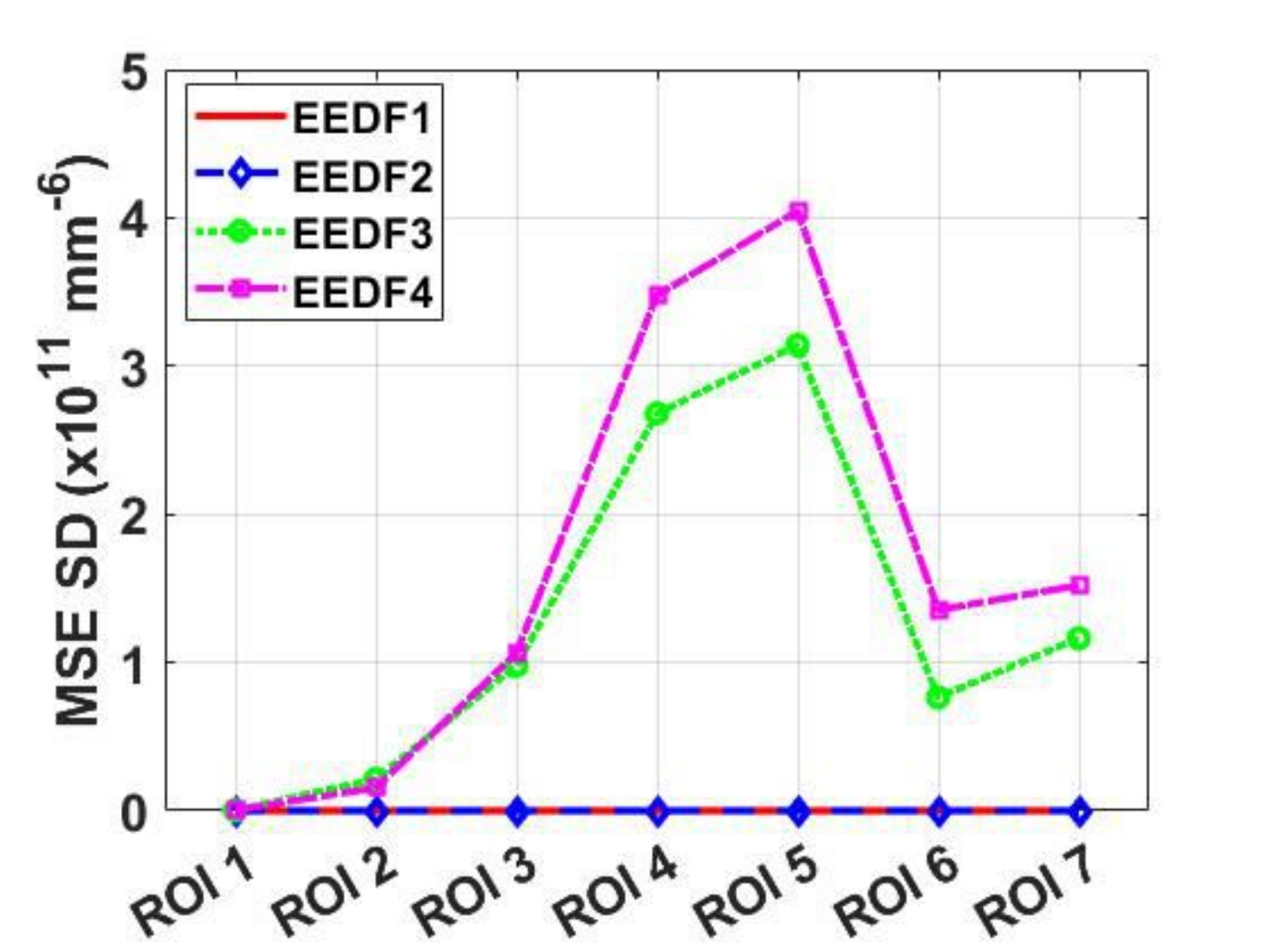}
  \end{subfigure} 
  \caption{ROI-averaged MSE for electron density (left) and SD of ROI-averaged MSE (right) in all ROIs, for $[0,2]\,\mathrm{keV}$ interval.}
  \label{fig4}
  \end{figure}

\begin{figure}
  \begin{subfigure}{0.49\columnwidth}
  \includegraphics[width=\textwidth]{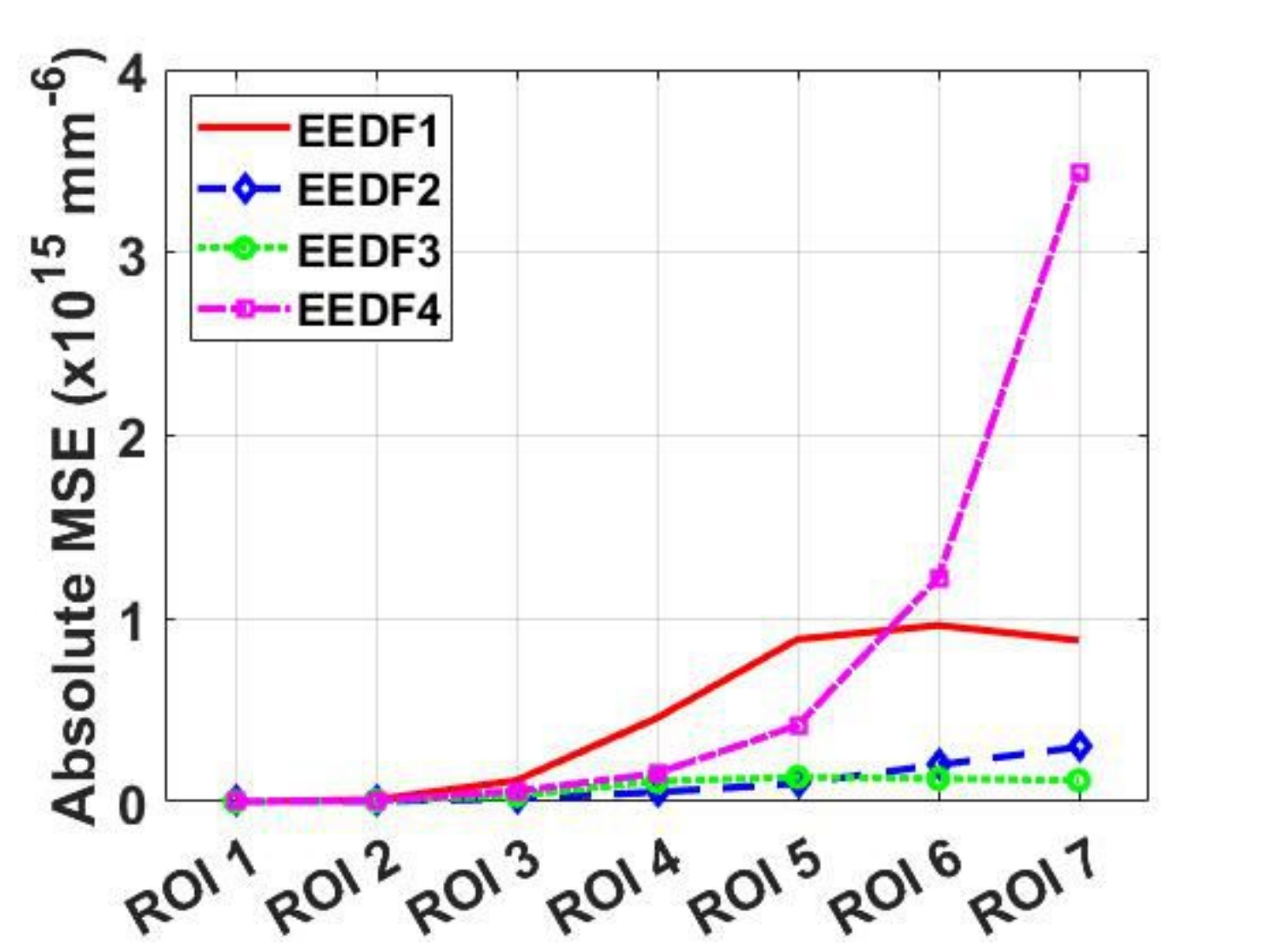}
  \end{subfigure}
  \hfill
  \begin{subfigure}{0.49\columnwidth}
  \includegraphics[width=\textwidth]{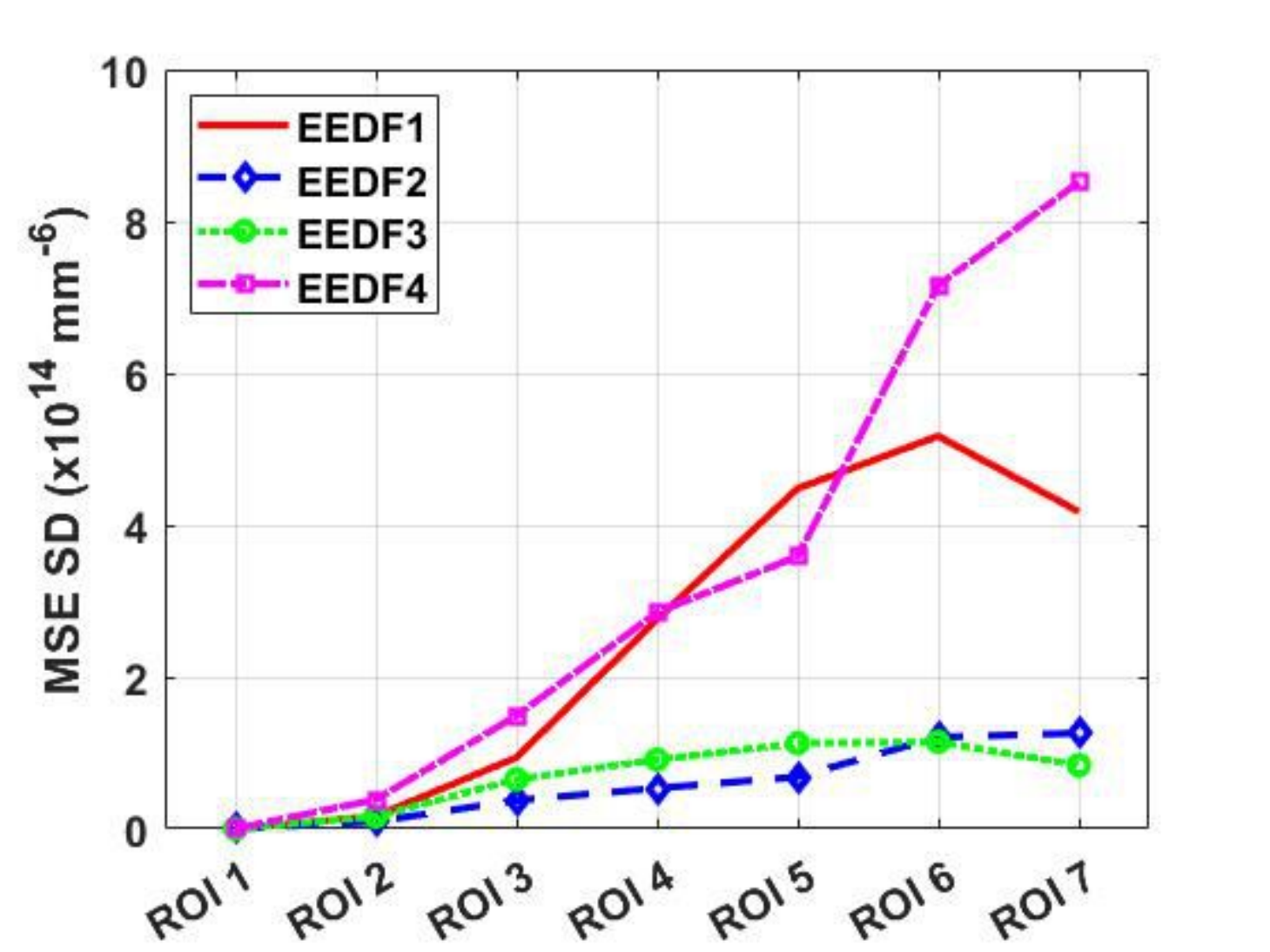}
  \end{subfigure} 
  \caption{ROI-averaged MSE for electron density (left) and SD of ROI-averaged MSE (right) in all ROIs, for $[2,\infty]\,\mathrm{keV}$ interval.}
  \label{fig5}
  \end{figure}

\begin{figure}
  \begin{subfigure}{0.49\columnwidth}
  \includegraphics[width=\textwidth]{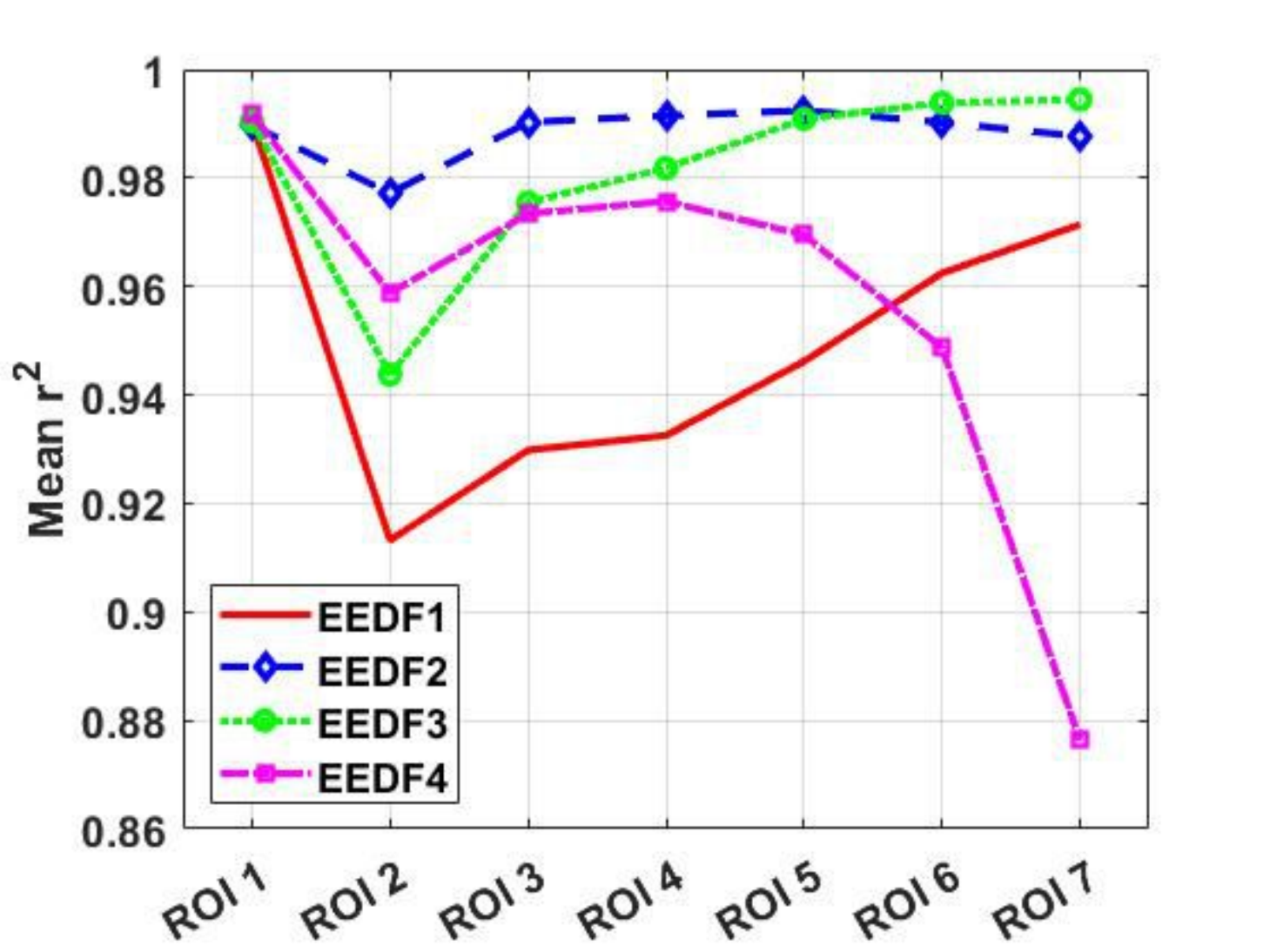}
  \end{subfigure}
  \hfill
  \begin{subfigure}{0.49\columnwidth}
  \includegraphics[width=\textwidth]{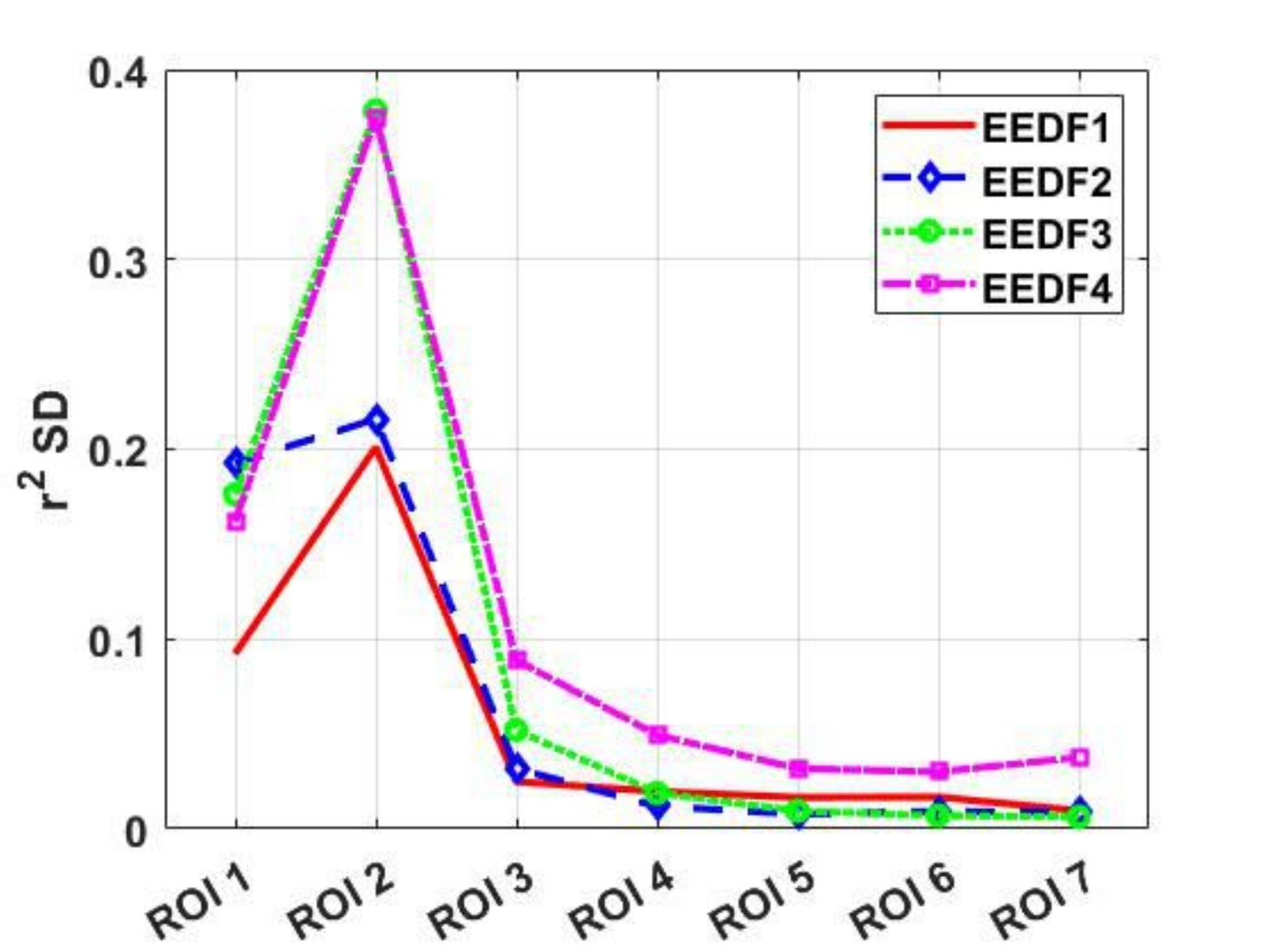}
  \end{subfigure} 
  \caption{ROI-averaged $r^{2}$-score for electron density (left) and SD of ROI-averaged $r^{2}$-score (right) in all ROIs, for $[2,\infty]\,\mathrm{keV}$ interval.}
  \label{fig6}
  \end{figure}

\begin{figure}
  \begin{subfigure}{0.49\columnwidth}
  \includegraphics[width=\textwidth]{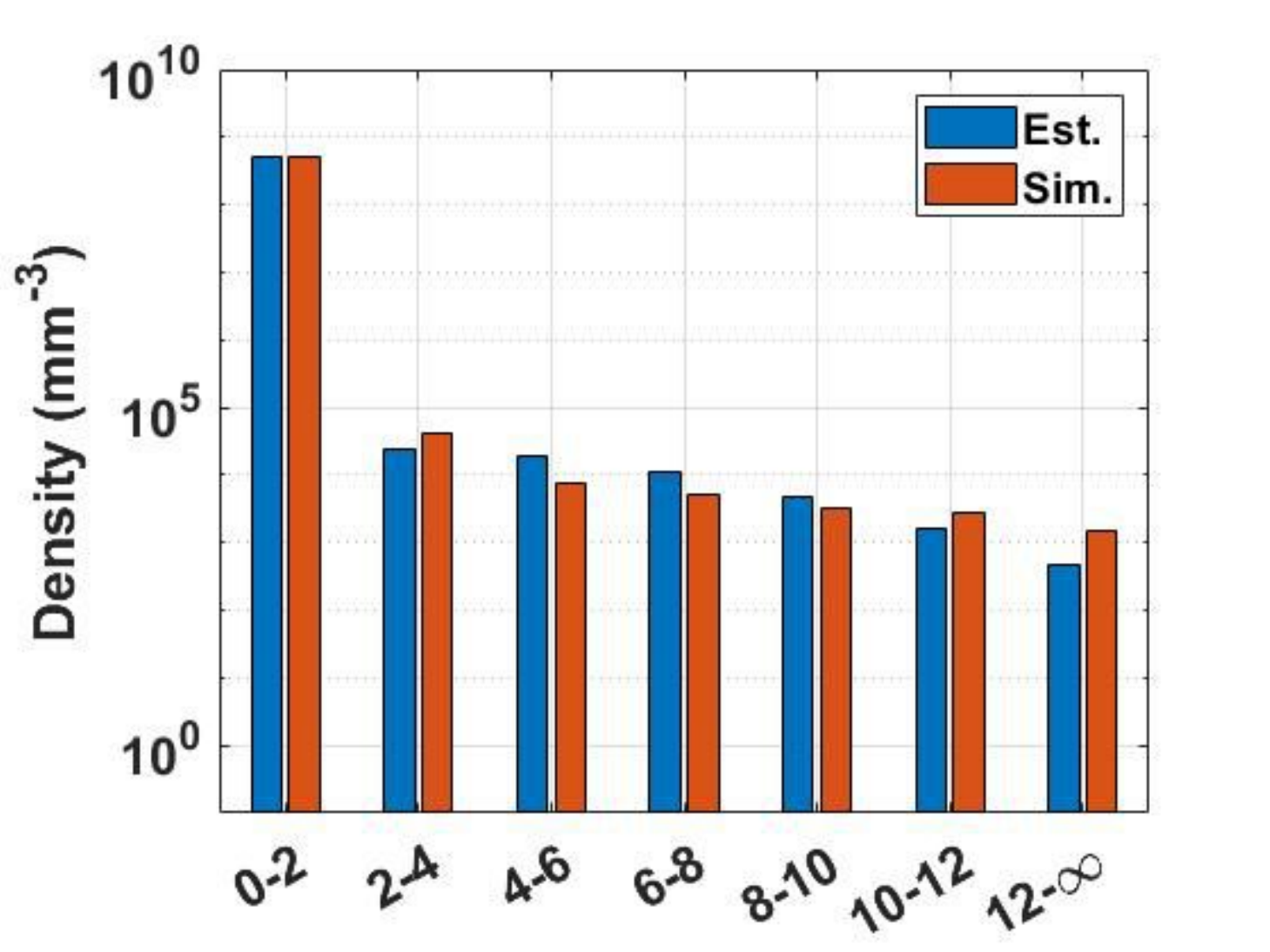}
  \end{subfigure}
  \hfill
  \begin{subfigure}{0.49\columnwidth}
  \includegraphics[width=\textwidth]{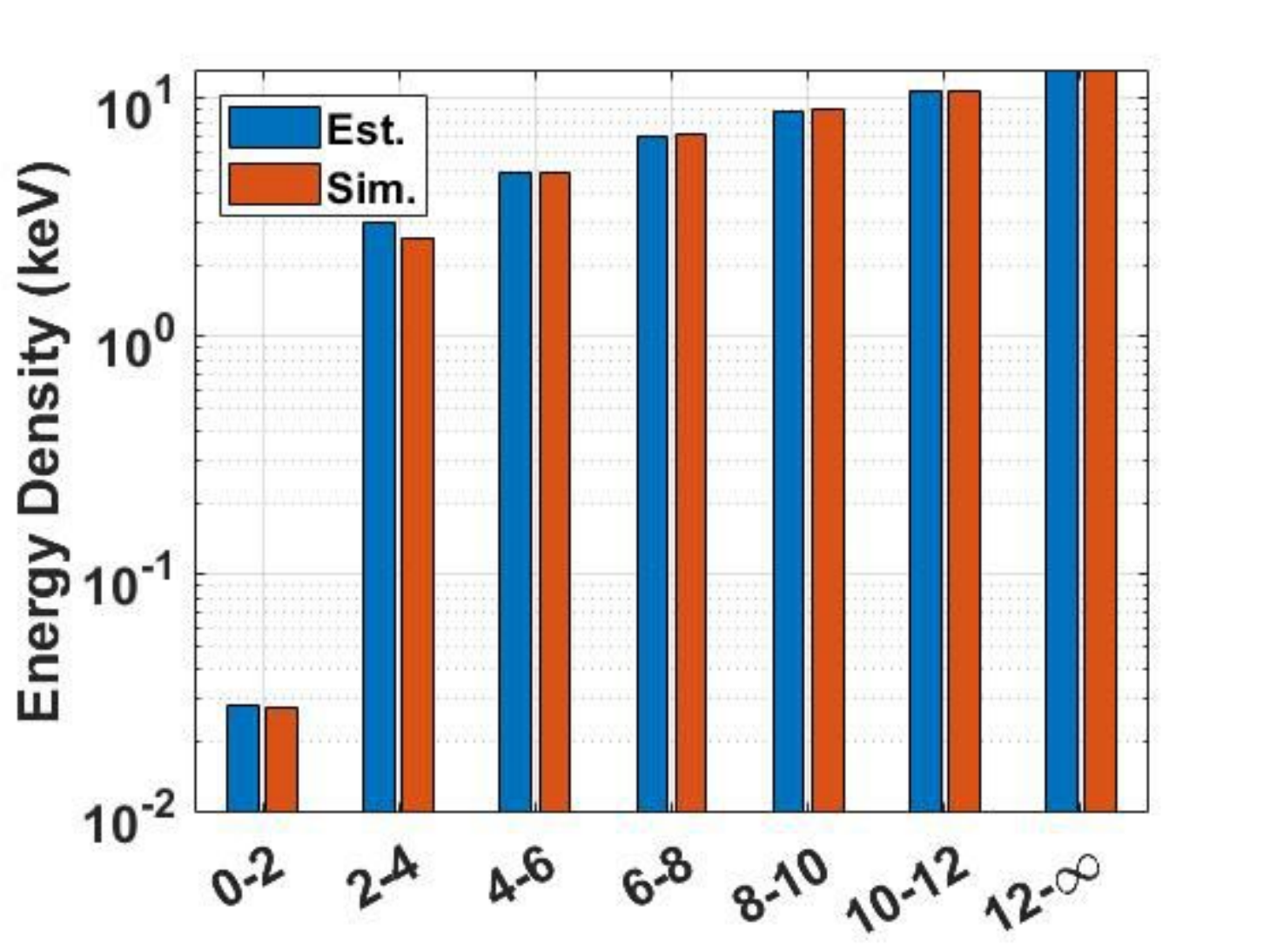}
  \end{subfigure} 
  \begin{subfigure}{0.49\columnwidth} 
  \includegraphics[width=\textwidth]{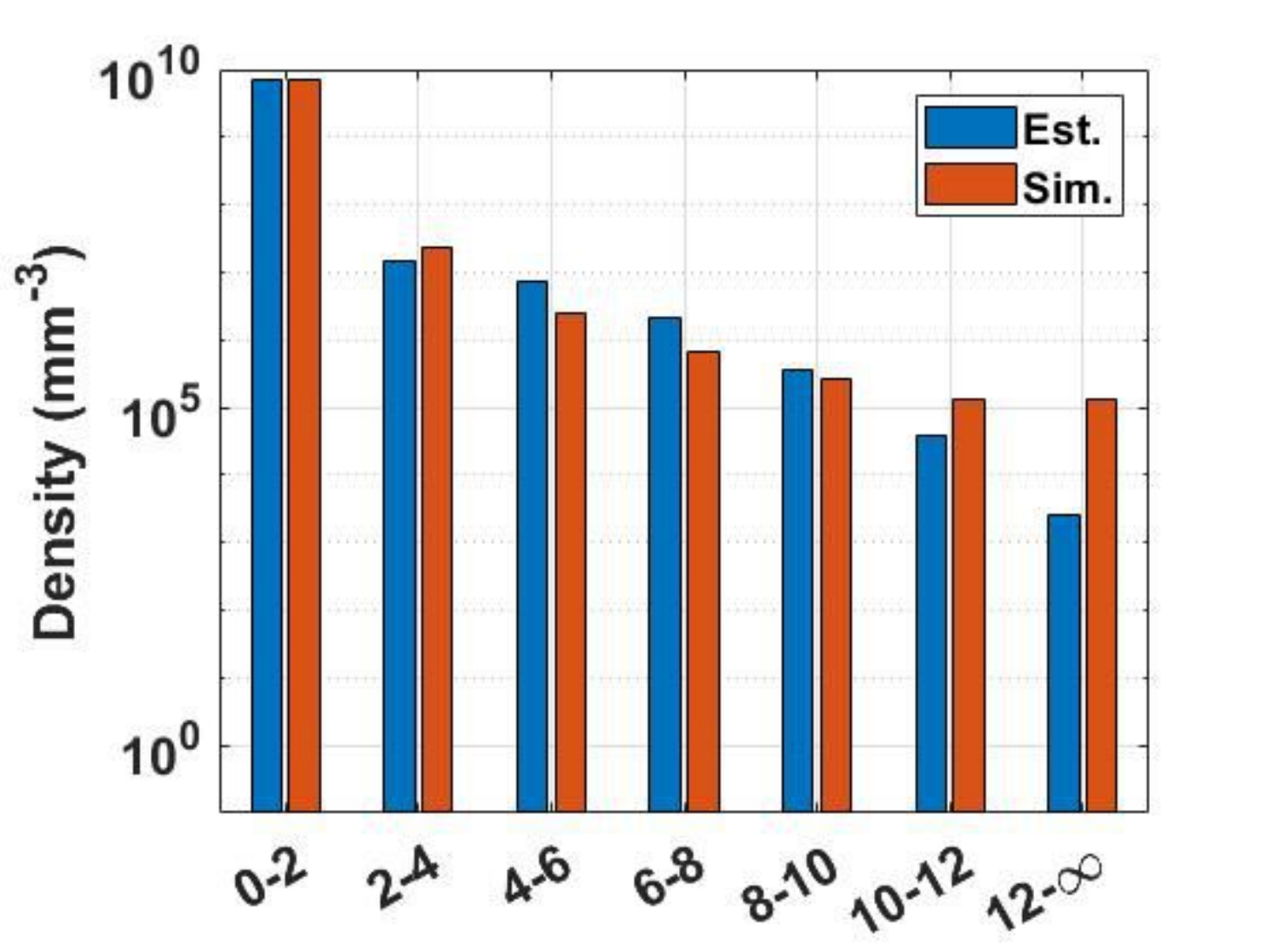} 
  \end{subfigure}  
  \hfill 
  \begin{subfigure}{0.49\columnwidth} 
  \includegraphics[width=\textwidth]{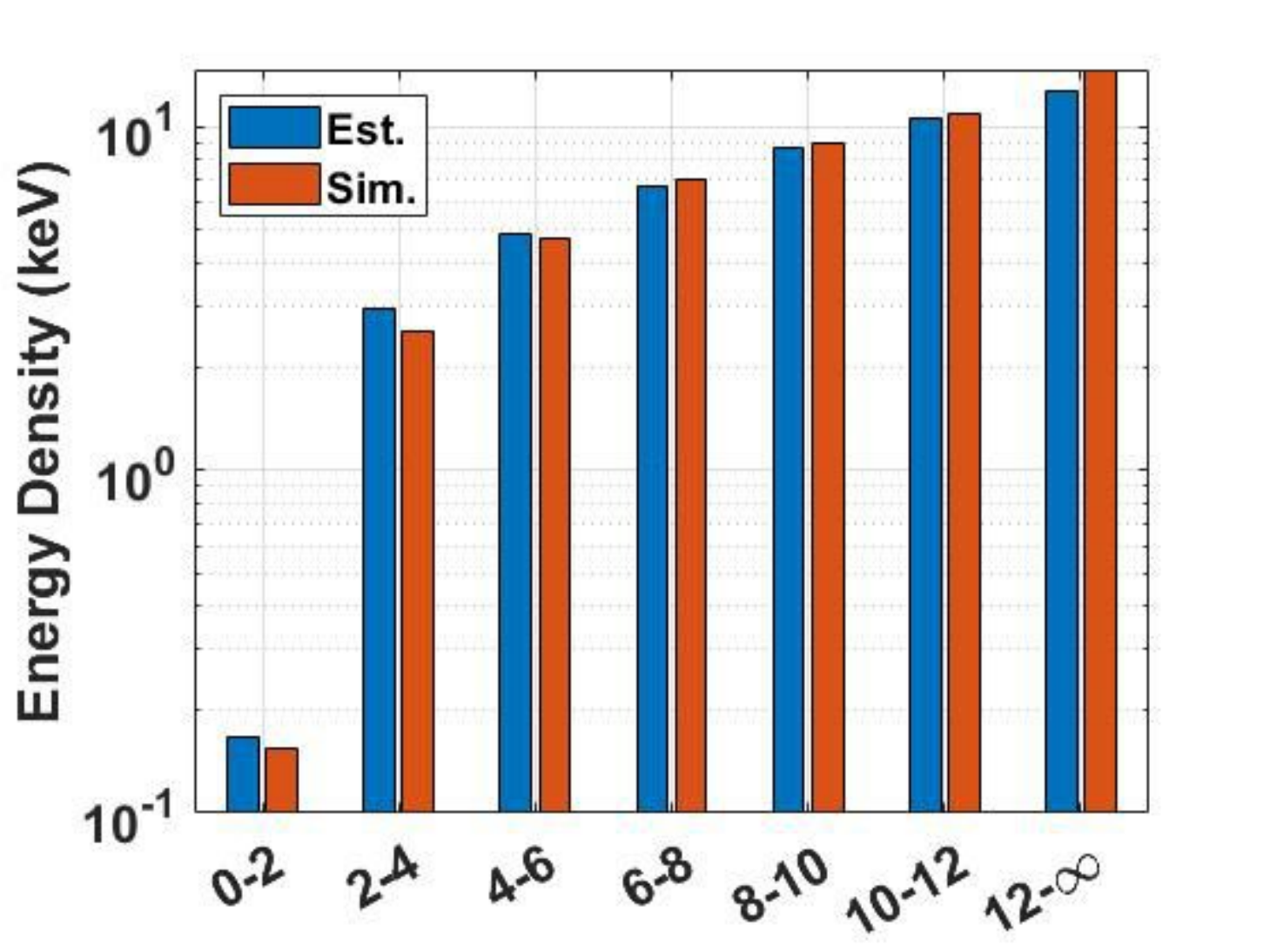}  
  \end{subfigure}
  \begin{subfigure}{0.49\columnwidth} 
  \includegraphics[width=\textwidth]{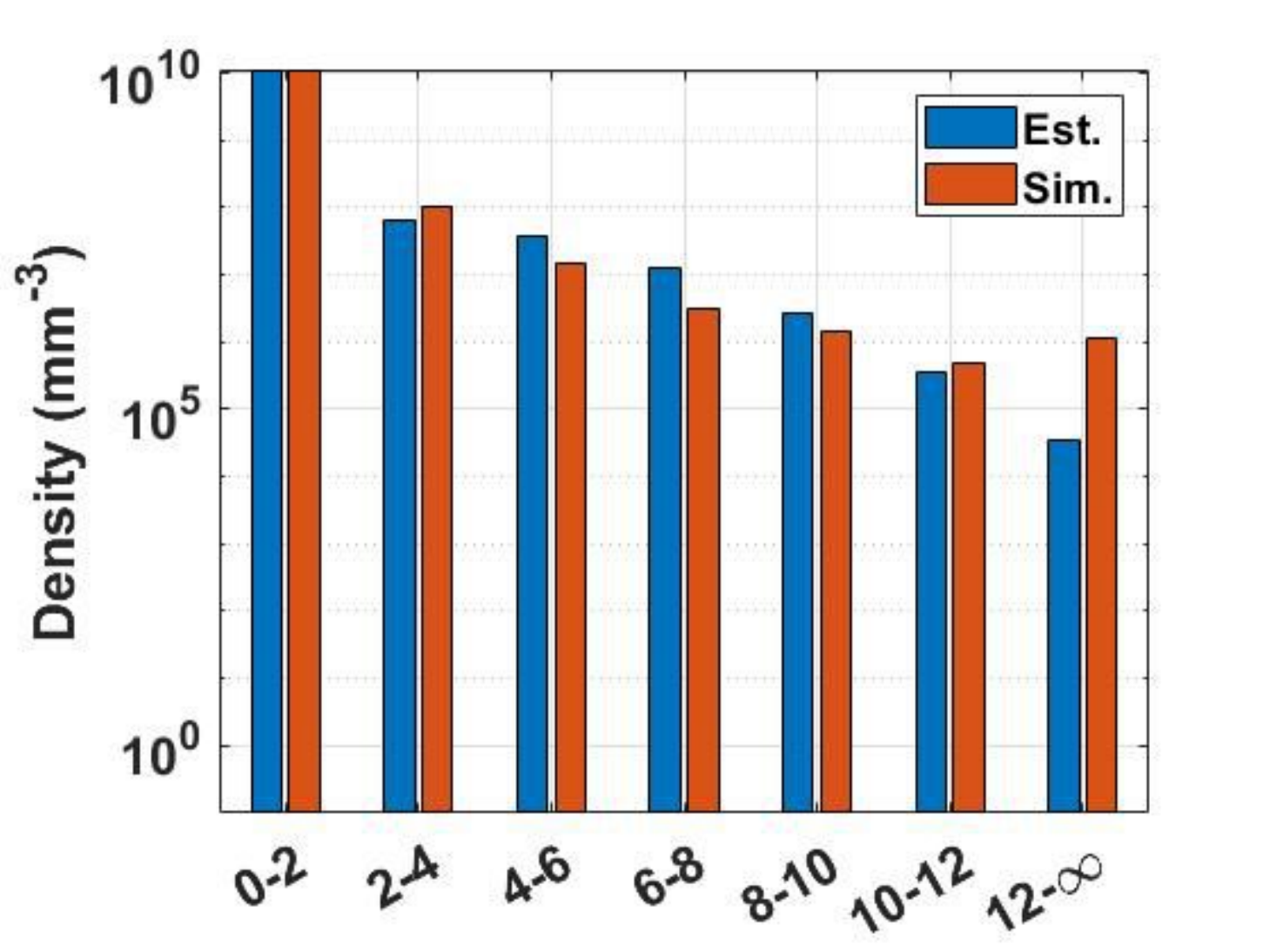} 
  \end{subfigure}  
  \hfill 
  \begin{subfigure}{0.49\columnwidth} 
  \includegraphics[width=\textwidth]{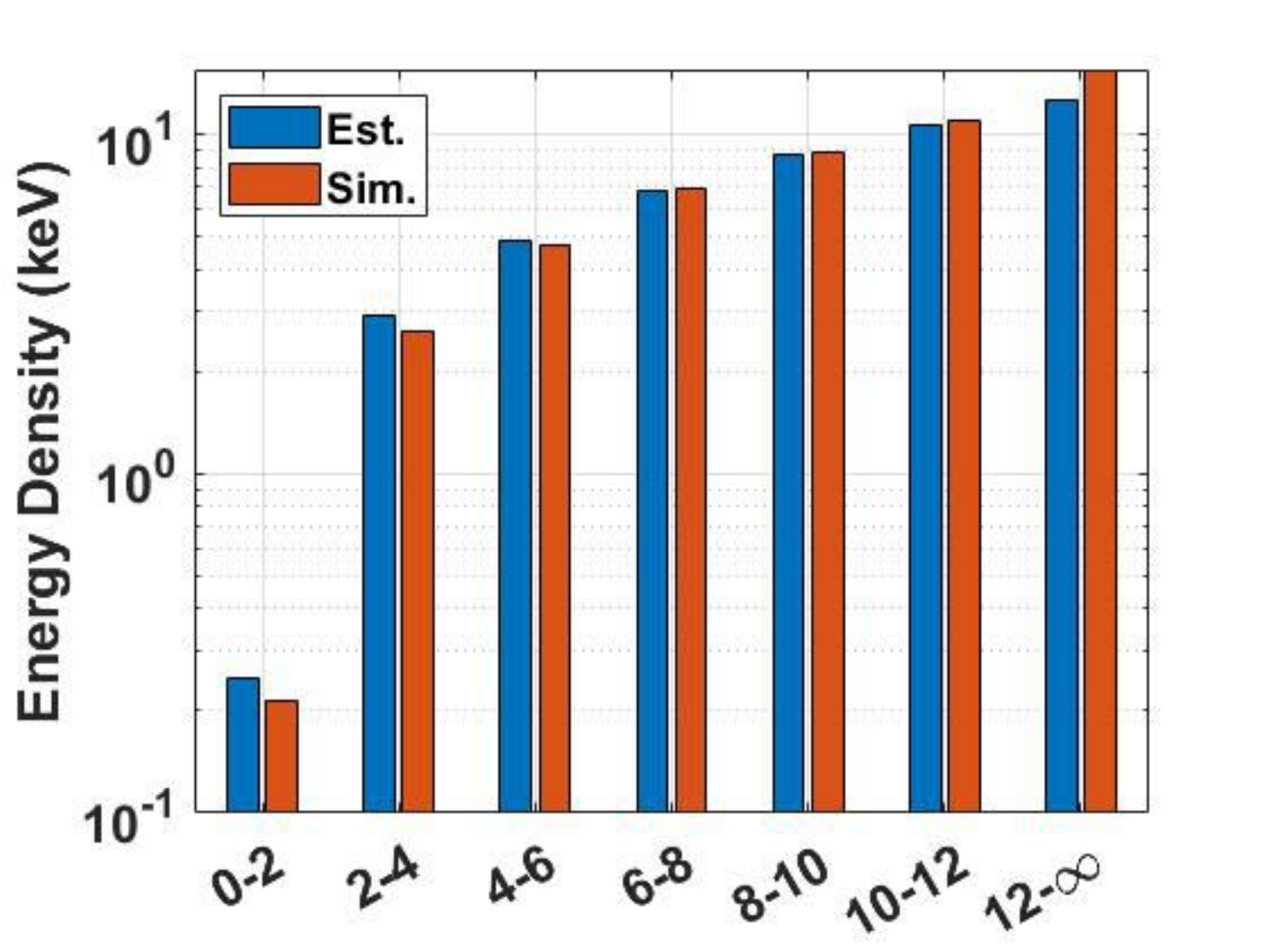}  
  \end{subfigure}
  \caption{ROI-averaged simulated (Sim.) electron density (left) and energy density (right) in the seven energy intervals for each of the ROIs displayed in Fig. \ref{fig:1}, against approximated results (Est.) numerically calculated using EEDF2.}
\label{fig:7}
 \end{figure}

Similarly, a three-component EEDF is taken to be of the form
\begin{align}
\label{eq16}
f(E;k_{B}T_{l},k_{B}T_{m},k_{B}T_{h})=A_{l}f_{l}(E;k_{B}T_{l})+\\\nonumber
A_{m}f_{m}(E;k_{B}T_{m})+A_{h}f_{h}({E;k_{B}T_{h}}),
\end{align}
where now we consider the electrons to be divided into \emph{three} separate populations, $[0,2]$, $[2,12]$ and $[12,\infty]$ $\mathrm{keV}$ intervals. Just like the two-component case, also here we have the same concerns with respect to the temperatures and normalization coefficients of the individual components. Thus, we first estimate the temperatures as 
\begin{align}
\label{eq17}
\mathbf{(k_{B}T_{\emph{l}})_{\emph{j}}}=\frac{2}{3}\mathbf{C_{\emph{j}}}\mathbf{E}_{1\emph{j}},\,\,\,&\mathbf{(k_{B}T_{\emph{m}})_{\emph{j}}}=\mathbf{S}_{1\emph{j}}\frac{\sum_{i=2}^{6}\bm{\rho}_{ij}\mathbf{E_{\emph{ij}}}}{\sum_{i=2}^{6}\bm{\rho}_{ij}},&\\\nonumber
& \mathbf{(k_{B}T_{\emph{h}})_{\emph{j}}}=\frac{2}{3}\mathbf{S}_{2\emph{j}}\mathbf{E}_{7\emph{j}}&,
\end{align}
with $\mathbf{C_{\emph{j}}}, \mathbf{S}_{1\emph{j}}$ and $\mathbf{S}_{2\emph{j}}$ being optimization factor matrices for ROI $j$.
Then the normalization coefficients are computed as,
\begin{equation}
\label{eq18}
\mathbf{A_{\emph{lj}}}=\frac{1}{\sum_{i=1}^{7}\bm{\rho}_{ij}}\bigg[\bm{\rho}_{1\emph{j}}-\frac{\mathbf{u_{\emph{j}}}}{\mathbf{p}_{1\emph{j}}\mathbf{q}_{2\emph{j}}-\mathbf{p}_{2\emph{j}}\mathbf{q}_{1\emph{j}}}(\mathbf{q}_{2\emph{j}}\sum_{i=2}^{6}\bm{\rho}_{ij}-\mathbf{p}_{2\emph{j}}\bm{\rho}_{7\emph{j}})\bigg],
\end{equation}
\begin{equation}
\label{eq19}
\mathbf{A_{\emph{mj}}}=\frac{1}{\sum_{i=1}^{7}\bm{\rho}_{ij}}\frac{\mathbf{q}_{2\emph{j}}\sum_{i=2}^{6}\bm{\rho}_{ij}-\mathbf{p}_{2\emph{j}}\bm{\rho}_{7\emph{j}}}{\mathbf{p}_{1\emph{j}}\mathbf{q}_{2\emph{j}}-\mathbf{p}_{2\emph{j}}\mathbf{q}_{1\emph{j}}},
\end{equation}
\begin{equation}
\label{eq20}
\mathbf{A_{\emph{hj}}}=\frac{1}{\sum_{i=1}^{7}\bm{\rho}_{ij}}\frac{\mathbf{p}_{1\emph{j}}\bm{\rho}_{7\emph{j}}-\mathbf{q}_{1\emph{j}}\sum_{i=2}^{6}\bm{\rho}_{ij}}{\mathbf{p}_{1\emph{j}}\mathbf{q}_{2\emph{j}}-\mathbf{p}_{2\emph{j}}\mathbf{q}_{1\emph{j}}}.
\end{equation}
The Eqs. (\ref{eq18}-\ref{eq20}) now involve multiple overlap factors because the medium- and high-energy distribution functions extend into each other's intervals, no matter how carefully the temperatures are chosen. These factors are calculated as 
\begin{align}
\mathbf{u_{\emph{j}}}=\int_{0}^{2}f_{\emph{m}}(E;\mathbf{(k_{B}T_{\emph{m}})_{\emph{j}}})\mathrm{d}E\,\,,\\\nonumber
\mathbf{p}_{1\emph{j}}=\int_{2}^{12}f_{\emph{m}}(E;\mathbf{(k_{B}T_{\emph{m}})_{\emph{j}}})\mathrm{d}E\,\,,\,\,& \mathbf{p}_{2\emph{j}}=\int_{2}^{12}f_{\emph{h}}(E;\mathbf{(k_{B}T_{\emph{h}})_{\emph{j}}})\mathrm{d}E,\\\nonumber
\mathbf{q}_{1\emph{j}}=\int_{12}^{\infty}f_{\emph{m}}(E;\mathbf{(k_{B}T_{\emph{m}})_{\emph{j}}})\mathrm{d}E\,\,,\,\,& \mathbf{q}_{2\emph{j}}=\int_{12}^{\infty}f_{\emph{h}}(E;\mathbf{(k_{B}T_{\emph{h}})_{\emph{j}}})\mathrm{d}E\,\,.\\\nonumber
\end{align}
The low-energy distribution's overlaps are neglected. The electron density and energy density are estimated in each interval using Eqs. (\ref{eq14},\ref{eq15}).
Therefore, each test EEDF generates the approximated data for each ROI. Once these data are provided, we need to determine \emph{quantitatively} the goodness-of-fit. This is accomplished by calculating the MSE and the $r^{2}$ coefficient, which are generally defined as 
\begin{equation}
\label{eq22}
MSE=\frac{1}{n}\sum_{p=1}^{n}(y_{p}-f_{p})^{2}\,,\,\,\,\,\,r^{2}=1-\frac{\sum_{p=1}^{n}(y_{p}-f_{p})^{2}}{\sum_{p=1}^{n}(y_{p}-\langle y \rangle)^{2}}
\end{equation}
where $\langle y \rangle=\frac{1}{n}\sum_{p=1}^{n}y_{p}$ is the mean of the \emph{n} \emph{true} data points $y_{p}$, and $f_{p}$ is the prediction for the same, made using a suitable model.
Usually when employing such statistics in data analytics, one has a large number of data points $y_{p}$, against which the model's prediction $f_{p}$ is tested, to calculate the cost function (most often the MSE), which is then minimized to deduce the best model parameters. While the general utility of the quantities is retained, an exact minimization process is not employed here due to lack of sufficient computational process to iteratively minimize a huge cost function composed of MSE values from each cell in any ROI. Instead, the performance of each EEDF is judged by first calculating the MSE and $r^{2}$-score of the EEDF fit in \emph{each} cell of the ROI, followed by calculation of the ROI-mean and standard deviation (SD) of the same, according to Eqs. (\ref{eq22} - \ref{eq24})
\begin{align}
\label{eq23}
\langle MSE \rangle_{j}=\frac{1}{N(j)}\sum_{k(j)=1}^{N(j)}MSE_{k(j)}\,\,\,\,\,, \\\nonumber
\sigma_{\langle MSE \rangle_{j}}=\frac{1}{\sqrt{N(j)-1}}\sqrt{\sum_{k(j)=1}^{N(j)}(MSE_{k(j)}-\langle MSE \rangle_{j})^{2}}
\end{align}
\begin{align}
\label{eq24}
\langle r^{2} \rangle_{j}=\frac{1}{N(j)}\sum_{k(j)=1}^{N(j)}(r^{2})_{k(j)}\,\,\,\,\,, \\\nonumber
\sigma_{\langle r^{2}\rangle_{j}}= \frac{1}{\sqrt{N(j)-1}}\sqrt{\sum_{k(j)=1}^{N(j)}((r^{2})_{k(j)}-\langle r^{2} \rangle_{j})^{2}}
\end{align}
where $k(j)$ is the index of any cell in the $j$-th ROI, $N(j)$ is the total number of cells in the same, $\langle MSE \rangle_{j}$ and $\langle r^{2} \rangle_{j}$ represent the ROI-averaged MSE and $r^{2}$-scores respectively, and $\sigma$ represents their respective SD. The optimization factors used in Eqs. (\ref{eq8},\ref{eq17}) are varied to obtain the best performance, which is marked by the lowest SD for both the MSE and $r^{2}$-score, and the highest and lowest values for the mean-$r^{2}$ and mean-MSE, respectively. This procedure is repeated for each ROI and with each EEDF. The resulting statistics is reported below in Table \ref{2and3EEDFs}, as well as plotted in Figs. \ref{fig4} - \ref{fig6}. All the above diagnostic quantities are calculated separately for the electron population in $[0,2]\,\mathrm{keV}$ and the remaining intervals, because the higher magnitude of electron density in the former interval largely obscures judgement of the fit in the others. Through this systematic analysis, a decent idea about the electron properties in different regions of the plasma is obtained.
From Figs. \ref{fig4} - \ref{fig6}, it is shown that EEDF2, i.e. the two-component function, made of a low-energy Maxwell and a high-energy Druyvesteyn distribution function, is the best one to describe the warm electron population in all ROIs considered, given this current scheme of plasma slicing. This result can be further appreciated in Fig. \ref{fig:7}, where we compare the same ROI-averaged data as in Eq. (\ref{eq3}) against the numerically approximated data of Eq. (\ref{eq5}), but with the single-component Maxwell distribution replaced by EEDF2.
It is worth noting that deeper in the plasma (ROIs $6$ and $7$), EEDF3 performs just as well if not better than EEDF2 because of improved statistics in the $[12,\infty]\,\mathrm{keV}$ interval (which the short tail of the DR distribution cannot fit in its entirety). But this result does not suggest some fundamental physics about ECR plasmas, at least not at this stage of the investigation.\\\\ The performance of each EEDF and subsequent selection based on statistical metrics is subject to changes in the raw electron data and the ROI generation scheme, making the functions purely phenomenological in nature. Here the raw data included are both cold ($0-2\,\mathrm{keV}$) and warm electrons ($2-20\,\mathrm{keV}$), with finer resolution in the latter for emphasis. This forced us to assume multi-component distribution functions to account for the large disparity in counts between the two sub-populations. The EEDFs may revert to single-components if more energetic electrons are simulated (see Sec. \ref{sec:5}).\\ Similarly, by adopting different plasma slicing schemes, one may obtain shifts in statistics between the various intervals, thus altering the EEDFs as well. The present tool in its current form is only a method to convert discrete data into continuous functions for precise calculations, and no further understanding about charge particle dynamics in ECR plasmas may be obtained without suitable experimental validation. However, with regards to analysis of warm (plus cold) electrons alone, the results underline the nature of ECR plasma anisotropy, in the sense that not only do the EEDF parameters change as a function of real-space position, but the EEDFs themselves change too.    

\section{Electron Density and Temperature from X-ray Spectroscopy}
\label{sec:5}

\begin{figure}
\resizebox{1.00\columnwidth}{!}{\includegraphics{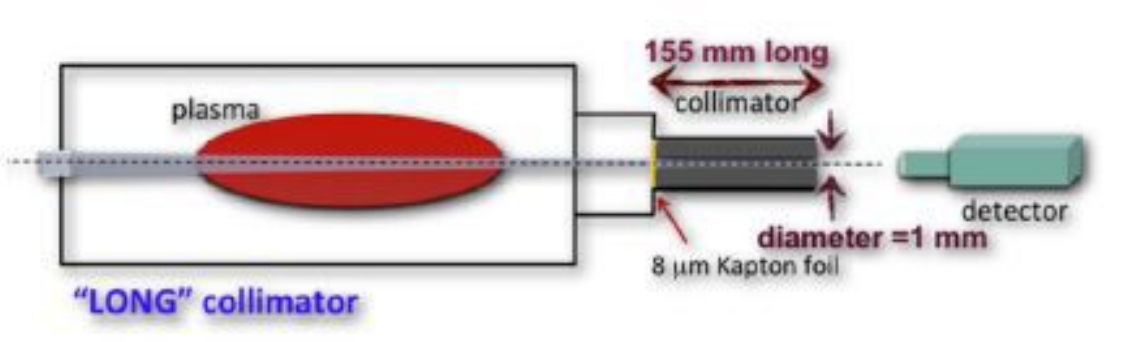}}
\caption{Sketch of the SDD long collimator configuration to filter X-rays from walls and extraction plate. Figure adapted from \cite{Ref17}.}
\label{fig:8}
\end{figure}

\begin{figure}
\resizebox{1.00\columnwidth}{!}{\includegraphics{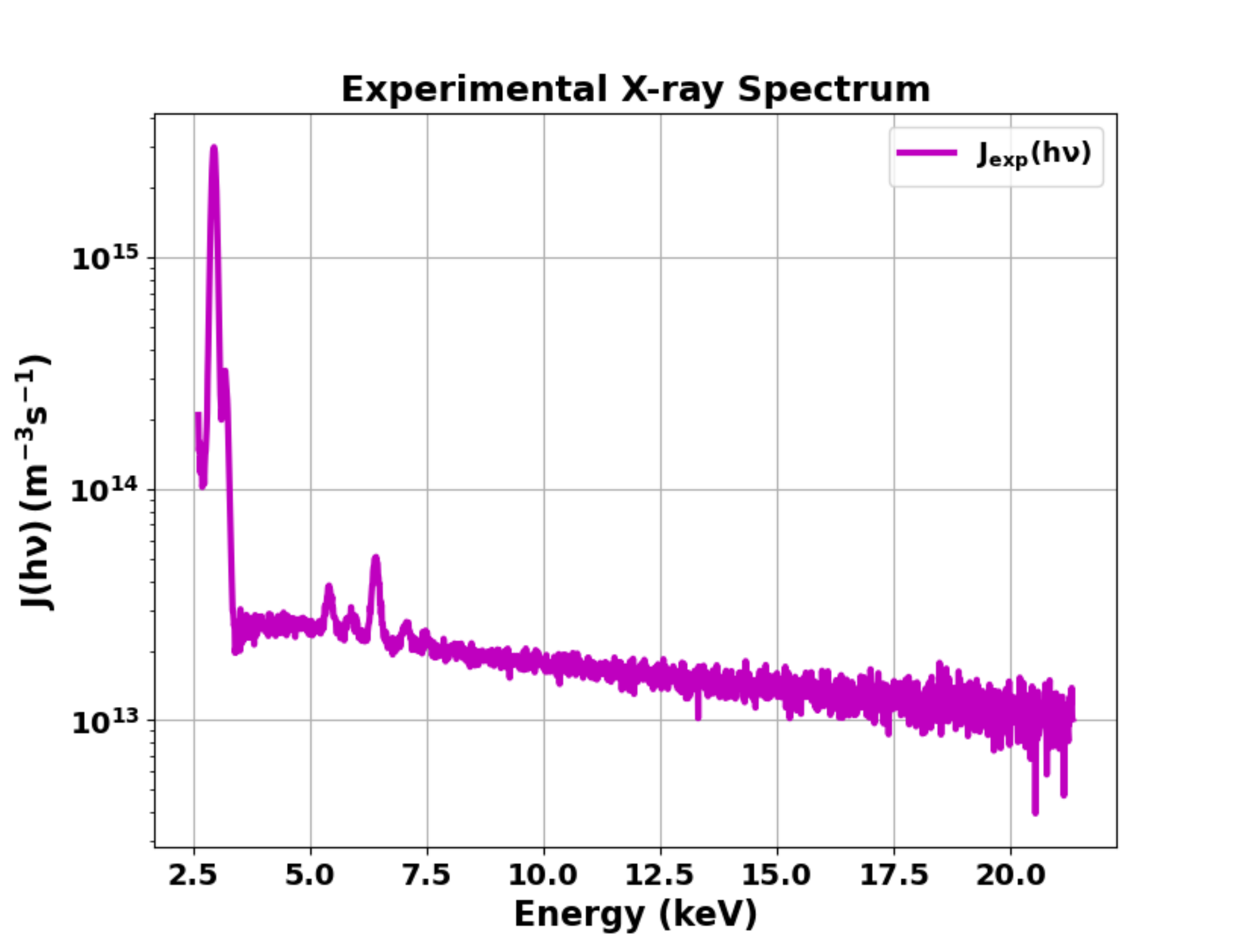}}  
\caption{Experimentally measured X-ray emissivity density of plasma.}
\label{fig:9}
\end{figure}

X-ray emission spectroscopy is a powerful passive diagnostic tool for deducing plasma properties because radiation emissivity densities can be directly correlated with charge particle reaction rates. The technique is particularly useful for ECR plasmas because it allows different regions to be probed by customising the experimental setup, helping resolve the inherent anisotropy. A complete qualitative and quantitative analysis of the spectra can reveal precious information about the relevant populations involved, including ionic charge state (distribution), inner-shell ionisation rate, electron and ion densities, and energy distribution functions (EDFs). Many groups have already worked on these ideas - Douysset \emph{et al} estimated ion densities of individual charge states $n_{i}^{q}$ and correlated them with extracted mass spectra using model-based prediction of K$\alpha$ intensities \cite{Ref30}, Santos \emph{et al} determined the ion CSD of a sulphur plasma by fitting the intensity of line spectra using suitable models \cite{Ref31} and Sakildien \emph{et al} studied the variation of K-shell ionisation rate of argon ions as a function of plasma parameters by calculating K$\alpha$ emissivity densities and comparing them with experimental results \cite{Ref32}.\\\\
In 2014, the INFN-LNS and ATOMKI groups performed a number of experiments on a state-of-the-art ECR trap capable of simultaneously recording output currents, high-energy bremsstrahlung ($>30$ keV) and mid-energy X-rays ($2-30$ keV). This enabled them to correlate the plasma density and temperature with output CSD and beam intensity for different plasma operating conditions \cite{Ref17}. The setup used a bending magnet with a Faraday cup for measuring the extracted currents, while the high and mid-energy X-rays were detected using an Hp\ce{{Ge}} and SDD detector in long collimator configuration respectively. The SDD could be switched with a CCD camera in pinhole setup for capturing the space-resolved X-ray images instead as well. Since bremsstrahlung above $30\,\mathrm{keV}$ can be generated by electrons having at least that much energy, only the SDD spectrum is used here for purposes of studying the warm and intermediate energy electrons. The long collimator setup is shown in Fig. \ref{fig:8}.\\
As already mentioned in the previous sections, the numerical method presented, and the estimated EEDFs thereof, need to be experimentally corroborated to obtain insight into the physics of ECR devices. For this purpose, and on the basis of the setup of \cite{Ref17} as shown in Fig. \ref{fig:8}, we perform a quantitative analysis of the measured X-ray spectrum taken from the argon plasma within the solid-angle subtended by the collimator. This will allow us to evaluate the contribution of the warm electrons to the spectrum (thus assessing the possibility for benchmarking), while also estimating the electron density for macroparticle scaling. Consequently, instead of employing the EEDF deduced from Sec. \ref{sec:4} right away, we first employ a standard single-component MB distribution to check the quality of fit to the spectrum, and proceed with the verification of EEDF2 only if the warm electron share is deemed substantial.\\
In order to facilitate analysis, the raw data from argon plasma is calibrated using iron lines, renormalised for photon detection probability (quantum efficiency renormalisation) and corrected for dead time (details in Ref. \cite{Ref17}). The final spectrum is converted from counts per energy $N^{p}$ into the emissivity density using the expression
\begin{equation}
\label{eq25}
J(h\nu)=h\nu\frac{N^{p}(h\nu)}{t}\frac{4\pi}{\Delta EV_{P}\Omega_{g}}
\end{equation}
where $t$ is the acquisition time, $\Delta E$ is the energy per channel, $h\nu$ is the photon energy, $V_{p}$ is the plasma emission volume and $\Omega_{g}$ is the geometrical efficiency of the collimator setup. Equation \ref{eq25} refers to the total energy emitted per unit time, volume and energy, through photons of energy $h\nu$. Considering the acquisition time as $900\,\mathrm{s}$, geometrical efficiency $1.16\times10^{-6}$ and the emission volume as $18\,\mathrm{cm^{3}}$, the experimental plasma emissivity density $J_{exp}(h\nu)$ (in units $m^{-3}s^{-1}$) is shown in Fig. \ref{fig:9}.
Using the fact that the plasma X-ray spectrum can be decomposed into continuous bremsstrahlung and discrete line emission, the analytical expression for the theoretical emissivity density $J_{theo}(h\nu)$ is given as a simple algebraic sum of the two in Eq. \ref{eq26} 
\begin{equation}
\label{eq26}
J_{theo}(h\nu)=J_{theo,brem}(h\nu)+J_{theo,line}(h\nu).
\end{equation}
The basic expression for the bremsstrahlung emissivity density is given as  
\begin{equation}
\label{eq27}
J(h\nu)=\rho_{e}\rho_{i}h\nu\int_{h\nu}^{\infty}\frac{\mathrm{d}\sigma_{K}(h\nu)}{\mathrm{d}h\nu}v_{e}(E)f(E)\mathrm{d}E
\end{equation}
where $\rho_{e}\rho_{i}$ is the product of the electron and ion density in the plasma, $h\nu$ is the photon energy, $v_{e}$ is the electron speed, $\mathrm{d}\sigma_{K}(h\nu)/\mathrm{d}h\nu$ is the differential cross-section for a photon to be emitted after an electron collides with a nucleus of charge $Z$, and $f(E)$ is the usual EEDF. Using Kramer's formula for the differential cross-section \cite{Ref33}
\begin{equation}
\label{eq28}
\frac{\mathrm{d}\sigma_{K}(h\nu)}{\mathrm{d}h\nu}=\frac{16\pi}{3\sqrt{3}}\alpha^{3}\bigg(\frac{\hbar}{m_{e}c}\bigg)^{2}\bigg(\frac{c}{v_{e}}\bigg)^{2}\frac{Z^{2}}{h\nu}
\end{equation}
and a MB EEDF for simplicity, the bremsstrahlung emissivity density is calculated as 
\begin{equation}
\label{eq29}
J_{theo,brem}(h\nu)=\rho_{e}\rho_{i}(Z\hbar)^{2}\bigg(\frac{4\alpha}{\sqrt{6m_{e}}}\bigg)^{3}\bigg(\frac{\pi}{k_{B}T_{e}}\bigg)^{1/2}\mathrm{e}^{(-h\nu/k_{B}T_{e})}
\end{equation}
In Eqs. \ref{eq28} and \ref{eq29}, $m_{e}$ is the electron mass, $c$ is the speed of light, and $\alpha$ is the fine structure constant. While electron distributions with any temperature can contribute to bremsstrahlung as long as they contain particles above the photon energy $h\nu$, in general contributions from more energetic EEDFs mask that from low-energy counterparts, and hence $k_{B}T_{e}$ is expected somewhere around $20\,\mathrm{keV}$ as deduced in Ref. \cite{Ref17} which is at the boundary between the warm and hot electron populations.

The line emissivity density is a slightly more complicated problem owing to greater number of inputs required to generate the model. Through qualitative inspection, the main peaks around $2.96$ and $3.19\,\mathrm{keV}$ in Fig. \ref{fig:9} are identified as the argon K$\alpha$ and K$\beta$ peaks respectively, while the smaller peaks between $5$ and $7\,\mathrm{keV}$ are the same characteristic radiation arising from chromium and iron atoms from the extraction plate (the area subtended by the collimator on the end plate together with the extraction hole produces a narrow ring-like active area, leading to small emission lines). The basic expression for the line emissivity density of the argon ions interacting with the confined electrons for the $nl\rightarrow nl'$ transition is 
\begin{equation}
\label{eq30}
J_{nl\rightarrow nl'}=\frac{h\nu_{nl\rightarrow nl'}}{\Delta E}\rho_{e}\rho_{i}\omega_{nl\rightarrow nl'}\int_{I}^{\infty}\sigma_{nl,ion}(E) v_{e}(E)f(E)\mathrm{d}E
\end{equation}
where $h\nu_{nl\rightarrow nl'}$ is the photon energy,$\omega_{nl\rightarrow nl'}$ is the fluorescence factor associated with the transition, $\Delta E$ is the energy per channel, $I$ is the binding energy of the $nl$ orbital, and $\sigma_{nl,ion}$ is the ionisation cross-section of the same. In case of the chromium and iron atoms, however, the same is given by the expression 
\begin{equation}
\label{eq30a}
J_{nl\rightarrow nl'}=\frac{h\nu_{nl\rightarrow nl'}}{\Delta EV_{P}}\rho_{e,loss}N_{i}\omega_{nl\rightarrow nl'}\int_{I}^{\infty}\sigma_{nl,ion}(E) v_{e}(E)f(E)\mathrm{d}E
\end{equation}
owing to the fact that instead of interacting with a gaseous species, the target atoms are now producing fluorescence due to collision with a beam of escaping electrons. Now $\rho_{e,loss}$ represents the loss electron density, $N_{i}$ is the number of target atoms and $V_{P}$ is introduced to convert the \emph{total} extraction plate fluorescence emissivity density into a volume-averaged value to fit Fig. \ref{fig:9}. Equation \ref{eq30a} still needs modification on account of factors like velocity distribution of beam electrons, constant loss of energy due to collisions and radiation inside the target material, and truncation of penetration depth, all of which together manifest in the final expression as
\begin{widetext}
\begin{equation}
J_{nl\rightarrow nl'}=\frac{h\nu_{nl\rightarrow nl'}}{\Delta EV_{P}}\rho_{e,loss}n_{i}\omega_{nl\rightarrow nl'}\bigg(4\pi\varepsilon_{g}l^{2}-\pi\frac{d^{2}}{4}\bigg)\int_{I}^{\infty}v_{e}(E)f(E)\int_{E}^{I}\frac{1}{S(E')}\sigma_{nl,ion}(E')\mathrm{d}E'\mathrm{d}E 
\label{eq30b}
\end{equation}
\end{widetext}
Here $n_{i}$ is the target atom number density, $\varepsilon_{g}=\Delta\Omega/4\pi$ is the geometrical efficiency of the collimator setup, $l$ is the separation between the extraction plate and detection cone vertex, $d$ is the diameter of the extraction hole and $S(E)$ is the total stopping power of the material. 

\begin{figure}
  \begin{subfigure}{0.49\columnwidth}
  \includegraphics[width=\textwidth]{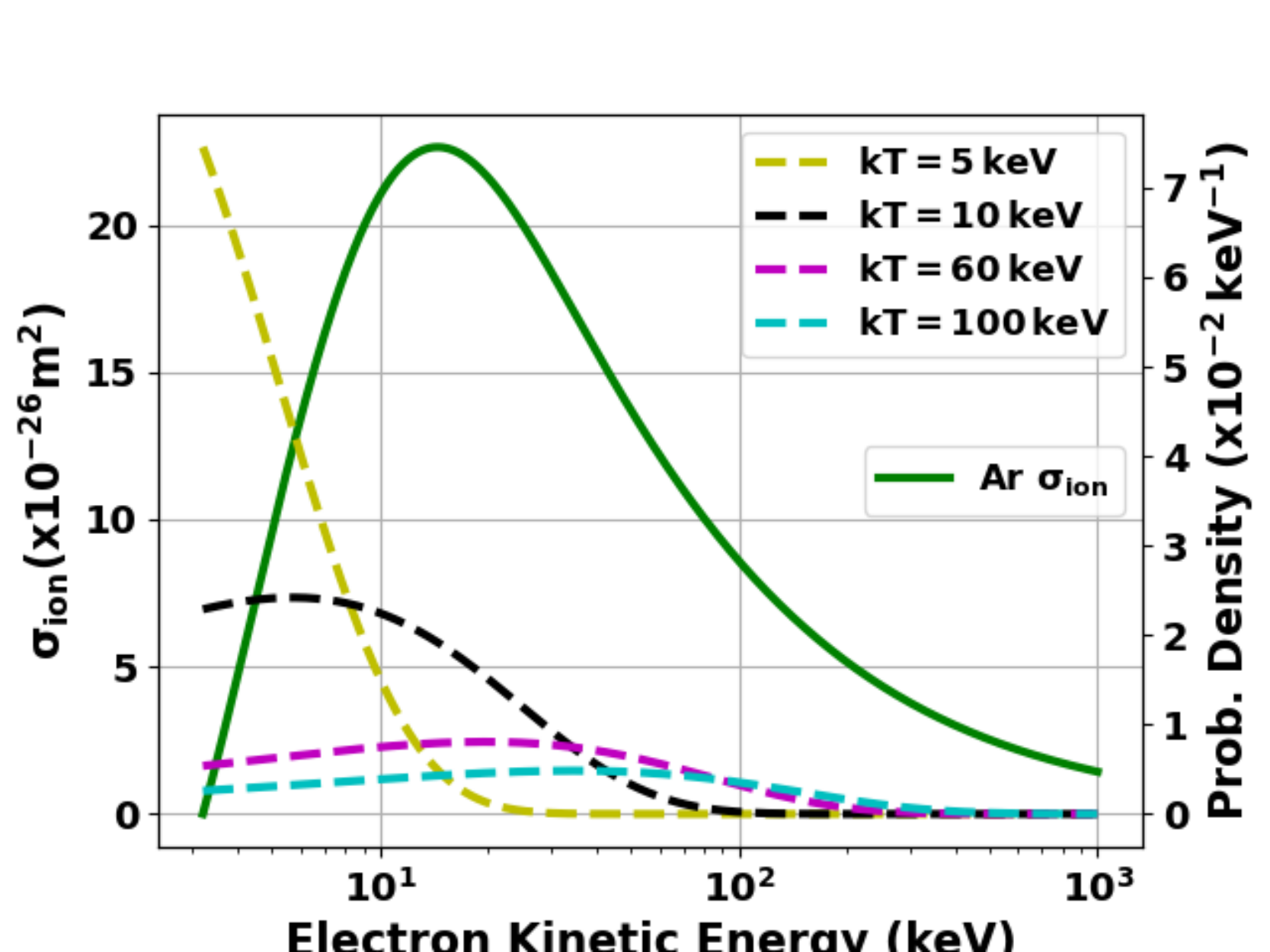}
  \end{subfigure}
  \hfill
  \begin{subfigure}{0.49\columnwidth}
  \includegraphics[width=\textwidth]{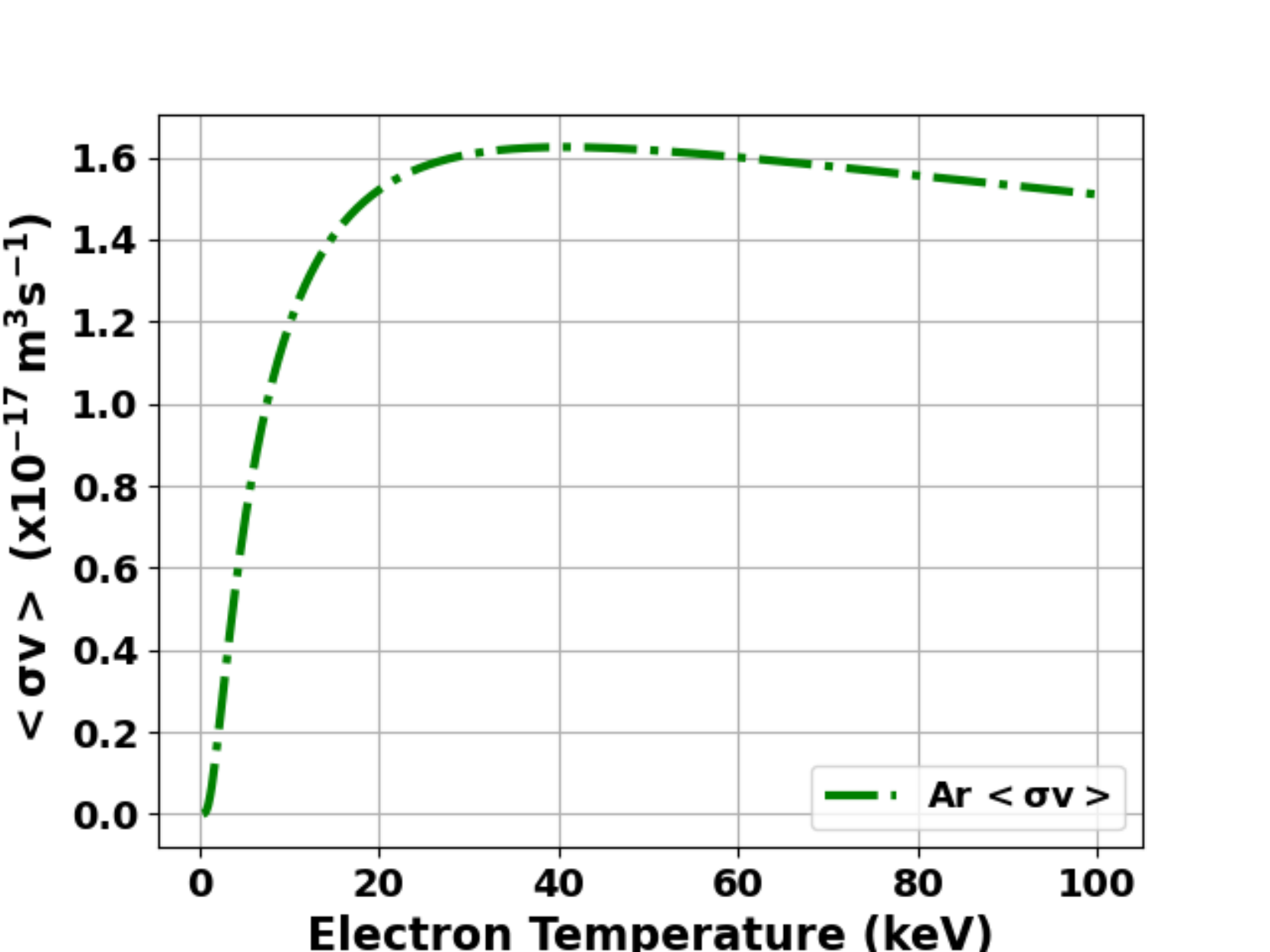}
  \end{subfigure} 
  \begin{subfigure}{0.49\columnwidth} 
  \includegraphics[width=\textwidth]{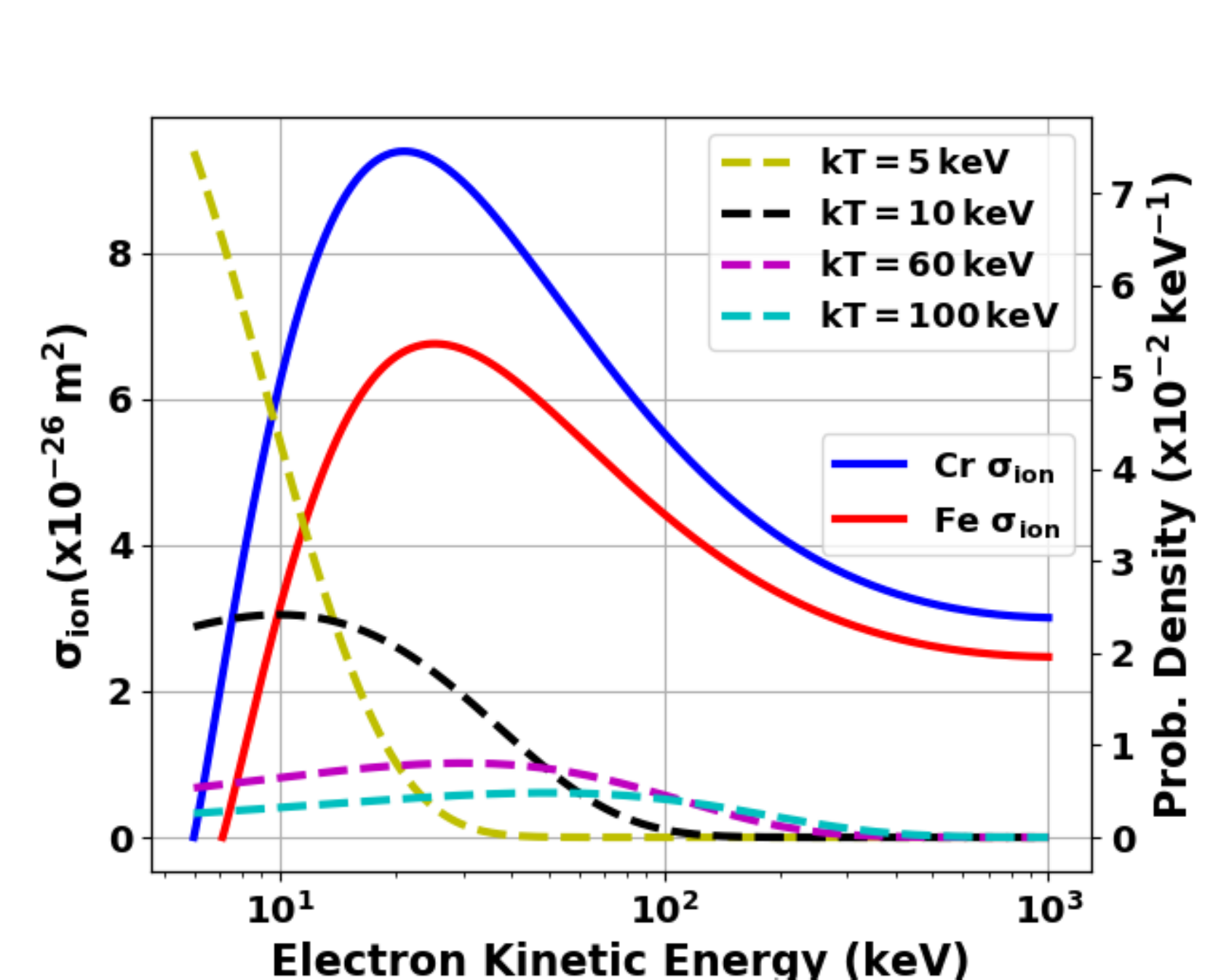} 
  \end{subfigure}  
  \hfill 
  \begin{subfigure}{0.49\columnwidth} 
  \includegraphics[width=\textwidth]{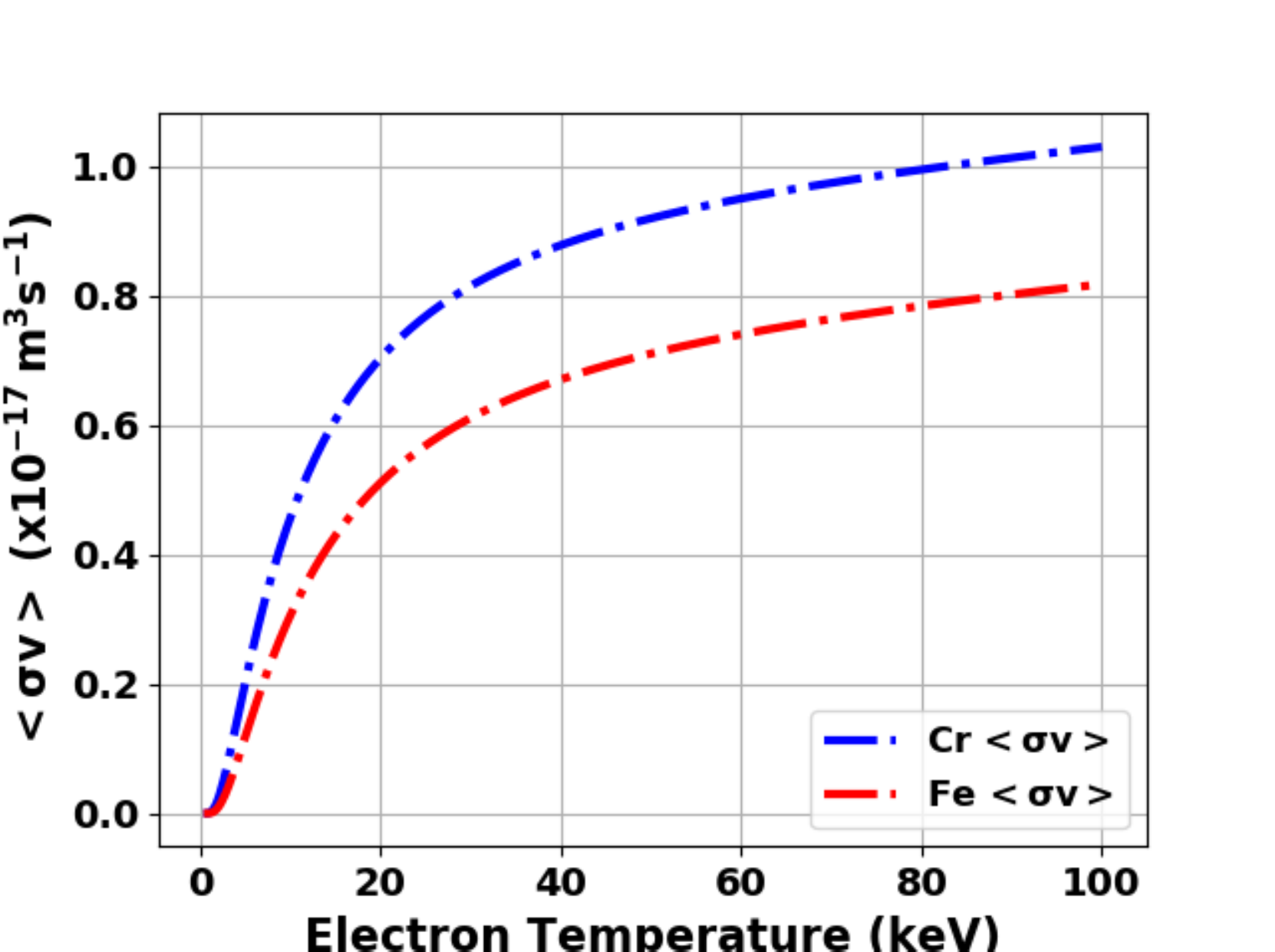}  
  \end{subfigure}
  \caption{(Top) Lotz cross-section for K-shell ionisation with different $k_{B}T_{e}$ MB EEDFs (left) and EEDF-averaged cross-section with upper limit of integral truncated to $2000\,\mathrm{keV}$ (right) and (bottom) the same for DM cross-section.}
\label{fig:10}
 \end{figure}

Just like bremsstrahlung, both low and high-temperature EEDFs can contribute to the radiation as long as there exist electrons with energy above $I$, but concrete conclusions can be made only after analysing the overlap between the energy-dependent $\sigma_{ion,nl\rightarrow nl'}$ and $f(E)$ which is simply the integral in Eq. \ref{eq30}. In case of argon, the semi-empirical Lotz cross-section is used to calculate ionisation rates from $1s$ (K) shell \cite{Ref34}
\begin{equation}
\label{eq31}
\sigma_{1s,ion}=a_{1s}q_{1s}\frac{\ln \varepsilon/I}{\varepsilon I}\{1-b_{1s}\mathrm{exp}[-c_{1s}(\varepsilon/I-1)]\}
\end{equation} 
where $a_{1s}$, $b_{1s}$ and $c_{1s}$ are constants which do not strongly depend on the charge state of the ion and are determined respectively as $4.0\times10^{-14}\,\mathrm{cm^{2}eV^{2}}$, $0.75$ and $0.5$, $I$ is the K-shell ionisation energy ($\sim3.205\,\mathrm{keV}$), $\varepsilon$ is the electron kinetic energy and $q_{1s}$ is the number of equivalent electrons in the K-shell ($=2$). Similarly for chromium and iron, an improved expression valid till relativistic energies is used, derived under the Deutsch-M\"{a}rk formalism \cite{Ref35}
\begin{equation}
\label{eq32}
\sigma_{1s,ion}=g_{1s}\pi(r_{1s})^{2}\xi_{1s}f(U)F(U)
\end{equation}
Here $g_{1s}$ is the weight factor, $r_{1s}$ is the radius of maximum radial density, $\xi_{1s}$ is the number of electrons in the K-shell ($=2$) and the functions $f(U),F(U)$ define the energy dependence of the cross-section. $U=E/E_{1s}$ is the reduced impact energy where $E_{1s}=I$ is the ionisation energy of the K-shell and the functions $f_{U},F_{U}$ are defined as
\begin{equation}
\label{eq33}
f(U)=d\frac{1}{U}\bigg[\frac{U-1}{U+1}\bigg]^{a}\{b+c\bigg[1-\frac{1}{2U}\bigg]\ln[2.7+(U-1)^{1/2}]\}
\end{equation}
\begin{equation}
\label{eq34}
F(U)=R(U)\bigg[1+2\frac{U^{1/4}}{J^{2}}\bigg]
\end{equation}
In Eq. \ref{eq33}, the parameters have predetermined values given as $a=1.06$, $b=0.23$, $c=1.00$ and $d=1.1$, while in Eq. \ref{eq34}, $F(U)$ is the relativistic correction factor with $J=(m_{e}c^{2})/E_{1s}$, $m_{e}c^{2}$ being the electron rest mass energy. Data on $r_{1s}$ and $g_{1s}$ in Eq. \ref{eq32} can be found respectively in Ref. \cite{Ref36} and \cite{Ref37} and $R(U)$ is given by Eq. \ref{eq35}
\begin{equation}
\label{eq35}
R(U)=\frac{1+2J}{U+2J}\bigg[\frac{U+J}{1+J}\bigg]^{2}\bigg[\frac{(1+U)(U+2J)(1+J)^{2}}{J^{2}(1+2J)+U(U+2J)(1+J)^{2}}\bigg]^{1.5}
\end{equation}
The visual representation of the cross-section and EEDF overlap for all these ions is shown in Fig. \ref{fig:10}, and it is clear that here too the contribution of EEDFs with $k_{B}T_{e}\sim20\,\mathrm{keV}$ will dominate the spectrum, in line with the bremsstrahlung analysis. We then conclude straight-away that analysis of the argon X-ray spectrum cannot be used as an experimental benchmark for the EEDFs and/or warm electrons obtained from the simulations, because their contribution is much too small as compared to a hotter electron population which exists for sure in the plasma. However, it is still worth proceeding with the analysis to extract crucial properties of these electrons.  
The final step is accounting for the broadening of the line emissivity density peak, and a pseudo-Voigt profile is used to model the same (Eq. \ref{eq36})
\begin{equation}
\label{eq36}
D_{P}(x-x_{0},f)=\eta L(x-x_{0},\tau_{L})+(1-\eta)G(x-x_{0},\sigma_{G})
\end{equation} 
where $0<\eta<1$ and is a function of the FWHM parameter $f$ calculated using the approximate expression
\begin{equation}
\label{eq37}
\eta=1.36603(f_{L}/f)-0.47719(f_{L}/f)^{2}+0.11116(f_{L}/f)^{3}
\end{equation}
and $f$ itself is evaluated as
\begin{multline}
\label{eq38}
f=[f_{G}^{5}+2.69269f_{G}^{4}f_{L}+2.42843f_{G}^{3}f_{L}^{2}+4.47163f_{G}^{2}f_{L}^{3}+ \\
     0.07842f_{G}f_{L}^{4}+f_{L}^{5}]^{1/5}
\end{multline}
Here $f_{G}$ and $f_{L}$ are, respectively, the FWHM of the Gaussian and Lorentzian distributions and are related to the SD of the same as $f_{G}=2\sqrt{2\ln2}\sigma_{G}$ and $f_{L}=2\tau_{L}$. The terms $L(x-x_{0},\tau_{L})$ and $G(x-x_{0},\sigma_{G})$ are the Cauchy-Lorentz and Gaussian distribution functions centred at $x_{0}$, respectively, given by the expressions $L(x-x_{0},\tau
_{L})=\frac{\tau_{L}}{\pi((x-x_{0})^{2}+\tau_{L}^{2})}$ and $G(x-x_{0},\sigma_{G})=\frac{\mathrm{e}^{-(x-x_{0})^{2}/(2\sigma_{G})^{2}}}{\sigma_{G}\sqrt{2\pi}}$. 
Using the definitions in Eq. \ref{eq30} and \ref{eq36}-\ref{eq38}, the line emissivity density is calculated as
\begin{multline}
\label{eq39}
J_{theo,line}(h\nu)=[J_{\ce{{Ar}},K\alpha}D_{P}(h\nu-2.96,f_{\ce{{Ar}},K\alpha}) + \\
J_{\ce{{Ar}},K\beta}D_{P}(h\nu-3.19,f_{\ce{{Ar}},K\beta})+J_{\ce{{Cr}},K\alpha}D_{P}(h\nu-5.41,f_{\ce{{Cr}},K\alpha}) + \\
J_{\ce{{Cr}},K\beta}D_{P}(h\nu-5.94,f_{\ce{{Cr}},K\beta})+J_{\ce{{Fe}},K\alpha}D_{P}(h\nu-6.40,f_{\ce{{Fe}},K\alpha}) + \\
J_{\ce{{Fe}},K\beta}D_{P}(h\nu-7.05,f_{\ce{{Ar}},K\beta})]\Delta E
\end{multline}
Here the K$\alpha$ and K$\beta$ represent, respectively, the $2p\rightarrow 1s$ and $3p\rightarrow 1s$ transition, according to the Siegbahn notation. 

\begin{figure}
\resizebox{1.00\columnwidth}{!}{\includegraphics{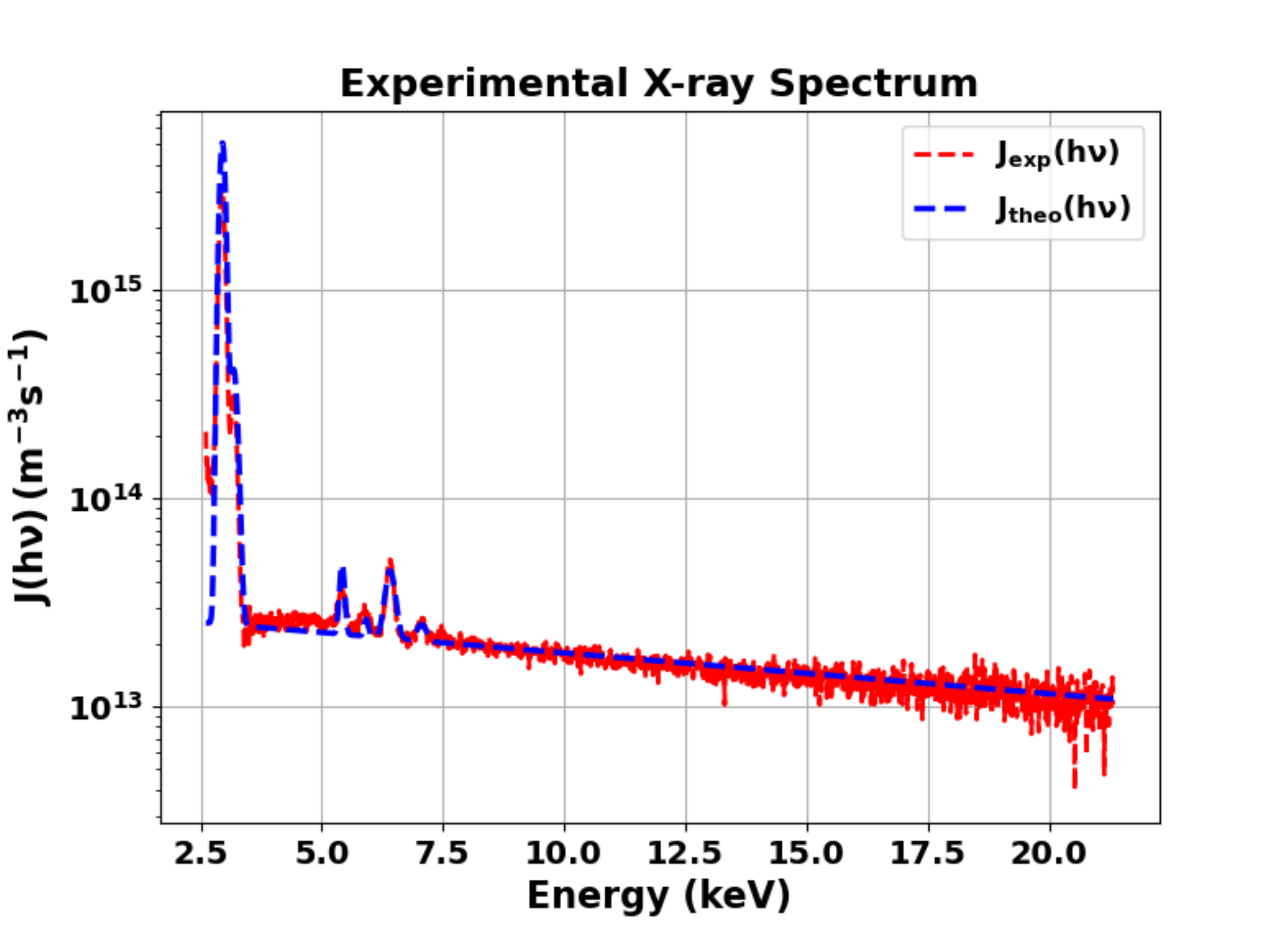}}  
\caption{$J_{exp}(h\nu)$ and analytical expression fit using Trust Region Selective least squares fitting method.}
\label{fig:11}
\end{figure}

By fitting Eq. \ref{eq26} to $J_{exp}(h\nu)$ from Eq. \ref{eq25}, the combined charge particle density $\rho_{e}\rho_{i}$ for argon is estimated around $1.36\times10^{32}\,\mathrm{m^{-6}}$ while the electron temperature is about $22.18$ keV. Additionally, rough estimates of loss electron density $\rho_{e,loss}$ are obtained at $10^{12}\,\mathrm{m^{-3}}$, corresponding to a current density of $2-5\,\mathrm{mA/cm^{2}}$ which is in keeping with recorded ion current densities \cite{Ref17}. Thus, a single-component MB distribution seems to work well enough to reproduce not only the emissions due to confined but also escaping electrons. The peaks are found to obey a simpler Gaussian rather than a Voigt profile, with standard deviation in the range $0.05-0.08\,\mathrm{keV}$. The degree of fit is shown in Fig. \ref{fig:11} and while some uncertainties still remain, they mainly stem from lack of information about contribution from hotter electrons whose spectra is recorded in the Hp\ce{{Ge}} detector.\\
As with the numerical tool, analysis of the experimental spectrum does provide useful information in the form of electron density, temperature and EEDF, but the results are sensitive to the photon energy being probed and region of emission. Here we find that a single MB distribution with $k_{B}T_{e}\sim22.18\,\mathrm{keV}$ is sufficient to reproduce the spectrum, but this may change if a different region of the plasma is probed, or if more energetic X-rays are analysed. With regards to usability of the spectrum for experimentally validating EEDF2 as estimated in Sec. \ref{sec:4}, nothing can be said owing to the negligible contribution of warm electrons, and so either a different diagnostic setup should be adapted for such electrons, or the simulations need to be updated to probe more energetic species.

\section{Conclusion and Future Perspectives}
\label{sec:6}

We have presented here a detailed numerical method to derive analytical functions representing anisotropic electron populations in an ECR plasma trap. The approach is based on first dividing the plasma volume into smaller regions based on electron energy content, and then quantitatively analysing the performance of several potential EEDFs in those regions using MSE and $r^{2}$ values. The method has been tested on rough data on warm electrons obtained from robust electron dynamics simulations, and spatially-anisotropic EEDFs have been deduced. In addition, experimental X-ray spectra have been analysed, providing crucial estimates about electron properties in the plasma interior.\\
Concerning the validity of our tool, several things can be said: as mentioned at the end of Sec. \ref{sec:4}, the numerical tool and its outputs are highly sensitive to the nature of input data (energy intervals) and plasma division (ROIs). On the other hand, as mentioned at the end of Sec. \ref{sec:5}, results from analysis of experimental spectra are also sensitive to the photon energies and emission regions. It is then evident that for a complete understanding of the plasma, simulations and experiments need to be coupled. Experiments need to be a multi-perspective setup capable of studying different regions of the plasma and different emission energies, and once the energies are fixed, simulations need to be run till steady-state, while accumulating electrons in compatible energy intervals. Our numerical tool is a general statistical method that can be used on any anisotropic population, but only after application to simulation data compatible with experiments, deeper physics of ECR plasmas can be inferred. We are currently working in this direction - updating our simulation models to probe warm to hot energy electrons and running them till steady-state, and analysing space-resolved 2D K$\alpha$ maps - and the results of this work will be published in another report. 

\section*{Data Availability Statement}

The data that support the findings of this study are available from the corresponding author upon reasonable request.

% If in two-column mode, this environment will change to single-column format so that long equations can be displayed. 
%\begin{widetext}
%$$\mbox{put long equation here}$$
%\end{widetext}

% Figures should be put into the text as floats. 
% Use the graphics or graphicx packages (distributed with LaTeX2e).
% Put the label that you will use with \ref{} command in the braces of the \label{} command.
%
% \begin{figure}
% \end{figure}

% \begin{table}
% \end{tabular}
% \end{table}

%\begin{acknowledgments}
% Put your acknowledgments here.
%\end{acknowledgments}

%\nocite{*}
\bibliography{Article1}

\providecommand{\noopsort}[1]{}\providecommand{\singleletter}[1]{#1}%
\begin{thebibliography}{36}%
\makeatletter
\providecommand \@ifxundefined [1]{%
 \@ifx{#1\undefined}
}%
\providecommand \@ifnum [1]{%
 \ifnum #1\expandafter \@firstoftwo
 \else \expandafter \@secondoftwo
 \fi
}%
\providecommand \@ifx [1]{%
 \ifx #1\expandafter \@firstoftwo
 \else \expandafter \@secondoftwo
 \fi
}%
\providecommand \natexlab [1]{#1}%
\providecommand \enquote  [1]{``#1''}%
\providecommand \bibnamefont  [1]{#1}%
\providecommand \bibfnamefont [1]{#1}%
\providecommand \citenamefont [1]{#1}%
\providecommand \href@noop [0]{\@secondoftwo}%
\providecommand \href [0]{\begingroup \@sanitize@url \@href}%
\providecommand \@href[1]{\@@startlink{#1}\@@href}%
\providecommand \@@href[1]{\endgroup#1\@@endlink}%
\providecommand \@sanitize@url [0]{\catcode `\\12\catcode `\$12\catcode
  `\&12\catcode `\#12\catcode `\^12\catcode `\_12\catcode `\%12\relax}%
\providecommand \@@startlink[1]{}%
\providecommand \@@endlink[0]{}%
\providecommand \url  [0]{\begingroup\@sanitize@url \@url }%
\providecommand \@url [1]{\endgroup\@href {#1}{\urlprefix }}%
\providecommand \urlprefix  [0]{URL }%
\providecommand \Eprint [0]{\href }%
\providecommand \doibase [0]{http://dx.doi.org/}%
\providecommand \selectlanguage [0]{\@gobble}%
\providecommand \bibinfo  [0]{\@secondoftwo}%
\providecommand \bibfield  [0]{\@secondoftwo}%
\providecommand \translation [1]{[#1]}%
\providecommand \BibitemOpen [0]{}%
\providecommand \bibitemStop [0]{}%
\providecommand \bibitemNoStop [0]{.\EOS\space}%
\providecommand \EOS [0]{\spacefactor3000\relax}%
\providecommand \BibitemShut  [1]{\csname bibitem#1\endcsname}%
\let\auto@bib@innerbib\@empty
%</preamble>
\bibitem [{\citenamefont {Pardo}(2016)}]{Ref1}%
  \BibitemOpen
  \bibinfo {editor} {\bibfnamefont {R.}~\bibnamefont {Pardo}},\ ed.,\
  \href@noop {} {\emph {\bibinfo {title} {$16^{th}$ International Conference on
  Ion Sources}}}\ (\bibinfo  {publisher} {Review of Scientific Instruments},\
  \bibinfo {year} {2016})\BibitemShut {NoStop}%
\bibitem [{Ref(2017)}]{Ref2}%
  \BibitemOpen
  \href@noop {} {\emph {\bibinfo {title} {$22^{nd}$ International Workshop on
  ECRIS}}}\ (\bibinfo {year} {2017})\ \bibinfo {note} {at press
  2017}\BibitemShut {NoStop}%
\bibitem [{\citenamefont {Mascali}\ \emph
  {et~al.}(2016{\natexlab{a}})\citenamefont {Mascali}, \citenamefont {Torrisi},
  \citenamefont {Leonardo}, \citenamefont {Sorbello}, \citenamefont {Castro},
  \citenamefont {Celona}, \citenamefont {Miracoli}, \citenamefont {Agnello},\
  and\ \citenamefont {Gammino}}]{Ref3}%
  \BibitemOpen
  \bibfield  {author} {\bibinfo {author} {\bibfnamefont {D.}~\bibnamefont
  {Mascali}}, \bibinfo {author} {\bibfnamefont {G.}~\bibnamefont {Torrisi}},
  \bibinfo {author} {\bibfnamefont {O.}~\bibnamefont {Leonardo}}, \bibinfo
  {author} {\bibfnamefont {G.}~\bibnamefont {Sorbello}}, \bibinfo {author}
  {\bibfnamefont {G.}~\bibnamefont {Castro}}, \bibinfo {author} {\bibfnamefont
  {L.}~\bibnamefont {Celona}}, \bibinfo {author} {\bibfnamefont
  {R.}~\bibnamefont {Miracoli}}, \bibinfo {author} {\bibfnamefont
  {R.}~\bibnamefont {Agnello}}, \ and\ \bibinfo {author} {\bibfnamefont
  {S.}~\bibnamefont {Gammino}},\ }\href@noop {} {\bibfield  {journal} {\bibinfo
   {journal} {Rev. Sci. Instrum.}\ }\textbf {\bibinfo {volume} {87}} (\bibinfo
  {year} {2016}{\natexlab{a}})}\BibitemShut {NoStop}%
\bibitem [{\citenamefont {Mascali}\ \emph {et~al.}(2014)\citenamefont
  {Mascali}, \citenamefont {Celona}, \citenamefont {Maimone}, \citenamefont
  {Maeder}, \citenamefont {Castro}, \citenamefont {Romano}, \citenamefont
  {Musumarra}, \citenamefont {Altana}, \citenamefont {Caliri}, \citenamefont
  {Torrisi}, \citenamefont {Neri}, \citenamefont {Gammino}, \citenamefont
  {Tinschert}, \citenamefont {Spaedtke}, \citenamefont {Rossbach},
  \citenamefont {Lang},\ and\ \citenamefont {Ciavola}}]{Ref4}%
  \BibitemOpen
  \bibfield  {author} {\bibinfo {author} {\bibfnamefont {D.}~\bibnamefont
  {Mascali}}, \bibinfo {author} {\bibfnamefont {L.}~\bibnamefont {Celona}},
  \bibinfo {author} {\bibfnamefont {F.}~\bibnamefont {Maimone}}, \bibinfo
  {author} {\bibfnamefont {J.}~\bibnamefont {Maeder}}, \bibinfo {author}
  {\bibfnamefont {G.}~\bibnamefont {Castro}}, \bibinfo {author} {\bibfnamefont
  {F.}~\bibnamefont {Romano}}, \bibinfo {author} {\bibfnamefont
  {A.}~\bibnamefont {Musumarra}}, \bibinfo {author} {\bibfnamefont
  {C.}~\bibnamefont {Altana}}, \bibinfo {author} {\bibfnamefont
  {C.}~\bibnamefont {Caliri}}, \bibinfo {author} {\bibfnamefont
  {G.}~\bibnamefont {Torrisi}}, \bibinfo {author} {\bibfnamefont
  {L.}~\bibnamefont {Neri}}, \bibinfo {author} {\bibfnamefont {S.}~\bibnamefont
  {Gammino}}, \bibinfo {author} {\bibfnamefont {K.}~\bibnamefont {Tinschert}},
  \bibinfo {author} {\bibfnamefont {K.}~\bibnamefont {Spaedtke}}, \bibinfo
  {author} {\bibfnamefont {J.}~\bibnamefont {Rossbach}}, \bibinfo {author}
  {\bibfnamefont {R.}~\bibnamefont {Lang}}, \ and\ \bibinfo {author}
  {\bibfnamefont {G.}~\bibnamefont {Ciavola}},\ }\href@noop {} {\bibfield
  {journal} {\bibinfo  {journal} {Rev. Sci. Instrum.}\ }\textbf {\bibinfo
  {volume} {85}} (\bibinfo {year} {2014})}\BibitemShut {NoStop}%
\bibitem [{\citenamefont {Biri}\ \emph {et~al.}(2004)\citenamefont {Biri},
  \citenamefont {Valek}, \citenamefont {Suta}, \citenamefont {Tak\'{a}cs},
  \citenamefont {Szab\'{o}}, \citenamefont {Hudson}, \citenamefont {Radics},
  \citenamefont {Imrek}, \citenamefont {Juh\'{a}sz},\ and\ \citenamefont
  {P\'{a}link\'{a}s}}]{Ref5}%
  \BibitemOpen
  \bibfield  {author} {\bibinfo {author} {\bibfnamefont {S.}~\bibnamefont
  {Biri}}, \bibinfo {author} {\bibfnamefont {A.}~\bibnamefont {Valek}},
  \bibinfo {author} {\bibfnamefont {T.}~\bibnamefont {Suta}}, \bibinfo {author}
  {\bibfnamefont {E.}~\bibnamefont {Tak\'{a}cs}}, \bibinfo {author}
  {\bibfnamefont {C.}~\bibnamefont {Szab\'{o}}}, \bibinfo {author}
  {\bibfnamefont {L.}~\bibnamefont {Hudson}}, \bibinfo {author} {\bibfnamefont
  {B.}~\bibnamefont {Radics}}, \bibinfo {author} {\bibfnamefont
  {J.}~\bibnamefont {Imrek}}, \bibinfo {author} {\bibfnamefont
  {B.}~\bibnamefont {Juh\'{a}sz}}, \ and\ \bibinfo {author} {\bibfnamefont
  {J.}~\bibnamefont {P\'{a}link\'{a}s}},\ }\href@noop {} {\bibfield  {journal}
  {\bibinfo  {journal} {Rev. Sci. Instrum.}\ }\textbf {\bibinfo {volume} {75}}
  (\bibinfo {year} {2004})}\BibitemShut {NoStop}%
\bibitem [{\citenamefont {R\'{a}cz}, \citenamefont {Biri},\ and\ \citenamefont
  {P\'{a}link\'{a}s}(2011)}]{Ref6}%
  \BibitemOpen
  \bibfield  {author} {\bibinfo {author} {\bibfnamefont {R.}~\bibnamefont
  {R\'{a}cz}}, \bibinfo {author} {\bibfnamefont {S.}~\bibnamefont {Biri}}, \
  and\ \bibinfo {author} {\bibfnamefont {J.}~\bibnamefont {P\'{a}link\'{a}s}},\
  }\href@noop {} {\bibfield  {journal} {\bibinfo  {journal} {Plasma Sources
  Sci. Technol.}\ }\textbf {\bibinfo {volume} {20}} (\bibinfo {year}
  {2011})}\BibitemShut {NoStop}%
\bibitem [{\citenamefont {Jauberteau}\ \emph {et~al.}(2016)\citenamefont
  {Jauberteau}, \citenamefont {Jauberteau}, \citenamefont {Cort\'{a}zar},\ and\
  \citenamefont {Meg\'{i}a-Mac\'{i}as}}]{Ref7}%
  \BibitemOpen
  \bibfield  {author} {\bibinfo {author} {\bibfnamefont {J.}~\bibnamefont
  {Jauberteau}}, \bibinfo {author} {\bibfnamefont {I.}~\bibnamefont
  {Jauberteau}}, \bibinfo {author} {\bibfnamefont {O.}~\bibnamefont
  {Cort\'{a}zar}}, \ and\ \bibinfo {author} {\bibfnamefont {A.}~\bibnamefont
  {Meg\'{i}a-Mac\'{i}as}},\ }\href@noop {} {\bibfield  {journal} {\bibinfo
  {journal} {Phys. Plasmas}\ }\textbf {\bibinfo {volume} {23}} (\bibinfo {year}
  {2016})}\BibitemShut {NoStop}%
\bibitem [{\citenamefont {Ken\'{e}z}\ \emph {et~al.}(2002)\citenamefont
  {Ken\'{e}z}, \citenamefont {Biri}, \citenamefont {Kar\'{a}csony},
  \citenamefont {Valek}, \citenamefont {Nakagawa}, \citenamefont {Stiebing},\
  and\ \citenamefont {Mironov}}]{Ref8}%
  \BibitemOpen
  \bibfield  {author} {\bibinfo {author} {\bibfnamefont {L.}~\bibnamefont
  {Ken\'{e}z}}, \bibinfo {author} {\bibfnamefont {S.}~\bibnamefont {Biri}},
  \bibinfo {author} {\bibfnamefont {J.}~\bibnamefont {Kar\'{a}csony}}, \bibinfo
  {author} {\bibfnamefont {A.}~\bibnamefont {Valek}}, \bibinfo {author}
  {\bibfnamefont {T.}~\bibnamefont {Nakagawa}}, \bibinfo {author}
  {\bibfnamefont {K.}~\bibnamefont {Stiebing}}, \ and\ \bibinfo {author}
  {\bibfnamefont {V.}~\bibnamefont {Mironov}},\ }\href@noop {} {\bibfield
  {journal} {\bibinfo  {journal} {Rev. Sci. Instrum.}\ }\textbf {\bibinfo
  {volume} {73}} (\bibinfo {year} {2002})}\BibitemShut {NoStop}%
\bibitem [{\citenamefont {Leitner}\ \emph {et~al.}(2002)\citenamefont
  {Leitner}, \citenamefont {Benitez}, \citenamefont {Lyneis},\ and\
  \citenamefont {Todd}}]{Ref9}%
  \BibitemOpen
  \bibfield  {author} {\bibinfo {author} {\bibfnamefont {D.}~\bibnamefont
  {Leitner}}, \bibinfo {author} {\bibfnamefont {J.}~\bibnamefont {Benitez}},
  \bibinfo {author} {\bibfnamefont {C.}~\bibnamefont {Lyneis}}, \ and\ \bibinfo
  {author} {\bibfnamefont {D.}~\bibnamefont {Todd}},\ }\href@noop {} {\bibfield
   {journal} {\bibinfo  {journal} {Rev. Sci. Instrum.}\ }\textbf {\bibinfo
  {volume} {79}} (\bibinfo {year} {2002})}\BibitemShut {NoStop}%
\bibitem [{\citenamefont {Gammino}\ \emph {et~al.}(2009)\citenamefont
  {Gammino}, \citenamefont {Mascali}, \citenamefont {Celona}, \citenamefont
  {Maimone},\ and\ \citenamefont {Ciavola}}]{Ref10}%
  \BibitemOpen
  \bibfield  {author} {\bibinfo {author} {\bibfnamefont {S.}~\bibnamefont
  {Gammino}}, \bibinfo {author} {\bibfnamefont {D.}~\bibnamefont {Mascali}},
  \bibinfo {author} {\bibfnamefont {L.}~\bibnamefont {Celona}}, \bibinfo
  {author} {\bibfnamefont {F.}~\bibnamefont {Maimone}}, \ and\ \bibinfo
  {author} {\bibfnamefont {G.}~\bibnamefont {Ciavola}},\ }\href@noop {}
  {\bibfield  {journal} {\bibinfo  {journal} {Plasma Sources Sci. Technol.}\
  }\textbf {\bibinfo {volume} {18}} (\bibinfo {year} {2009})}\BibitemShut
  {NoStop}%
\bibitem [{\citenamefont {Baru\`{e}}\ \emph {et~al.}(1994)\citenamefont
  {Baru\`{e}}, \citenamefont {Lamoureux}, \citenamefont {Briand}, \citenamefont
  {Girard},\ and\ \citenamefont {Melin}}]{Ref11}%
  \BibitemOpen
  \bibfield  {author} {\bibinfo {author} {\bibfnamefont {C.}~\bibnamefont
  {Baru\`{e}}}, \bibinfo {author} {\bibfnamefont {M.}~\bibnamefont
  {Lamoureux}}, \bibinfo {author} {\bibfnamefont {P.}~\bibnamefont {Briand}},
  \bibinfo {author} {\bibfnamefont {A.}~\bibnamefont {Girard}}, \ and\ \bibinfo
  {author} {\bibfnamefont {G.}~\bibnamefont {Melin}},\ }\href@noop {}
  {\bibfield  {journal} {\bibinfo  {journal} {J. Appl. Phys.}\ }\textbf
  {\bibinfo {volume} {76}} (\bibinfo {year} {1994})}\BibitemShut {NoStop}%
\bibitem [{\citenamefont {Tuske}(2004)}]{Ref12}%
  \BibitemOpen
  \bibfield  {author} {\bibinfo {author} {\bibfnamefont {O.~e.}\ \bibnamefont
  {Tuske}},\ }\href@noop {} {\bibfield  {journal} {\bibinfo  {journal} {Rev.
  Sci. Instrum.}\ }\textbf {\bibinfo {volume} {75}} (\bibinfo {year}
  {2004})}\BibitemShut {NoStop}%
\bibitem [{\citenamefont {Bibinov}\ \emph {et~al.}(2005)\citenamefont
  {Bibinov}, \citenamefont {Bratsev}, \citenamefont {Kokh}, \citenamefont
  {Ochkur},\ and\ \citenamefont {Weisemann}}]{Ref13}%
  \BibitemOpen
  \bibfield  {author} {\bibinfo {author} {\bibfnamefont {N.}~\bibnamefont
  {Bibinov}}, \bibinfo {author} {\bibfnamefont {V.}~\bibnamefont {Bratsev}},
  \bibinfo {author} {\bibfnamefont {D.}~\bibnamefont {Kokh}}, \bibinfo {author}
  {\bibfnamefont {V.}~\bibnamefont {Ochkur}}, \ and\ \bibinfo {author}
  {\bibfnamefont {K.}~\bibnamefont {Weisemann}},\ }\href@noop {} {\bibfield
  {journal} {\bibinfo  {journal} {Plasma Sources Sci. Technol.}\ }\textbf
  {\bibinfo {volume} {14}} (\bibinfo {year} {2005})}\BibitemShut {NoStop}%
\bibitem [{\citenamefont {Tak\'{a}cs}\ \emph {et~al.}(2005)\citenamefont
  {Tak\'{a}cs}, \citenamefont {Radics}, \citenamefont {Szab\'{o}},
  \citenamefont {Biri}, \citenamefont {Hudson}, \citenamefont {Imrek},
  \citenamefont {Suta}, \citenamefont {Valek},\ and\ \citenamefont
  {Pálinkás}}]{Ref15}%
  \BibitemOpen
  \bibfield  {author} {\bibinfo {author} {\bibfnamefont {E.}~\bibnamefont
  {Tak\'{a}cs}}, \bibinfo {author} {\bibfnamefont {B.}~\bibnamefont {Radics}},
  \bibinfo {author} {\bibfnamefont {C.}~\bibnamefont {Szab\'{o}}}, \bibinfo
  {author} {\bibfnamefont {S.}~\bibnamefont {Biri}}, \bibinfo {author}
  {\bibfnamefont {L.}~\bibnamefont {Hudson}}, \bibinfo {author} {\bibfnamefont
  {B.}~\bibnamefont {Imrek}, \bibfnamefont {J.and~Juh\'{a}sz}}, \bibinfo
  {author} {\bibfnamefont {T.}~\bibnamefont {Suta}}, \bibinfo {author}
  {\bibfnamefont {A.}~\bibnamefont {Valek}}, \ and\ \bibinfo {author}
  {\bibfnamefont {J.}~\bibnamefont {Pálinkás}},\ }\href@noop {} {\bibfield
  {journal} {\bibinfo  {journal} {Nucl. Instrum. Methods Phys. Res. Sect. B}\
  }\textbf {\bibinfo {volume} {235}},\ \bibinfo {pages} {120--125} (\bibinfo
  {year} {2005})}\BibitemShut {NoStop}%
\bibitem [{\citenamefont {R\'{a}cz}\ \emph {et~al.}(2016)\citenamefont
  {R\'{a}cz}, \citenamefont {Biri}, \citenamefont {P\'{a}link\'{a}s},
  \citenamefont {Mascali}, \citenamefont {Castro}, \citenamefont {Caliri},
  \citenamefont {Romano},\ and\ \citenamefont {Gammino}}]{Ref16}%
  \BibitemOpen
  \bibfield  {author} {\bibinfo {author} {\bibfnamefont {R.}~\bibnamefont
  {R\'{a}cz}}, \bibinfo {author} {\bibfnamefont {S.}~\bibnamefont {Biri}},
  \bibinfo {author} {\bibfnamefont {J.}~\bibnamefont {P\'{a}link\'{a}s}},
  \bibinfo {author} {\bibfnamefont {D.}~\bibnamefont {Mascali}}, \bibinfo
  {author} {\bibfnamefont {G.}~\bibnamefont {Castro}}, \bibinfo {author}
  {\bibfnamefont {C.}~\bibnamefont {Caliri}}, \bibinfo {author} {\bibfnamefont
  {F.~P.}\ \bibnamefont {Romano}}, \ and\ \bibinfo {author} {\bibfnamefont
  {S.}~\bibnamefont {Gammino}},\ }\href@noop {} {\bibfield  {journal} {\bibinfo
   {journal} {Rev. Sci. Instrum.}\ }\textbf {\bibinfo {volume} {87}} (\bibinfo
  {year} {2016})}\BibitemShut {NoStop}%
\bibitem [{\citenamefont {Mascali}\ \emph
  {et~al.}(2016{\natexlab{b}})\citenamefont {Mascali}, \citenamefont {Castro},
  \citenamefont {Biri}, \citenamefont {R\'{a}cz}, \citenamefont
  {P\'{a}link\'{a}s}, \citenamefont {Caliri}, \citenamefont {Celona},
  \citenamefont {Neri}, \citenamefont {Romano}, \citenamefont {Torrisi},\ and\
  \citenamefont {Gammino}}]{Ref17}%
  \BibitemOpen
  \bibfield  {author} {\bibinfo {author} {\bibfnamefont {D.}~\bibnamefont
  {Mascali}}, \bibinfo {author} {\bibfnamefont {G.}~\bibnamefont {Castro}},
  \bibinfo {author} {\bibfnamefont {S.}~\bibnamefont {Biri}}, \bibinfo {author}
  {\bibfnamefont {R.}~\bibnamefont {R\'{a}cz}}, \bibinfo {author}
  {\bibfnamefont {J.}~\bibnamefont {P\'{a}link\'{a}s}}, \bibinfo {author}
  {\bibfnamefont {C.}~\bibnamefont {Caliri}}, \bibinfo {author} {\bibfnamefont
  {L.}~\bibnamefont {Celona}}, \bibinfo {author} {\bibfnamefont
  {L.}~\bibnamefont {Neri}}, \bibinfo {author} {\bibfnamefont {F.}~\bibnamefont
  {Romano}}, \bibinfo {author} {\bibfnamefont {G.}~\bibnamefont {Torrisi}}, \
  and\ \bibinfo {author} {\bibfnamefont {S.}~\bibnamefont {Gammino}},\
  }\href@noop {} {\bibfield  {journal} {\bibinfo  {journal} {Rev. Sci.
  Instrum.}\ }\textbf {\bibinfo {volume} {87}} (\bibinfo {year}
  {2016}{\natexlab{b}})}\BibitemShut {NoStop}%
\bibitem [{\citenamefont {R\'{a}cz}\ \emph {et~al.}(2017)\citenamefont
  {R\'{a}cz}, \citenamefont {Mascali}, \citenamefont {Biri}, \citenamefont
  {Caliri}, \citenamefont {Castro}, \citenamefont {Galat\`{a}}, \citenamefont
  {Gammino}, \citenamefont {Neri}, \citenamefont {P\'{a}link\'{a}s},
  \citenamefont {Romano},\ and\ \citenamefont {Torrisi}}]{Ref18}%
  \BibitemOpen
  \bibfield  {author} {\bibinfo {author} {\bibfnamefont {R.}~\bibnamefont
  {R\'{a}cz}}, \bibinfo {author} {\bibfnamefont {D.}~\bibnamefont {Mascali}},
  \bibinfo {author} {\bibfnamefont {S.}~\bibnamefont {Biri}}, \bibinfo {author}
  {\bibfnamefont {C.}~\bibnamefont {Caliri}}, \bibinfo {author} {\bibfnamefont
  {G.}~\bibnamefont {Castro}}, \bibinfo {author} {\bibfnamefont
  {A.}~\bibnamefont {Galat\`{a}}}, \bibinfo {author} {\bibfnamefont
  {S.}~\bibnamefont {Gammino}}, \bibinfo {author} {\bibfnamefont
  {L.}~\bibnamefont {Neri}}, \bibinfo {author} {\bibfnamefont {J.}~\bibnamefont
  {P\'{a}link\'{a}s}}, \bibinfo {author} {\bibfnamefont {F.}~\bibnamefont
  {Romano}}, \ and\ \bibinfo {author} {\bibfnamefont {G.}~\bibnamefont
  {Torrisi}},\ }\href@noop {} {\bibfield  {journal} {\bibinfo  {journal}
  {Plasma Sources Sci. Technol.}\ }\textbf {\bibinfo {volume} {26}} (\bibinfo
  {year} {2017})}\BibitemShut {NoStop}%
\bibitem [{\citenamefont {Galat\`{a}}\ \emph {et~al.}(2020)\citenamefont
  {Galat\`{a}}, \citenamefont {Mascali}, \citenamefont {Gallo},\ and\
  \citenamefont {Torrisi}}]{Ref23}%
  \BibitemOpen
  \bibfield  {author} {\bibinfo {author} {\bibfnamefont {A.}~\bibnamefont
  {Galat\`{a}}}, \bibinfo {author} {\bibfnamefont {D.}~\bibnamefont {Mascali}},
  \bibinfo {author} {\bibfnamefont {C.~S.}\ \bibnamefont {Gallo}}, \ and\
  \bibinfo {author} {\bibfnamefont {G.}~\bibnamefont {Torrisi}},\ }\href@noop
  {} {\bibfield  {journal} {\bibinfo  {journal} {Rev. Sci. Instrum.}\ }\textbf
  {\bibinfo {volume} {91}} (\bibinfo {year} {2020})}\BibitemShut {NoStop}%
\bibitem [{\citenamefont {Mascali}\ \emph {et~al.}(2015)\citenamefont
  {Mascali}, \citenamefont {Torrisi}, \citenamefont {Neri}, \citenamefont
  {Sorbello}, \citenamefont {Castro}, \citenamefont {Celona},\ and\
  \citenamefont {Gammino}}]{Ref19}%
  \BibitemOpen
  \bibfield  {author} {\bibinfo {author} {\bibfnamefont {D.}~\bibnamefont
  {Mascali}}, \bibinfo {author} {\bibfnamefont {G.}~\bibnamefont {Torrisi}},
  \bibinfo {author} {\bibfnamefont {L.}~\bibnamefont {Neri}}, \bibinfo {author}
  {\bibfnamefont {G.}~\bibnamefont {Sorbello}}, \bibinfo {author}
  {\bibfnamefont {G.}~\bibnamefont {Castro}}, \bibinfo {author} {\bibfnamefont
  {L.}~\bibnamefont {Celona}}, \ and\ \bibinfo {author} {\bibfnamefont
  {S.}~\bibnamefont {Gammino}},\ }\href@noop {} {\bibfield  {journal} {\bibinfo
   {journal} {Eur. Phys. J. D.}\ }\textbf {\bibinfo {volume} {69}} (\bibinfo
  {year} {2015})}\BibitemShut {NoStop}%
\bibitem [{\citenamefont {Torrisi}\ \emph {et~al.}(2014)\citenamefont
  {Torrisi}, \citenamefont {Mascali}, \citenamefont {Sorbello}, \citenamefont
  {Neri}, \citenamefont {Celona}, \citenamefont {Castro}, \citenamefont
  {Isernia},\ and\ \citenamefont {Gammino}}]{Ref25}%
  \BibitemOpen
  \bibfield  {author} {\bibinfo {author} {\bibfnamefont {G.}~\bibnamefont
  {Torrisi}}, \bibinfo {author} {\bibfnamefont {D.}~\bibnamefont {Mascali}},
  \bibinfo {author} {\bibfnamefont {G.}~\bibnamefont {Sorbello}}, \bibinfo
  {author} {\bibfnamefont {L.}~\bibnamefont {Neri}}, \bibinfo {author}
  {\bibfnamefont {L.}~\bibnamefont {Celona}}, \bibinfo {author} {\bibfnamefont
  {G.}~\bibnamefont {Castro}}, \bibinfo {author} {\bibfnamefont
  {T.}~\bibnamefont {Isernia}}, \ and\ \bibinfo {author} {\bibfnamefont
  {S.}~\bibnamefont {Gammino}},\ }\href@noop {} {\bibfield  {journal} {\bibinfo
   {journal} {Journal of Electromagnetic Waves and Applications}\ }\textbf
  {\bibinfo {volume} {28}} (\bibinfo {year} {2014})}\BibitemShut {NoStop}%
\bibitem [{\citenamefont {Galat\`{a}}\ \emph {et~al.}(2016)\citenamefont
  {Galat\`{a}}, \citenamefont {Mascali}, \citenamefont {Neri},\ and\
  \citenamefont {Celona}}]{Ref26}%
  \BibitemOpen
  \bibfield  {author} {\bibinfo {author} {\bibfnamefont {A.}~\bibnamefont
  {Galat\`{a}}}, \bibinfo {author} {\bibfnamefont {D.}~\bibnamefont {Mascali}},
  \bibinfo {author} {\bibfnamefont {L.}~\bibnamefont {Neri}}, \ and\ \bibinfo
  {author} {\bibfnamefont {L.}~\bibnamefont {Celona}},\ }\href@noop {}
  {\bibfield  {journal} {\bibinfo  {journal} {Plasma Sources Sci. Technol.}\
  }\textbf {\bibinfo {volume} {25}} (\bibinfo {year} {2016})}\BibitemShut
  {NoStop}%
\bibitem [{\citenamefont {Tarvainen}\ \emph {et~al.}(2016)\citenamefont
  {Tarvainen}, \citenamefont {Koivisto}, \citenamefont {Galat\`{a}},
  \citenamefont {Angot}, \citenamefont {Lamy}, \citenamefont {Thuillier},
  \citenamefont {Delahaye}, \citenamefont {Maunoury}, \citenamefont {Mascali},\
  and\ \citenamefont {Neri}}]{Ref27}%
  \BibitemOpen
  \bibfield  {author} {\bibinfo {author} {\bibfnamefont {O.}~\bibnamefont
  {Tarvainen}}, \bibinfo {author} {\bibfnamefont {H.}~\bibnamefont {Koivisto}},
  \bibinfo {author} {\bibfnamefont {A.}~\bibnamefont {Galat\`{a}}}, \bibinfo
  {author} {\bibfnamefont {J.}~\bibnamefont {Angot}}, \bibinfo {author}
  {\bibfnamefont {T.}~\bibnamefont {Lamy}}, \bibinfo {author} {\bibfnamefont
  {T.}~\bibnamefont {Thuillier}}, \bibinfo {author} {\bibfnamefont
  {P.}~\bibnamefont {Delahaye}}, \bibinfo {author} {\bibfnamefont
  {L.}~\bibnamefont {Maunoury}}, \bibinfo {author} {\bibfnamefont
  {D.}~\bibnamefont {Mascali}}, \ and\ \bibinfo {author} {\bibfnamefont
  {L.}~\bibnamefont {Neri}},\ }\href@noop {} {\bibfield  {journal} {\bibinfo
  {journal} {Phys. Rev. Accel. Beams}\ }\textbf {\bibinfo {volume} {19}}
  (\bibinfo {year} {2016})}\BibitemShut {NoStop}%
\bibitem [{\citenamefont {Galat\`{a}}\ \emph {et~al.}(2019)\citenamefont
  {Galat\`{a}}, \citenamefont {Gallo}, \citenamefont {Mascali},\ and\
  \citenamefont {Torrisi}}]{Ref24}%
  \BibitemOpen
  \bibfield  {author} {\bibinfo {author} {\bibfnamefont {A.}~\bibnamefont
  {Galat\`{a}}}, \bibinfo {author} {\bibfnamefont {C.~S.}\ \bibnamefont
  {Gallo}}, \bibinfo {author} {\bibfnamefont {D.}~\bibnamefont {Mascali}}, \
  and\ \bibinfo {author} {\bibfnamefont {G.}~\bibnamefont {Torrisi}},\
  }\href@noop {} {}\bibinfo {howpublished} {e-print arXiv:1912.01988} (\bibinfo
  {year} {2019})\BibitemShut {NoStop}%
\bibitem [{\citenamefont {Turner}, \citenamefont {Doyle},\ and\ \citenamefont
  {Hopkins}(1993)}]{Ref20}%
  \BibitemOpen
  \bibfield  {author} {\bibinfo {author} {\bibfnamefont {M.}~\bibnamefont
  {Turner}}, \bibinfo {author} {\bibfnamefont {R.}~\bibnamefont {Doyle}}, \
  and\ \bibinfo {author} {\bibfnamefont {M.}~\bibnamefont {Hopkins}},\
  }\href@noop {} {\bibfield  {journal} {\bibinfo  {journal} {Appl. Phys.
  Lett.}\ }\textbf {\bibinfo {volume} {62}} (\bibinfo {year}
  {1993})}\BibitemShut {NoStop}%
\bibitem [{\citenamefont {Godyak}, \citenamefont {Piejak},\ and\ \citenamefont
  {Alexandrovich}(2002)}]{Ref21}%
  \BibitemOpen
  \bibfield  {author} {\bibinfo {author} {\bibfnamefont {V.}~\bibnamefont
  {Godyak}}, \bibinfo {author} {\bibfnamefont {R.}~\bibnamefont {Piejak}}, \
  and\ \bibinfo {author} {\bibfnamefont {B.}~\bibnamefont {Alexandrovich}},\
  }\href@noop {} {\bibfield  {journal} {\bibinfo  {journal} {Plasma Phys. Sci.
  Technol.}\ }\textbf {\bibinfo {volume} {11}} (\bibinfo {year}
  {2002})}\BibitemShut {NoStop}%
\bibitem [{\citenamefont {Girard}\ \emph {et~al.}(1994)\citenamefont {Girard},
  \citenamefont {Briand}, \citenamefont {Gaudart}, \citenamefont {Klein},
  \citenamefont {Bourg}, \citenamefont {Debernardi}, \citenamefont {Mathonet},
  \citenamefont {Melin},\ and\ \citenamefont {Su}}]{Ref22}%
  \BibitemOpen
  \bibfield  {author} {\bibinfo {author} {\bibfnamefont {A.}~\bibnamefont
  {Girard}}, \bibinfo {author} {\bibfnamefont {P.}~\bibnamefont {Briand}},
  \bibinfo {author} {\bibfnamefont {G.}~\bibnamefont {Gaudart}}, \bibinfo
  {author} {\bibfnamefont {J.}~\bibnamefont {Klein}}, \bibinfo {author}
  {\bibfnamefont {F.}~\bibnamefont {Bourg}}, \bibinfo {author} {\bibfnamefont
  {J.}~\bibnamefont {Debernardi}}, \bibinfo {author} {\bibfnamefont
  {J.}~\bibnamefont {Mathonet}}, \bibinfo {author} {\bibfnamefont
  {G.}~\bibnamefont {Melin}}, \ and\ \bibinfo {author} {\bibfnamefont
  {Y.}~\bibnamefont {Su}},\ }\href@noop {} {\bibfield  {journal} {\bibinfo
  {journal} {Rev. Sci. Instrum.}\ }\textbf {\bibinfo {volume} {65}} (\bibinfo
  {year} {1994})}\BibitemShut {NoStop}%
\bibitem [{\citenamefont {Fujita}\ and\ \citenamefont
  {Yamazaki}(1990)}]{Ref28}%
  \BibitemOpen
  \bibfield  {author} {\bibinfo {author} {\bibfnamefont {F.}~\bibnamefont
  {Fujita}}\ and\ \bibinfo {author} {\bibfnamefont {H.}~\bibnamefont
  {Yamazaki}},\ }\href@noop {} {\bibfield  {journal} {\bibinfo  {journal}
  {Japanese Journal of Applied Physics}\ }\textbf {\bibinfo {volume} {29}}
  (\bibinfo {year} {1990})}\BibitemShut {NoStop}%
\bibitem [{\citenamefont {Adams}, \citenamefont {Miles},\ and\ \citenamefont
  {Demidov}(2017)}]{Ref29}%
  \BibitemOpen
  \bibfield  {author} {\bibinfo {author} {\bibfnamefont {S.~F.}\ \bibnamefont
  {Adams}}, \bibinfo {author} {\bibfnamefont {J.~A.}\ \bibnamefont {Miles}}, \
  and\ \bibinfo {author} {\bibfnamefont {V.~I.}\ \bibnamefont {Demidov}},\
  }\href@noop {} {\bibfield  {journal} {\bibinfo  {journal} {Phys. Plasmas}\
  }\textbf {\bibinfo {volume} {24}} (\bibinfo {year} {2017})}\BibitemShut
  {NoStop}%
\bibitem [{\citenamefont {Douysset}\ \emph {et~al.}(2000)\citenamefont
  {Douysset}, \citenamefont {Khodja}, \citenamefont {Girard},\ and\
  \citenamefont {Briand}}]{Ref30}%
  \BibitemOpen
  \bibfield  {author} {\bibinfo {author} {\bibfnamefont {G.}~\bibnamefont
  {Douysset}}, \bibinfo {author} {\bibfnamefont {H.}~\bibnamefont {Khodja}},
  \bibinfo {author} {\bibfnamefont {A.}~\bibnamefont {Girard}}, \ and\ \bibinfo
  {author} {\bibfnamefont {J.}~\bibnamefont {Briand}},\ }\href@noop {}
  {\bibfield  {journal} {\bibinfo  {journal} {Phys. Rev. E.}\ }\textbf
  {\bibinfo {volume} {61}} (\bibinfo {year} {2000})}\BibitemShut {NoStop}%
\bibitem [{\citenamefont {Santos}\ \emph {et~al.}(2010)\citenamefont {Santos},
  \citenamefont {Costa}, \citenamefont {Marques}, \citenamefont {Martins},
  \citenamefont {Indelicato},\ and\ \citenamefont {Parente}}]{Ref31}%
  \BibitemOpen
  \bibfield  {author} {\bibinfo {author} {\bibfnamefont {J.}~\bibnamefont
  {Santos}}, \bibinfo {author} {\bibfnamefont {A.~M.}\ \bibnamefont {Costa}},
  \bibinfo {author} {\bibfnamefont {J.}~\bibnamefont {Marques}}, \bibinfo
  {author} {\bibfnamefont {M.}~\bibnamefont {Martins}}, \bibinfo {author}
  {\bibfnamefont {P.}~\bibnamefont {Indelicato}}, \ and\ \bibinfo {author}
  {\bibfnamefont {F.}~\bibnamefont {Parente}},\ }\href@noop {} {\bibfield
  {journal} {\bibinfo  {journal} {Phys. Rev. A.}\ }\textbf {\bibinfo {volume}
  {82}} (\bibinfo {year} {2010})}\BibitemShut {NoStop}%
\bibitem [{\citenamefont {Sakildien}\ \emph {et~al.}(2018)\citenamefont
  {Sakildien}, \citenamefont {Kronholm}, \citenamefont {Tarvainen},
  \citenamefont {Kalvas}, \citenamefont {Jones}, \citenamefont {Thomae},\ and\
  \citenamefont {Koivisto}}]{Ref32}%
  \BibitemOpen
  \bibfield  {author} {\bibinfo {author} {\bibfnamefont {M.}~\bibnamefont
  {Sakildien}}, \bibinfo {author} {\bibfnamefont {R.}~\bibnamefont {Kronholm}},
  \bibinfo {author} {\bibfnamefont {O.}~\bibnamefont {Tarvainen}}, \bibinfo
  {author} {\bibfnamefont {T.}~\bibnamefont {Kalvas}}, \bibinfo {author}
  {\bibfnamefont {P.}~\bibnamefont {Jones}}, \bibinfo {author} {\bibfnamefont
  {R.}~\bibnamefont {Thomae}}, \ and\ \bibinfo {author} {\bibfnamefont
  {H.}~\bibnamefont {Koivisto}},\ }\href@noop {} {\bibfield  {journal}
  {\bibinfo  {journal} {Nuclear Inst. and Methods in Physics Research A}\
  }\textbf {\bibinfo {volume} {900}} (\bibinfo {year} {2018})}\BibitemShut
  {NoStop}%
\bibitem [{\citenamefont {Kramers}(1923)}]{Ref33}%
  \BibitemOpen
  \bibfield  {author} {\bibinfo {author} {\bibfnamefont {H.}~\bibnamefont
  {Kramers}},\ }\href@noop {} {\bibfield  {journal} {\bibinfo  {journal} {The
  London, Edinburgh and Dublin Philosophical Magazine and Journal of Science}\
  }\textbf {\bibinfo {volume} {46:275}},\ \bibinfo {pages} {836--871} (\bibinfo
  {year} {1923})}\BibitemShut {NoStop}%
\bibitem [{\citenamefont {Lotz}(1968)}]{Ref34}%
  \BibitemOpen
  \bibfield  {author} {\bibinfo {author} {\bibfnamefont {W.}~\bibnamefont
  {Lotz}},\ }\href@noop {} {\bibfield  {journal} {\bibinfo  {journal} {Z.
  Physik}\ }\textbf {\bibinfo {volume} {216}},\ \bibinfo {pages} {241--247}
  (\bibinfo {year} {1968})}\BibitemShut {NoStop}%
\bibitem [{\citenamefont {Deutsch}\ \emph {et~al.}(2002)\citenamefont
  {Deutsch}, \citenamefont {Becker}, \citenamefont {Gstis},\ and\ \citenamefont
  {Mark}}]{Ref35}%
  \BibitemOpen
  \bibfield  {author} {\bibinfo {author} {\bibfnamefont {H.}~\bibnamefont
  {Deutsch}}, \bibinfo {author} {\bibfnamefont {K.}~\bibnamefont {Becker}},
  \bibinfo {author} {\bibfnamefont {B.}~\bibnamefont {Gstis}}, \ and\ \bibinfo
  {author} {\bibfnamefont {T.}~\bibnamefont {Mark}},\ }\href@noop {} {\bibfield
   {journal} {\bibinfo  {journal} {Int. J. Mass Spectrom.}\ }\textbf {\bibinfo
  {volume} {213}},\ \bibinfo {pages} {5--8} (\bibinfo {year}
  {2002})}\BibitemShut {NoStop}%
\bibitem [{\citenamefont {Desclaux}(1973)}]{Ref36}%
  \BibitemOpen
  \bibfield  {author} {\bibinfo {author} {\bibfnamefont {J.}~\bibnamefont
  {Desclaux}},\ }\href@noop {} {\bibfield  {journal} {\bibinfo  {journal}
  {Atomic Data and Nuclear Data Tables}\ }\textbf {\bibinfo {volume} {12}},\
  \bibinfo {pages} {311--406} (\bibinfo {year} {1973})}\BibitemShut {NoStop}%
\bibitem [{\citenamefont {Deutsch}, \citenamefont {Margreiter},\ and\
  \citenamefont {Mark}(1994)}]{Ref37}%
  \BibitemOpen
  \bibfield  {author} {\bibinfo {author} {\bibfnamefont {H.}~\bibnamefont
  {Deutsch}}, \bibinfo {author} {\bibfnamefont {D.}~\bibnamefont {Margreiter}},
  \ and\ \bibinfo {author} {\bibfnamefont {T.}~\bibnamefont {Mark}},\
  }\href@noop {} {\bibfield  {journal} {\bibinfo  {journal} {Z. Physik D.}\
  }\textbf {\bibinfo {volume} {29}},\ \bibinfo {pages} {31--37} (\bibinfo
  {year} {1994})}\BibitemShut {NoStop}%
\end{thebibliography}%

\pagebreak

\begin{table*}
\setlength\tabcolsep{3pt}
\caption{\footnotesize{Fit parameters of 2-component EEDF1 and EEDF2, and of 3-components EEDF3 and EEDF4. Parameters $(\mathbf{k_{B}T_{\emph{l}}})_{\emph{j}}$ is given in eV, while $(\mathbf{k_{B}T_{\emph{m,h}}})_{\emph{j}}$ are given in keV.}\label{2and3EEDFs}}
\begin{adjustbox}{angle=90, scale=0.99,center}
\noindent\makebox[\textheight]{
\begin{tabular}{@{\extracolsep{\fill}}ll@{}ll@{}ll@{}ll@{}ll@{}ll@{}ll@{}ll@{}ll@{}ll@{}ll@{}ll@{}ll@{}ll@{}ll@{}ll@{}ll@{}ll@{}ll@{}ll@{}ll@{}ll@{}ll@{}ll@{}ll@{}ll@{}ll@{}}\toprule
\textbf{ROI} & \multicolumn{3}{c}{$\mathbf{A_{\emph{lj}}}$} & \multicolumn{3}{c}{$\mathbf{A_{\emph{hj}}}$} &\multicolumn{3}{c}{$(\mathbf{k_{B}T_{\emph{l}}})_{\emph{j}}$} &
\multicolumn{3}{c}{$(\mathbf{k_{B}T_{\emph{h}}})_{\emph{j}}$} &
\multicolumn{3}{c}{$\mathbf{C_{\emph{j}}}$} & \multicolumn{3}{c}{$\mathbf{S_{\emph{j}}}$} &
\multicolumn{2}{c}{$\langle MSE \rangle_{j}$} &
\multicolumn{2}{c}{$\sigma_{\langle MSE \rangle_{j}}$} &
\multicolumn{2}{c}{$\langle r^{2} \rangle_{j}$} &
\multicolumn{2}{c}{$\sigma_{\langle r^{2}\rangle_{j}}$}\\\toprule
\multicolumn{27}{c}{\textbf{EEDF1}} \\ \midrule
$j=1$  & \multicolumn{3}{c}{$[0.9797,1.0]$}     &  \multicolumn{3}{c}{$[0.0,0.0203]$}       & \multicolumn{3}{c}{$[0.0086,30]$} &
\multicolumn{3}{c}{$[1.1004,8.3002]$} & \multicolumn{3}{c}{$0.45$} & \multicolumn{3}{c}{$0.55$} & \multicolumn{2}{c}{$6.88E{+}09$} & \multicolumn{2}{c}{$1.42E{+}11$} & \multicolumn{2}{c}{$0.9904$} & \multicolumn{2}{c}{$0.0923$}\\
$j=2$  & \multicolumn{3}{c}{$[0.9493,1.0]$}     &  \multicolumn{3}{c}{$[0.0,0.0507]$}       & \multicolumn{3}{c}{$[3.6,32.8]$}    &
\multicolumn{3}{c}{$[0.9024,4.0956]$} & \multicolumn{3}{c}{$0.25$} & \multicolumn{3}{c}{$0.45$} & \multicolumn{2}{c}{$1.08E{+}13$}  & \multicolumn{2}{c}{$1.67E{+}13$} & \multicolumn{2}{c}{$0.9132$} & \multicolumn{2}{c}{$0.2010$}\\
$j=3$  & \multicolumn{3}{c}{$[0.9153,0.9978]$} &  \multicolumn{3}{c}{$[0.0022,0.0847]$} & \multicolumn{3}{c}{$[8.9,46.3]$}    &
\multicolumn{3}{c}{$[1.0532,2.9176]$} & \multicolumn{3}{c}{$0.25$} & \multicolumn{3}{c}{$0.45$} & \multicolumn{2}{c}{$1.15E{+}14$}  &  \multicolumn{2}{c}{$9.39E{+}13$} & \multicolumn{2}{c}{$0.9298$} & \multicolumn{2}{c}{$0.0249$}\\
$j=4$  & \multicolumn{3}{c}{$[0.8844,0.9740]$} & \multicolumn{3}{c}{ $[0.0260,0.1156]$} & \multicolumn{3}{c}{$[13.5,45.7]$}  &
\multicolumn{3}{c}{$[1.1666,2.5533]$} & \multicolumn{3}{c}{$0.20$} & \multicolumn{3}{c}{$0.45$} & \multicolumn{2}{c}{$4.57E{+}14$}  & \multicolumn{2}{c}{$2.76E{+}14$} & \multicolumn{2}{c}{$0.9325$} & \multicolumn{2}{c}{$0.0197$}\\
$j=5$  & \multicolumn{3}{c}{$[0.8586,0.9450]$} &  \multicolumn{3}{c}{$[0.0550,0.1414]$} & \multicolumn{3}{c}{$[19.3,54.3]$}  &
\multicolumn{3}{c}{$[1.2521,2.2049]$} & \multicolumn{3}{c}{$0.20$} & \multicolumn{3}{c}{$0.45$} & \multicolumn{2}{c}{$8.87E{+}14$}  & \multicolumn{2}{c}{$4.48E{+}14$} & \multicolumn{2}{c}{$0.9462$} & \multicolumn{2}{c}{$0.0166$}\\
$j=6$  &\multicolumn{3}{c}{ $[0.8366,0.9038]$} & \multicolumn{3}{c}{ $[0.0962,0.1634]$} & \multicolumn{3}{c}{$[27.9,56.1]$}  &
\multicolumn{3}{c}{$[1.3095,2.1365]$} & \multicolumn{3}{c}{$0.20$} & \multicolumn{3}{c}{$0.45$} & \multicolumn{2}{c}{$9.63E{+}14$}  & \multicolumn{2}{c}{$5.17E{+}14$} & \multicolumn{2}{c}{$0.9625$} & \multicolumn{2}{c}{$0.0168$}\\
$j=7$  & \multicolumn{3}{c}{$[0.8319,0.8763]$} & \multicolumn{3}{c}{ $[0.1327,0.1681]$} & \multicolumn{3}{c}{$[35.4,54.7]$}  &
\multicolumn{3}{c}{$[1.0092,1.9694]$} & \multicolumn{3}{c}{$0.20$} & \multicolumn{3}{c}{$0.45$} & \multicolumn{2}{c}{$8.81E{+}14$}  & \multicolumn{2}{c}{$4.18E{+}14$} & \multicolumn{2}{c}{$0.9714$} & \multicolumn{2}{c}{$0.0096$}\\ 
\midrule
\multicolumn{27}{c}{\textbf{EEDF2}} \\ \midrule
$j=1$ & \multicolumn{3}{c}{$[0.9728,1.0]$}	           &    \multicolumn{3}{c}{ $[0.0,0.0272]$} &	\multicolumn{3}{c}{$[0.0086,30]$}   &	 \multicolumn{3}{c}{ $[1.1004,8.3002]$} & \multicolumn{3}{c}{$0.45$}  &	
\multicolumn{3}{c}{$0.55$} & \multicolumn{2}{c}{$5.72E{+}09$} & \multicolumn{2}{c}{$1.88E{+}11$} & \multicolumn{2}{c}{$0.9897$} & \multicolumn{2}{c}{$0.1927$}\\
$j=2$ & \multicolumn{3}{c}{$[0.9548,1.0]$}	           &    \multicolumn{3}{c}{ $[0.0,0.0452]$}&	\multicolumn{3}{c}{$[3.6,32.8]$}	     &    \multicolumn{3}{c}{ $[1.3035,5.9194]$} & \multicolumn{3}{c}{$0.25$}  &	
\multicolumn{3}{c}{$0.65$} & \multicolumn{2}{c}{$2.06E{+}12$} & \multicolumn{2}{c}{$9.67E{+}12$} & \multicolumn{2}{c}{$0.9772$ }& \multicolumn{2}{c}{$0.2159$}\\
$j=3$ & \multicolumn{3}{c}{$[0.9332,0.9979]$}	&   \multicolumn{3}{c}{ $[0.0021,0.0768]$}&	\multicolumn{3}{c}{$[8.9,46.3]$}     &	\multicolumn{3}{c}{  $[1.5213,4.2143]$} & \multicolumn{3}{c}{$0.25$}  &	
\multicolumn{3}{c}{$0.65$} & \multicolumn{2}{c}{$1.56E{+}13$} & \multicolumn{2}{c}{$3.76E{+}13$} & \multicolumn{2}{c}{$0.9903$} & \multicolumn{2}{c}{$0.0317$}\\
$j=4$ & \multicolumn{3}{c}{$[0.8960,0.9754]$}	&   \multicolumn{3}{c}{ $[0.0246,0.1040]$}&	\multicolumn{3}{c}{$[13.5,45.7]$}   &	 \multicolumn{3}{c}{ $[1.6851,3.6882]$} & \multicolumn{3}{c}{$0.20$}  &\multicolumn{3}{c}{	
$0.65$} & \multicolumn{2}{c}{$5.19E{+}13$} & \multicolumn{2}{c}{$5.35E{+}13$} & \multicolumn{2}{c}{$0.9915$} & \multicolumn{2}{c}{$0.0120$}\\
$j=5$ & \multicolumn{3}{c}{$[0.8732,0.9510]$}	&  \multicolumn{3}{c}{  $[0.0490,0.1368]$}&	\multicolumn{3}{c}{$[19.3,54.3]$}   &\multicolumn{3}{c}{	  $[1.8086,3.1849]$} & \multicolumn{3}{c}{$0.20$}  &	
\multicolumn{3}{c}{$0.65$} & \multicolumn{2}{c}{$9.91E{+}13$} & \multicolumn{2}{c}{$6.85E{+}13$} & \multicolumn{2}{c}{$0.9924$} & \multicolumn{2}{c}{$0.0078$}\\
$j=6$ & \multicolumn{3}{c}{$[0.8525,0.9122]$}	&  \multicolumn{3}{c}{  $[0.0878,0.1475]$}&	\multicolumn{3}{c}{$[27.9,56.1]$}   &\multicolumn{3}{c}{      $[1.8915,3.0860]$} & \multicolumn{3}{c}{$0.65$}  &	
	\multicolumn{3}{c}{$0.20$} & \multicolumn{2}{c}{$2.01E{+}14$} &\multicolumn{2}{c}{$1.21E{+}14$} & \multicolumn{2}{c}{$0.9902$} & \multicolumn{2}{c}{$0.0092$}\\
$j=7$ & \multicolumn{3}{c}{$[0.8492,0.8812]$}	& \multicolumn{3}{c}{   $[0.1188,0.1508]$}&	\multicolumn{3}{c}{$[35.4,54.7]$}   & \multicolumn{3}{c}{     $[2.1430,2.8447]$} & \multicolumn{3}{c}{$0.65$ }  &	
\multicolumn{3}{c}{$0.20$} & \multicolumn{2}{c}{$2.98E{+}14$} & \multicolumn{2}{c}{$1.26E{+}14$} & \multicolumn{2}{c}{$0.9877$} & \multicolumn{2}{c}{$0.0087$}\\
\toprule
\textbf{ROI} & \multicolumn{2}{c}{$\mathbf{A_{\emph{lj}}}$} & \multicolumn{2}{c}{$\mathbf{A_{\emph{mj}}}$} & \multicolumn{2}{c}{$\mathbf{A_{\emph{hj}}}$} &
\multicolumn{2}{c}{$(\mathbf{k_{B}T_{\emph{l}}})_{\emph{j}}$} &
\multicolumn{2}{c}{$(\mathbf{k_{B}T_{\emph{m}}})_{\emph{j}}$} &
\multicolumn{2}{c}{$(\mathbf{k_{B}T_{\emph{h}}})_{\emph{j}}$} &
\multicolumn{2}{c}{$\mathbf{C_{\emph{j}}}$} & \multicolumn{2}{c}{$\mathbf{S_{1\emph{j}}}$} &\multicolumn{2}{c}{ $\mathbf{S_{2\emph{j}}}$} &
\multicolumn{2}{c}{$\langle MSE \rangle_{j}$} &
\multicolumn{2}{c}{$\sigma_{\langle MSE \rangle_{j}}$} &
\multicolumn{2}{c}{$\langle r^{2} \rangle_{j}$} &
\multicolumn{2}{c}{$\sigma_{\langle r^{2}\rangle_{j}}$}\\\toprule
\multicolumn{27}{c}{\textbf{EEDF3}} \\ \midrule
$j=1$ & \multicolumn{2}{c}{$[0.9672,1.0]$}	           &    \multicolumn{2}{c}{ $[0.0,0.0327]$}         &    \multicolumn{2}{c}{$[0.0,0.0013]$}   &	\multicolumn{2}{c}{$[0.0086,30]$}   &	  \multicolumn{2}{c}{$[0.7002,4.1957]$} &   
\multicolumn{2}{c}{$[9.0140,11.3505]$}   & \multicolumn{2}{c}{$0.45$}   &	\multicolumn{2}{c}{$0.35$} &  \multicolumn{2}{c}{ $0.75$}  &   \multicolumn{2}{c}{$5.32E{+}09$} &   \multicolumn{2}{c}{$1.75E{+}11$} &   \multicolumn{2}{c}{$0.9905$} &   \multicolumn{2}{c}{$0.1762$}   \\
$j=2$ & \multicolumn{2}{c}{$[0.9338,1.0]$}	           &     \multicolumn{2}{c}{$[0.0,0.0659]$}         &   \multicolumn{2}{c}{ $[0.0,0.002]$ }    &	\multicolumn{2}{c}{$[3.6,32.8]$}	     &     \multicolumn{2}{c}{$[0.7019,2.9986]$} & 
\multicolumn{2}{c}{$[9.0679,18.3101]$}   & \multicolumn{2}{c}{$0.25$}   &	\multicolumn{2}{c}{$0.35$} & \multicolumn{2}{c}{  $0.75$ } & \multicolumn{2}{c}{   $5.16E{+}12$}& \multicolumn{2}{c}{   $1.64E{+}13$}&  \multicolumn{2}{c}{  $0.9438$}&  \multicolumn{2}{c}{  $0.3784$ } \\
$j=3$ & \multicolumn{2}{c}{$[0.8853,0.9967]$}	&    \multicolumn{2}{c}{$[0.0033,0.1147]$  }   & \multicolumn{2}{c}{   $[0.0,0.0022]$ }   &\multicolumn{2}{c}{	$[8.9,46.3]$}     &\multicolumn{2}{c}{	  $[0.8192,2.1409]$} & \multicolumn{2}{c}{
$[9.0634,19.2783]$  } & \multicolumn{2}{c}{$0.25$ }   &\multicolumn{2}{c}{	$0.35$} &  \multicolumn{2}{c}{ $0.80$}  &    \multicolumn{2}{c}{$3.33E{+}13$}&   \multicolumn{2}{c}{ $6.52E{+}13$}&  \multicolumn{2}{c}{  $0.9755$}&  \multicolumn{2}{c}{  $0.0521$}  \\
$j=4$ &\multicolumn{2}{c}{ $[0.8468,0.9623]$	}&  \multicolumn{2}{c}{  $[0.0377,0.1533]$}     & \multicolumn{2}{c}{   $[0.0,0.0029]$ }   &	\multicolumn{2}{c}{$[13.5,45.7]$}   &	 \multicolumn{2}{c}{ $[0.9073,1.8902]$} & \multicolumn{2}{c}{
$[10.2056,20.3005]$ }& \multicolumn{2}{c}{$0.20$}   &	\multicolumn{2}{c}{$0.35$} & \multicolumn{2}{c}{  $0.85$}  &    \multicolumn{2}{c}{$1.11E{+}14$}&  \multicolumn{2}{c}{  $9.12E{+}14$}& \multicolumn{2}{c}{   $0.9818$}& \multicolumn{2}{c}{   $0.0189$ } \\
$j=5$ & \multicolumn{2}{c}{$[0.8143,0.9207]$	}& \multicolumn{2}{c}{   $[0.0775,0.1857]$}     & \multicolumn{2}{c}{   $[0.0,0.0034]$}    &	\multicolumn{2}{c}{$[19.3,54.3]$}   &	\multicolumn{2}{c}{  $[0.9738,1.6653]$ }& \multicolumn{2}{c}{
$[10.2431,14.1762]$} &\multicolumn{2}{c}{ $0.20$}   &	\multicolumn{2}{c}{$0.35$} & \multicolumn{2}{c}{  $0.85$}  &    \multicolumn{2}{c}{$1.35E{+}14$}& \multicolumn{2}{c}{   $1.13E{+}14$}& \multicolumn{2}{c}{   $0.9909$}& \multicolumn{2}{c}{   $0.0095$}  \\
$j=6$ & \multicolumn{2}{c}{$[0.7815,0.8670]$}	& \multicolumn{2}{c}{   $[0.1329,0.2185]$}     & \multicolumn{2}{c}{   $[0.0,0.0021]$}    &	\multicolumn{2}{c}{$[27.9,56.1]$}   & \multicolumn{2}{c}{     $[1.0185,1.5996]$} & \multicolumn{2}{c}{
$[10.4559,11.5533]$} & \multicolumn{2}{c}{$0.20$}   &	\multicolumn{2}{c}{$0.35$} & \multicolumn{2}{c}{  $0.85$}  &    \multicolumn{2}{c}{$1.26E{+}14$}& \multicolumn{2}{c}{   $1.15E{+}14$}& \multicolumn{2}{c}{   $0.9939$}& \multicolumn{2}{c}{   $0.0069$}  \\
$j=7$ & \multicolumn{2}{c}{$[0.7776,0.8252]$}	& \multicolumn{2}{c}{   $[0.1742,0.2222]$}     & \multicolumn{2}{c}{   $[0.0,0.0015]$ }   &	\multicolumn{2}{c}{$[35.4,54.7]$}   & \multicolumn{2}{c}{     $[1.1523,1.4958]$ }& \multicolumn{2}{c}{
$[11.0432,11.2715]$} & \multicolumn{2}{c}{$0.20$}   &	\multicolumn{2}{c}{$0.35$} & \multicolumn{2}{c}{  $0.85$}  &\multicolumn{2}{c}{    $1.61E{+}14$}& \multicolumn{2}{c}{   $8.39E{+}13$}& \multicolumn{2}{c}{   $0.9945$}& \multicolumn{2}{c}{   $0.0063$}  \\ 
\midrule
\multicolumn{27}{c}{\textbf{EEDF4}} \\ \midrule
$j=1$ & \multicolumn{2}{c}{$[0.9723,1.0]$}	           &    \multicolumn{2}{c}{ $[0.0,0.0276]$}         & \multicolumn{2}{c}{   $[0.0,0.0017]$ }  &\multicolumn{2}{c}{	$[0.0086,30]$}  &	\multicolumn{2}{c}{  $[1.004,6.5933]$} & \multicolumn{2}{c}{     $[9.6150,12.1072]$ }  & \multicolumn{2}{c}{$0.45$}  & \multicolumn{2}{c}{ $0.55$}  &  \multicolumn{2}{c}{$0.80$}  & \multicolumn{2}{c}{ $5.50E{+}09$}&    \multicolumn{2}{c}{$1.83E{+}11$}&  \multicolumn{2}{c}{  $0.9918$}&    \multicolumn{2}{c}{$0.1613$}  \\
$j=2$ & \multicolumn{2}{c}{$[0.9448,1.0]$}	           & \multicolumn{2}{c}{    $[0.0,0.0551]$}         & \multicolumn{2}{c}{   $[0.0,0.0023]$ }  &\multicolumn{2}{c}{	$[3.6,32.8]$	}    & \multicolumn{2}{c}{     $[1.1030,4.7121]$} & \multicolumn{2}{c}{  $[10.2769,20.7514]$}   & \multicolumn{2}{c}{$0.25$ } & \multicolumn{2}{c}{ $0.55$  }&  \multicolumn{2}{c}{$0.85$}  & \multicolumn{2}{c}{ $6.63E{+}12$}&    \multicolumn{2}{c}{$3.84E{+}13$}& \multicolumn{2}{c}{   $0.9588$}&    \multicolumn{2}{c}{$0.3739$}  \\
$j=3$ & \multicolumn{2}{c}{$[0.9032,0.9970]$}	&  \multicolumn{2}{c}{   $[0.0030,0.0968]$}   &  \multicolumn{2}{c}{  $[0.0,0.0029]$}   &\multicolumn{2}{c}{	$[8.9,46.3]$}     &	\multicolumn{2}{c}{  $[1.2873,3.3643]$ }& \multicolumn{2}{c}{  $[9.6034,19.2783]$}   & \multicolumn{2}{c}{$0.25$}  & \multicolumn{2}{c}{ $0.55$}  & \multicolumn{2}{c}{ $0.80$}  & \multicolumn{2}{c}{ $5.73E{+}13$}&    \multicolumn{2}{c}{$1.49E{+}14$}& \multicolumn{2}{c}{   $0.9734$}&    \multicolumn{2}{c}{$0.0889$}  \\
$j=4$ & \multicolumn{2}{c}{$[0.8718,0.9665]$}	& \multicolumn{2}{c}{$[0.0322,0.1285]$}   & \multicolumn{2}{c}{$[0.0,0.0030]$}   &	\multicolumn{2}{c}{$[13.5,45.7]$}   &\multicolumn{2}{c}{ $[1.4258,2.9704]$} &  \multicolumn{2}{c}{ $[9.6053,19.1064]$}   &\multicolumn{2}{c}{ $0.20$}  & \multicolumn{2}{c}{ $0.55$}  & \multicolumn{2}{c}{ $0.80$ } &\multicolumn{2}{c}{  $1.54E{+}14$}&    \multicolumn{2}{c}{$2.85E{+}14$}&  \multicolumn{2}{c}{  $0.9757$}&  \multicolumn{2}{c}{  $0.0494$} \\
$j=5$ & \multicolumn{2}{c}{$[0.8443,0.9336]$}	& \multicolumn{2}{c}{    $[0.0648,0.1560]$}   & \multicolumn{2}{c}{   $[0.0,0.0035]$}   &\multicolumn{2}{c}{	$[19.3,54.3]$ }  &	\multicolumn{2}{c}{  $[1.5303,2.6169]$} & \multicolumn{2}{c}{  $[9.6406,13.3423]$ }  &\multicolumn{2}{c}{ $0.20$}  &  \multicolumn{2}{c}{ $0.55$}      &   \multicolumn{2}{c}{$0.80$}  & \multicolumn{2}{c}{ $4.16E{+}14$}&    \multicolumn{2}{c}{$3.59E{+}14$}&  \multicolumn{2}{c}{  $0.9695$}&    \multicolumn{2}{c}{$0.0318$}  \\
$j=6$ &\multicolumn{2}{c}{$[0.8165,0.8882]$}	&  \multicolumn{2}{c}{   $[0.1096,0.1839]$}   & \multicolumn{2}{c}{   $[0.0,0.0027]$}   &	\multicolumn{2}{c}{$[27.9,56.1]$}   &  \multicolumn{2}{c}{    $[1.6005,2.5137]$} & \multicolumn{2}{c}{  $[9.8408,10.8737]$}   & \multicolumn{2}{c}{$0.20$}  & \multicolumn{2}{c}{ $0.55$}  & \multicolumn{2}{c}{  $0.80$ } & \multicolumn{2}{c}{ $1.22E+15$}&    \multicolumn{2}{c}{$7.16E{+}14$}& \multicolumn{2}{c}{   $0.9487$}&    \multicolumn{2}{c}{$0.0300$}  \\
$j=7$  & \multicolumn{2}{c}{$[0.7833,0.8338]$}	& \multicolumn{2}{c}{    $[0.1649,0.2172]$}   & \multicolumn{2}{c}{   $[0.0,0.0019]$}   &	\multicolumn{2}{c}{$[35.4,54.7]$}   & \multicolumn{2}{c}{     $[1.6461,2.1369]$} & \multicolumn{2}{c}{  $[10.3936,10.6085]$}   &\multicolumn{2}{c}{ $0.20$}  & \multicolumn{2}{c}{ $0.50$}  &   \multicolumn{2}{c}{$0.80$ } & \multicolumn{2}{c}{ $3.44E{+}15$}&    \multicolumn{2}{c}{$8.52E{+}14$}& \multicolumn{2}{c}{   $0.8764$}&    \multicolumn{2}{c}{$0.0376$}  \\
\toprule
\end{tabular}}
\end{adjustbox}
\end{table*}

\end{document}